\documentclass[a4paper, amsfonts, amssymb, amsmath, reprint, showkeys, nofootinbib, twoside]{revtex4-2}
\usepackage[english]{babel}
\usepackage{array}
\usepackage[utf8]{inputenc}
\usepackage{amssymb}
\usepackage{subcaption}
\usepackage[colorinlistoftodos, color=green!40, prependcaption]{todonotes}
\usepackage{amsthm}
\usepackage{mathtools}
\usepackage{physics}
\usepackage{xcolor}
\usepackage{graphicx}
\usepackage[left=23mm,right=13mm,top=35mm,columnsep=15pt]{geometry} 
\usepackage{adjustbox}
\usepackage{placeins}
\usepackage[T1]{fontenc}
\usepackage{lipsum}
\usepackage{csquotes}
\usepackage[pdftex, pdftitle={Article}, pdfauthor={Author}]{hyperref} 
\usepackage{graphicx}
\usepackage{upgreek}

\begin{document}
\title{\bf Numerical Simulation of Superparamagnetic Nanoparticle Motion in Blood Vessels for Magnetic Drug Delivery}

\author{Matthew Lee\textsuperscript{1},
Aditya Shelke\textsuperscript{2},
Saloni Singh\textsuperscript{3},
Jenny Fan\textsuperscript{4},
Philip Zaleski\textsuperscript{5},
Shahriar Afkhami\textsuperscript{5,6}}
    
\date{\today} 
\footnotetext[1]{East Brunswick High School, East Brunswick, NJ 08816, USA}
\footnotetext[2]{Middlesex County Academy for SMET, Edison, NJ 08837, USA}
\footnotetext[3]{High Tech High School, Secaucus, NJ 07094, USA}
\footnotetext[4]{Princeton Day School, Princeton, NJ 08540, USA}
\footnotetext[5]{Department of Mathematical Sciences, New Jersey Institute of Technology, Newark, NJ 07102, USA}

\footnotetext[6]{Correspondence email: shahriar.afkhami@njit.edu}
\begin{abstract}

A numerical model is developed for the motion of superparamagnetic nanoparticles in a non-Newtonian blood flow under the influence of a magnetic field. The rheological properties of blood are modeled by the Carreau flow and viscosity, and the stochastic effects of Brownian motion and red blood cell collisions are considered. The model is validated with existing data and good agreement with experimental results is shown. The effectiveness of magnetic drug delivery in various blood vessels is assessed and found to be most successful in arterioles and capillaries. A range of magnetic field strengths are modeled using equations for both a bar magnet and a point dipole: it is shown that the bar magnet is effective at capturing nanoparticles in limited cases while the point dipole is highly effective across a range of conditions. A parameter study is conducted to show the effects of changing the dipole moment, the distance from the magnet to the blood vessel, and the initial release point of the nanoparticles. The distance from the magnet to the blood vessel is shown to play a significant role in determining nanoparticle capture rate. The optimal initial release position is found to be located within the tumor radius in capillaries and arterioles to prevent rapid diffusion to the edges of the blood vessel prior to arriving at the tumor, and near the edge of the magnet when a bar magnet is used.
\end{abstract}

\keywords{Magnetic drug targeting, non-Newtonian blood flow, stochastic forces}

\maketitle

\section{Introduction}
\label{sec:introduction}

Magnetic drug delivery is a promising method for noninvasive, targeted treatment for certain diseases, including cancer \cite{mefford2007field, izci2021alternate, mody2014magnetic, liu2017magnetic}. This is achieved by injecting superparamagnetic iron oxide nanoparticles containing therapeutic drugs into the bloodstream and using an external magnet to guide these nanoparticles to the tumor site \cite{yue2012motion,rukshin2017modeling, grief2005mathematical, sharma2015mathematical, dadfar2019iron}. Superparamagnetic nanoparticles can orient themselves to an applied magnetic field, and randomly re-orient themselves once the field is removed due to their single domain structure \cite{rosensweig1985ferrohydrodynamics}. Once the magnetic field is removed, superparamagnetic nanoparticles are unable to agglomerate, avoiding vessel occlusion. Unlike current cancer treatments such as chemotherapy and radiation, which are systemic treatments that destroy both healthy and cancerous cells, magnetic drug delivery has the potential to target topical tumors and localize drug delivery without circulating throughout the body \cite{yue2012motion}. 

Recent reviews present several advantages of magnetic drug delivery \cite{dobson2006magnetic, mody2014magnetic, ulbrich2016targeted, liu2017magnetic,mukherjee2020recent,tran2017cancer}. 
Mody et al.~\cite{mody2014magnetic} highlight the practicality of magnetic drug delivery for targeting tumors due to its simplicity and adaptability to a broad range of applications. They note its potential to treat tumors in low-oxygen environments where traditional cancer treatments are not as effective, as well as its ability to minimize invasive procedures and reduce harmful effects to healthy tissues \cite{mody2014magnetic}. Liu et al.~also observe that magnetic drug delivery itself has no known side effects in the human body, although side effects may result from the specific drug used during treatment \cite{liu2017magnetic}. 
 
Clinical trials in mice have shown that using an external or subdermal magnet results in a higher concentration of drug-infused nanoparticles at the tumor site and a low concentration in healthy organs  \cite{kheirkhah2018magnetic, ak2018vitro, elbialy2015doxorubicin}. These \textit{in vivo} studies, however, either released nanoparticles in locations other than the bloodstream \cite{kheirkhah2018magnetic}, or far from the tumor site \cite{ak2018vitro, elbialy2015doxorubicin}, leading to lower capture rates. While these studies provide evidence that magnetic drug delivery can be effective, further studies are required to increase the nanoparticle capture rate before moving on to human clinical trials. 

However, many challenges must be overcome before magnetic drug delivery becomes a successful clinical treatment. Nanoparticle biocompatibility and potential long-term toxicity remains unstudied \cite{mody2014magnetic, ulbrich2016targeted}. Additional research on nanoparticle size, shape, stiffness, material, and surface coating is also necessary to optimize nanoparticle design and avoid rapid phagocytosis by leukocytes \cite{ye2018manipulating, ulbrich2016targeted, mukherjee2020recent}. Other obstacles include the need for a strong magnetic gradient, potential nanoparticle accumulation in smaller vessels and blocking blood flow, and the spatial geometry and limitations on the strength and gradient of the field \cite{dobson2006magnetic}. Above all, optimizing the conditions for magnetic drug delivery is critical for the success of the procedure and cannot easily be studied through \textit{in vivo} studies due to the extensive number of factors and practical considerations of cost and time. The current lack of optimal conditions for magnetic drug delivery leads studies to underestimate the procedure's effectiveness, and the variety of experimental designs for the procedure makes subsequent results difficult to compare directly. 

Past research has determined that external magnetic fields are unable to treat tumors within internal organs \cite{grief2005mathematical}. Similarly, nanoparticles released in large blood vessels such as the aorta are unable to withstand the comparatively vast distance and high blood velocity \cite{cherry2014comprehensive}. Therefore, superficial tumors, such as melanomas, that surround smaller vessels are the focus of this study. Fanelli et al.~\cite{fanelli2021magnetic} recently showed the commonly used Newtonian model overemphasizes the effect of the magnetic field, while the Carreau and Ellis models of non-Newtonian blood flow are more accurate. Previous research such as Yue et al.~\cite{yue2012motion} assumes blood to be a Newtonian fluid, while Rukshin et al.~\cite{rukshin2017modeling} considers various viscosity models but retains the traditional parabolic velocity profile; this research incorporates both viscosity and velocity components of non-Newtonian blood flow. Additionally, nanoparticles experience substantial levels of Brownian motion in the bloodstream, resulting in unstable and random trajectories \cite{yue2012motion}. Red blood cells (RBCs) and their collisions with nanoparticles also have significant impacts on nanoparticle trajectories \cite{rukshin2017modeling}. The extent of RBC collisions depends on various factors such as the particle volume fraction and the local shear rate \cite{zydney1987augmented}. Due to the importance of RBC collisions as shown in \cite{rukshin2017modeling} and the various factors that influence their magnitude \cite{zydney1987augmented}, we vary the amount of RBC collisions in our model. We also study individual particle trajectories, unlike \cite{grief2005mathematical, nacev2010magnetic} which use an advection-diffusion model.

This study uses a stochastic ordinary differential equation model to simulate nanoparticle trajectories in Carreau blood flow and to determine the optimal conditions for magnetic drug delivery. We consider the effects of Brownian motion, Stokes drag force, and varying levels of RBC collisions. We first validate our model with experimental results and show that it approximates nanoparticle motion well. We then examine nanoparticle motion in capillaries, arterioles, and arteries to determine which vessel types are best suited for the procedure. We consider the effects of changing the initial release point of the nanoparticles and the distance between the magnet and the vessel. Additionally, we study nanoparticle motion under magnetic fields generated by both a bar magnet and a much stronger point dipole to assess the effect of the magnetic field strength. This research combines key properties of blood flow into a single model and provides insight into how to optimize magnetic drug delivery for greatest nanoparticle capture, which can assist the design of future experimental studies and clinical trials.

\section{Governing Equations}

A schematic of our model is shown in Figure \ref{Schematic}. 
\subsection{Carreau Flow}

Here we model the blood vessel as a cylindrical pipe with radius $R$. Letting $\boldsymbol{u}$ denote the blood flow velocity and ignoring the effects that the particles have on the flow, we obtain the Navier--Stokes equations for an incompressible flow, 
\begin{equation}
\label{Navier}
    \rho \left [ \frac{d \boldsymbol{u}}{dt}+  (\boldsymbol{u} \cdot \nabla)\boldsymbol{u} \right ] =- \nabla p+ \nabla \cdot \boldsymbol{\uptau}
\end{equation}
\begin{equation}
    \nabla \cdot \boldsymbol{u} =0,
\end{equation}
where $\rho$ is fluid density, and $p$ is pressure. Furthermore, $\boldsymbol{\uptau}$ is the stress tensor defined as 
\begin{equation}
 \boldsymbol{  \uptau}= \eta(\dot{\gamma})   \boldsymbol{\dot{\gamma}},
\end{equation}
where $\eta$ is viscosity,
\begin{equation}
     \boldsymbol{\dot{\gamma}} =  \nabla \boldsymbol{u} +  \nabla\boldsymbol{u}^ T,
\end{equation}
and
\begin{equation}
    \dot{\gamma}= \left [\frac{1}{2} \left (\frac{\partial \boldsymbol{u}_i}{\partial x_j} +\frac{\partial \boldsymbol{u}_j}{\partial x_i} \right) \left (\frac{\partial \boldsymbol{u}_i}{\partial x_j} +\frac{\partial \boldsymbol{u}_j}{\partial x_i} \right) \right ]^{\frac{1}{2}}.
\end{equation}
To account for the non-Newtonian effects of blood, we use the Carreau viscosity model, which models both viscosity plateaus at high and low shear rates \cite{fanelli2021magnetic}. In the Carreau model, viscosity is given by 
\begin{equation}
    \eta(\dot{\gamma})=\eta_{\infty} + (\eta_{0}-\eta_{\infty}) \Big (1+ \lambda^2 \dot{\gamma}^2 \Big)^{ \frac{n_c-1}{2}},
\end{equation}
where $\eta_{\infty} $ is the viscosity at high shear rates, $\eta_0$ is the viscosity at low shear rates, $\lambda$ is the Carreau coefficient, and $n_c$ is the Carreau  exponent. Their values and those of other constants used throughout this study are given in Table \ref{table_of_constants}. Note that when $\eta_{\infty}=\eta_{0}$, we have that viscosity is constant and the model reduces to Newtonian flow.
\par
In addition, letting $\boldsymbol{u}= [u, v, w]^T$, we impose the following no-slip conditions at the boundary given by 
\begin{equation}
\boldsymbol{u}(0,0,-R)=\boldsymbol{u}(0,0,R)=[0, 0,0]^T,
\end{equation}
and we also prescribe the maximum flow velocity $u_{max}$ inside the  blood vessel which gives us the following condition 
\begin{equation}
\label{U_max_con}
    \boldsymbol{u}(0,0,0)=[ u_{max},0 ,0]^T.
\end{equation}
\par
Assuming a constant pressure gradient in the $x$ direction, i.e. $\boldsymbol{u}=[u, 0, 0]^T$ and converting the flow equations into polar coordinates, we obtain

\begin{equation}
    \dot{\gamma}= |u^\prime(r)|,
\end{equation}
and 

\begin{equation}
     \eta(\dot{\gamma})  =\eta_{\infty} + (\eta_{0}-\eta_{\infty}) \Big (1+ \lambda^2 \Big(u^\prime (r) \Big)^2  \Big)^{ \frac{n_c-1}{2}},
\end{equation}
where
\begin{equation}
        r= \sqrt{y^2+z^2}.
\end{equation}

Thus, from \eqref{Navier} we obtain the following Ordinary Differential Equation (ODE) for $u(r)$
\begin{multline}
\label{u_equ}
      \frac{\partial p}{ \partial x} = \Big (\eta_{\infty} \frac{u^\prime(r)}{r} + (\eta_{0}-\eta_{\infty} ) \frac{u^\prime(r)}{r} (1 + \lambda^2 (u^\prime(r))^2)^{\frac{n_c-1}{2}}  \Big) + \\  u^{\prime \prime}(r) \Big ( \eta_{\infty} +(\eta_{0} - \eta_{\infty}) (1 + \lambda^2 (u^\prime(r))^2)^{\frac{n_c-1}{2}}  +  \\ (\eta_{0} - \eta_{\infty}) (n_c-1) \lambda^2 (u^\prime(r))^2  (1 + \lambda^2 (u^\prime(r))^2)^{\frac{n_c-3}{2}} \Big),
\end{multline}
where $\frac{\partial p}{ \partial x} $ is a constant chosen to satisfy \eqref{U_max_con}.
\par
Lastly, note that when $\eta_{\infty}= \eta_0 =\eta $, viscosity is constant and we obtain the following parabolic flow
\begin{equation}
    u(r)= u_{max} \left (1 -\frac{r^2}{R^2} \right).
\end{equation}
\par
\begin{table}[]
    \centering
    \begin{tabular}{|c|c|m{3cm}|}
        \hline
        Notation & Value & Definition \\
        \hline
         $\mu_0$ & $4\times10^{-7}\pi$ N / m\textsuperscript{2} &  Permeability of free space \\
         \hline
         $k_b$ & $1.3806503 \times 10^{-23}$ J / K & Boltzmann constant \\
         \hline
         $T$  & $312$ K & Temperature\\
         \hline
         $a_p$ & $1\times 10^{-7}$ m & Nanoparticle radius 
         
         (unless otherwise specified)\\
         \hline
         $\chi$ & $0.2$ & Magnetic susceptibility\\
         \hline
         $r_{rbc}$ & $4\times10^{-6}$ m& RBC radius\\
         \hline
         $\eta_\infty$ & 0.00345 Pa s & Carreau viscosity at high shear rate \\
         \hline
         $\eta_0$ & 0.056 Pa s & Carreau viscosity at low shear rate \\
         \hline
         $\lambda$ & 3.313 & Carreau coefficient \\
         \hline
         $n_c$ & 0.3568 & Carreau exponent\\
         \hline
         $B_r$ & 1.48 T & Magnetic remanence\\
         \hline
         $2a$ & $0.018$ m & Bar magnet width\\
         \hline
         $2b$ & $0.08$ m & Bar magnet length \\
         \hline
         $L$ & $0.1$ m & Bar magnet height \\
         \hline 
         $r_t$ & 0.01 m & Tumor radius \\
        \hline
    \end{tabular}
    \caption{Table of Constants. Note that $\chi$, $\lambda$, $n_c$ are dimensionless. Values for $R$ and $u_{max}$ are given in Sec.~(\ref{results}) for the blood vessel radius and maximum flow velocity. Carreau constants are from \cite{fanelli2021magnetic}.}
    \label{table_of_constants}
\end{table}
\subsection{Magnetic Field }
In this study, we consider both the magnetic field generated by a magnetic dipole approximation denoted by $\boldsymbol{H}_d$ and the magnetic field generated by a bar magnet denoted by $\boldsymbol{H}_b$. For the dipole field, the external magnetic field potential around the nanoparticle  is given by 
\begin{equation}
    \phi_{d}(\boldsymbol{r_p}) = -\frac{1}{4 \pi} \frac{\boldsymbol{m} \cdot \boldsymbol{r_p}}{r_p^3},
\end{equation}
where $\boldsymbol{m}$ is the magnetic dipole moment, $\boldsymbol{r_p} $ is the vector from the dipole to the particle, and $r_p=|\boldsymbol{r_p}|$. From here we obtain the following magnetic field,
\begin{equation}
    \boldsymbol{H}_d(\boldsymbol{r_p}) = \nabla \phi_{d}=- \frac{1}{4 \pi} \frac{\boldsymbol{m}}{r_p^3} + \frac{3}{4 \pi} \frac{\boldsymbol{m} \cdot \boldsymbol{r_p}}{ r_p^4} \frac{\boldsymbol{r_p}}{r_p}.
    \label{dipoleMagneticField}
\end{equation}
In addition, the equations for the magnetic field $\boldsymbol{H}_b$ surrounding a bar magnet with  width  $2a$, length $2b$, height $L$, and residual magnetization $B_r$ are given in Appendix A. 
\subsection{Equations of Motion}
Here we let the vector $\boldsymbol{x}(t)$ denote the nanoparticle position in space.
The two main forces acting on the nanoparticle are the viscous drag force and the force from the magnetic field, which we denote by $\boldsymbol{F}_v$ and $\boldsymbol{F}_m$, respectively. We approximate the nanoparticle as a sphere of radius $a_p$, in which case the Stokes drag force is given by,
\begin{equation}
\label{Drag}
    \boldsymbol{F}_v= -D \boldsymbol{v}_s ,
\end{equation}
where 
\begin{equation}
    D= 6 \pi \eta(\boldsymbol{x})  a_p,
\end{equation}
and $\boldsymbol{v}_s$ is the slip velocity given by 
\begin{equation}
  \boldsymbol{  v}_s = \frac{d \boldsymbol{x}}{dt}- \boldsymbol{u}(\boldsymbol{x}).
\end{equation}

In general the magnetic force on an object in the external magnetic field $\boldsymbol{H}_e$ is given by
\begin{equation}
\label{F_M}
    \boldsymbol{F}_m= \int_{V} \mu_0 ( \boldsymbol{M} \cdot \nabla) \boldsymbol{H}_e dV.
\end{equation}
Here $V$ is the set occupied by the object and $\boldsymbol{M}=\chi \boldsymbol{H}$ is the magnetization, where $\chi$ denotes the magnetic susceptibility of the object (which we assume to be constant), and $ \boldsymbol{H}$  denotes the true field in the presence of the magnetic object. 
\par
Since the magnetic particles are small, the external magnetic field is approximately uniform throughout the particle. In this case, the magnetization inside the particle is given by the Clausius-Mossotti formula \cite{clausiusmossotti}

\begin{equation}
\label{magnetization}
    \boldsymbol{M}= \frac{3 \chi}{3+ \chi} \boldsymbol{H}_e.
\end{equation}

Thus, from \eqref{F_M} and \eqref{magnetization} we obtain that the magnetic force on a particle in an external magnetic field $\boldsymbol{H}_e$, is given by 
\begin{equation}
\label{F_mag}
    \boldsymbol{F}_m=\frac{4 \pi a_p^3 \mu_0 \chi}{3+ \chi}  (\boldsymbol{H}_e \cdot \nabla) \boldsymbol{H}_e.
\end{equation}
Now since the particle is small, we consider the force balance given by 
\begin{equation}
\label{Force_B}
    \boldsymbol{F}_v+\boldsymbol{F}_m=0 .
\end{equation}
Thus, from \eqref{Drag}, \eqref{F_mag}, and \eqref{Force_B} we obtain the following ODE,
\begin{equation}
    \frac{d \boldsymbol{x}}{dt}= \boldsymbol{u}(\boldsymbol{x})+ \frac{2a_p^2 \mu_0 \chi }{3 (3+ \chi )   \eta(\boldsymbol{x})}  (\boldsymbol{H}_e \cdot \nabla) \boldsymbol{H}_e.
\end{equation}
\par 

\par
\begin{figure*}[!htb]
    \centering
   \includegraphics[scale=0.35]{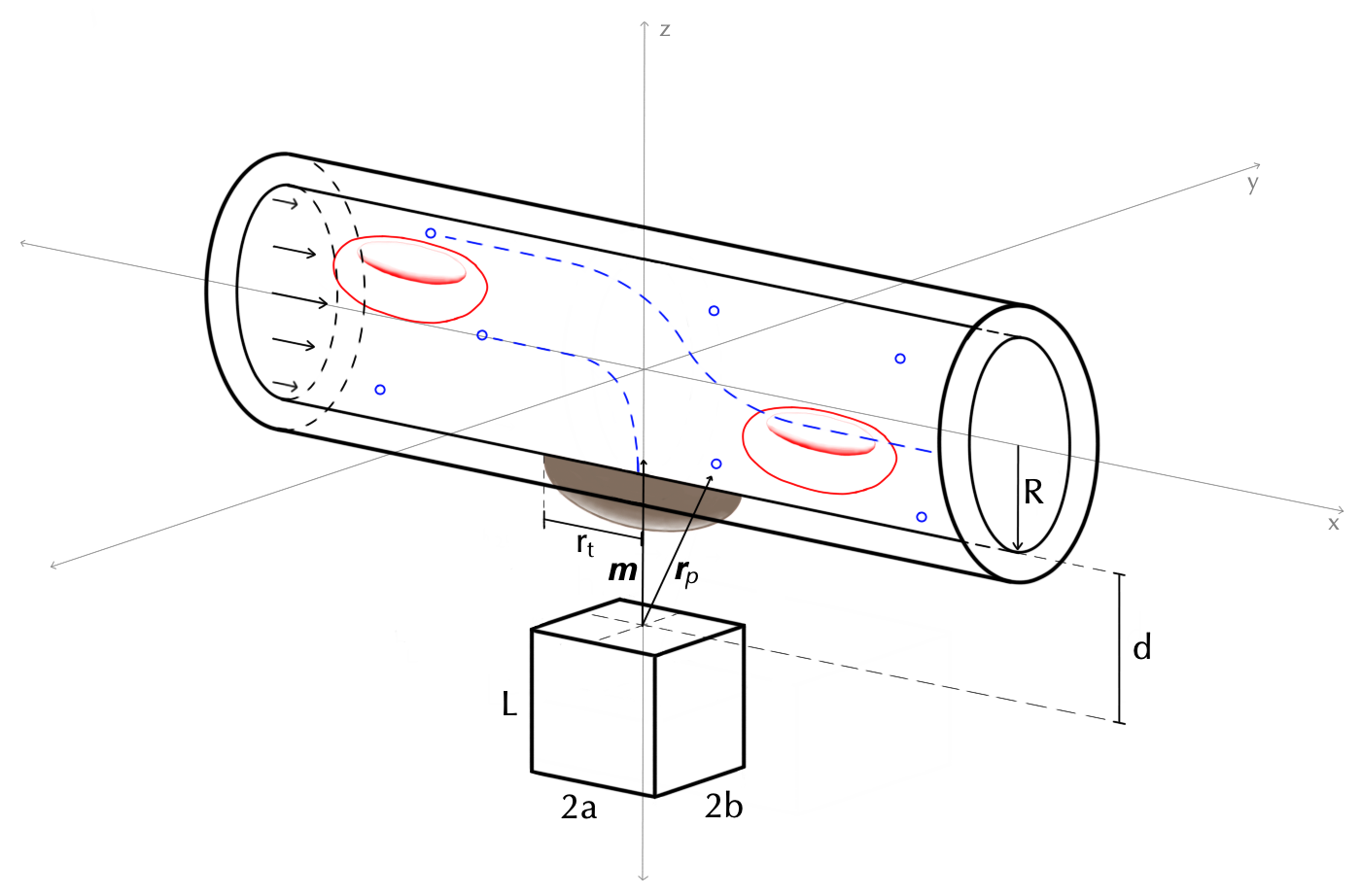}
    \caption{Schematic of both the bar magnet and point dipole model. The blood vessel with inner radius $R$ is centered around the $x$-axis with flow direction toward the positive $x$-axis. Relative blood flow velocity shown by the parallel black arrows. Nanoparticles represented by blue circles, with two sample trajectories shown by dashed lines. Two red blood cells provided for reference. The tumor is centered at $(0, 0, -R)$ with a radius of $r_t = 0.01$ m. In the bar magnet model, the magnet has width 
    $2a$, length $2b$ by height $L$ with the top of the magnet located at $(0, 0, -R -d)$, where $d$ is the distance from the inner vessel wall to the top of the magnet.  In the point dipole model, magnetic moment $\textit{\textbf{m}}$ points vertically along the the $z$ axis from the point dipole, located at $(0, 0, -R-d)$ where $d$ is the distance from the inner vessel wall to the point dipole, and $\textit{\textbf{r}}_p$ is the vector from the dipole to the particle.
    }
    \label{Schematic}
    
\end{figure*}

 
Finally, to account for Brownian motion and for the shear induced diffusion that the red blood cells (RBC) create \cite{grief2005mathematical}, we add the following stochastic term,

\begin{multline}
\label{Gov_eq}
         \frac{d \boldsymbol{x}}{dt}= \boldsymbol{u}(\boldsymbol{x})+ \frac{2a_p^2 \mu_0 \chi }{3 (3+ \chi )   \eta(\boldsymbol{x})}  (\boldsymbol{H}_e \cdot \nabla) \boldsymbol{H}_e+ \\ \sqrt{ \frac{2 k_b T}{ D} + 2 K_{sh} r_{rbc}^2 \dot{\gamma}} \frac{ \boldsymbol{N}(0,1)}{ \sqrt{ d t }},
\end{multline}
where $k_b$ is the Boltzmann constant, $T$ is temperature, $r_{rbc}$ is the red blood cell radius, $K_{sh}$ is a physical parameter, and $\boldsymbol{N}(0,1)$ is a vector of three independently generated standard normal random variables. Note that in the dipole model we have that 
\begin{equation}
    \boldsymbol{H}_e(x,y,z)= \boldsymbol{H}_{d}(x,y,z+ d+R),
\end{equation}
and in the bar magnet model we have that 
\begin{equation}
    \boldsymbol{H}_e(x,y,z)= \boldsymbol{H}_{b}(x,y,z+ d+R),
\end{equation}
where in the dipole case $d$ is the distance from the center of the dipole to $(0,0,-R)$, and in the bar magnet case $d$ is the distance from the top of the magnet to $(0,0,-R)$ (see Figure \ref{Schematic}). 
\par

From above, we nondimensionalize by letting,
\begin{equation}
  \boldsymbol{\Bar{x}}  = \frac{\boldsymbol{x}}{R},
\end{equation}
and 
\begin{equation}
    \Bar{t}=\frac{u_{max}}{R} t.
\end{equation}
In the dipole case, with this nondimensionalization, we obtain the following equation
\begin{equation}
    \frac{d \Bar{\boldsymbol{x}}}{d \Bar{t}} = \Bar{\boldsymbol{u}}(\Bar{\boldsymbol{x}})+ \frac{C_D}{\bar{\eta}(\bar{\boldsymbol{x}})} (\Bar{\boldsymbol{H}_e} \cdot \nabla ) \Bar{\boldsymbol{H}_e}+ \sqrt{ \frac{C_T}{ \Bar{\eta}(\bar{\boldsymbol{x}})} +C_{rbc} \Bar{\dot{\gamma}} } \frac{ \boldsymbol{N}(0,1)}{ \sqrt{ d \Bar{t}}},
\end{equation}
where 
\begin{equation}
\Bar{\boldsymbol{H}_e} (\boldsymbol{\bar{x}})=\Bar{\boldsymbol{H}_d}
    \left(\bar{x},\bar{y},\bar{z}+ \frac{d}{R}+1 \right),  
\end{equation} 
\begin{equation}
  \Bar{\boldsymbol{H}_d} (\boldsymbol{\bar{x}})=  - \frac{1}{4 \pi} \frac{\boldsymbol{m}}{m |\boldsymbol{\bar{x}}|^3} + \frac{3}{4 \pi} \frac{\boldsymbol{m} \cdot \boldsymbol{\bar{x}}}{ m |\boldsymbol{\bar{x}}|^4} \frac{\boldsymbol{\Bar{x}}}{|\boldsymbol{\bar{x}}|},
\end{equation}
where $m=\boldsymbol{|m|}$ and
\begin{equation}
    C_D= \frac{2 a_p^2 \mu_0 \chi m^2}{3 (3+ \chi)  R^7 \eta_\infty u_{max}},
\end{equation}
\begin{equation}
    C_T= \frac{k_b T}{3 \pi a_p \eta_{\infty} R u_{max}},
\end{equation}
\begin{equation}
    C_{rbc}= \frac{2 K_{sh} r_{rbc}^2}{R^2},
\end{equation}
\begin{equation}
    \Bar{\eta}(\Bar{\dot{\gamma}})=1 + (\frac{\eta_0}{\eta_{\infty}}-1)(1+ \frac{\lambda^2 u_{max}^2}{R^2} \Bar{\dot{\gamma}}^2)^{\frac{n_c-1}{2}},
\end{equation}
\begin{equation}
    \Bar{\dot{\gamma}}= |\Bar{u}^\prime(\Bar{r})|,
\end{equation}
and 
\begin{equation}
    \boldsymbol{ \Bar{u}}(\Bar{r})=\frac{ \boldsymbol{ u}(R \bar{r})}{u_{max}}.
\end{equation}
\par

\section{Numerical Methods}
In this section we give a brief overview of the numerical methods used to solve the governing equations. Firstly, to solve for the velocity profile given in \eqref{u_equ} we use Matlab's bvp5c function. Furthermore, the constant pressure gradient is an unknown, which is determined by using a bisection method to satisfy \eqref{U_max_con}. 
\par 
In addition, the stochastic differential equation given in \eqref{Gov_eq}, can be expressed in the general format as 
\begin{equation}
\label{general_equ}
    d \boldsymbol{x}=f(\boldsymbol{x}) dt + g(\boldsymbol{x}) d \boldsymbol{W}_t,
\end{equation}
where $\boldsymbol{W}_t$ is the standard Wiener process, i.e. $ d \boldsymbol{W}_t$ is a vector of three non-correlated random numbers $\sqrt{d t} N(0, 1)$,
\begin{equation}
  f(\boldsymbol{x}) =  \boldsymbol{u}(\boldsymbol{x})+ \frac{2a_p^2 \mu_0 \chi }{3 (3+ \chi )   \eta(\boldsymbol{x})}  (\boldsymbol{H}_e \cdot \nabla) \boldsymbol{H}_e,
\end{equation}
and
\begin{equation}
    g(\boldsymbol{x}) =  \sqrt{ \frac{2 k_b T}{ D} + 2 K_{sh} r_{rbc}^2 \dot{\gamma}}.
\end{equation}
Lastly, to solve \eqref{general_equ} we use the standard first order Euler–Maruyama method \cite{picchini2007sde, yue2012motion}.

\section{Model Validation}
In this section we compare our model with the experimental results given in \cite{lim2011magnetophoresis}. In \cite{lim2011magnetophoresis}, Lim et al.~placed superparamagnetic nanoparticles in deionized water with no background flow and analyzed the balance between the Brownian motion and the magnetic force. To attract the nanoparticles, Lim et al.~used a magnetic tweezer, which generates a large magnetic field gradient of up to $3000$ T/m at the tweezer tip. Because of this, our bar magnet model does not apply and we only use the dipole approximation. 
\par
Since there was no background flow or red blood cell collisions in the experiments given in \cite{lim2011magnetophoresis}, our dipole model in equation \eqref{Gov_eq} reduces to 
\begin{equation}
     \frac{d \boldsymbol{x}}{dt}=  \frac{2a_p^2 \mu_0 \chi }{3 (3+ \chi )   \eta}  (\boldsymbol{H}_e \cdot \nabla) \boldsymbol{H}_e+ \\ \sqrt{ \frac{2 k_b T}{ D} } \frac{ \boldsymbol{N}(0,1)}{ \sqrt{ d t }},
\end{equation}
where 
\begin{equation}
    \boldsymbol{H}_e(\boldsymbol{x})= - \frac{1}{4 \pi} \frac{\boldsymbol{m}}{|\boldsymbol{x}|^3} + \frac{3}{4 \pi} \frac{\boldsymbol{m} \cdot \boldsymbol{x}}{ |\boldsymbol{x}|^4} \frac{\boldsymbol{x}}{|\boldsymbol{x}|}.
\end{equation}
Note that here we set $d= 0$ m, $\eta =0.00089 $ Pa s, $a_p=18.8$ nm,  $\chi= 0.2$, and $T=298$ K to match the parameters given in \cite{lim2011magnetophoresis}. Furthermore, we let the magnetic dipole moment $\boldsymbol{m}=[0, 0, 4.5 \times 10^{-7}]$  A m\textsuperscript{2}, which was obtained by matching the field gradient given in  \cite{lim2011magnetophoresis} at a distance of $100$ $\mu$m from the tweezer tip. 
\par
\begin{figure}
   \centering
    \includegraphics[scale=.65]{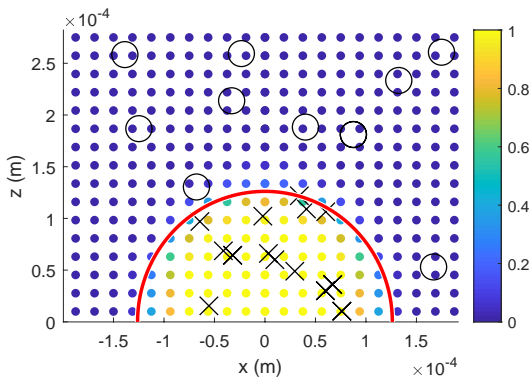}
    \caption{The probability of hitting the magnetic dipole is plotted for varying $x$ and $z$ values. Yellow represents a capture rate of one and blue represents a capture rate of zero. The experimental data from \cite{lim2011magnetophoresis} is also plotted, where X stands for a target hit and O stands for a target miss. Lastly, the radius of the red circle is given analytically by \eqref{R_max}. }
    \label{Verification}
\end{figure}

 
Since the magnetic field surrounding the dipole is rotationally symmetric we only look at a two dimensional slice of the problem and set $y=0$. We then run $100$ simulations to calculate the probability of the nanoparticle hitting the target after $25$ s. In Figure \ref{Verification}, we vary $x$ and $z$ and plot the probability of hitting the target on a color scale. Furthermore, the experimental data is also plotted where an X stands for a target hit and an O stands for a target miss. Here we see a good agreement between the experimental data and the numerical data. 
\par
Furthermore, from Figure \ref{Verification},  we see a region of attraction that lies inside the red circle. An approximation to the radius of this region of attraction can be expressed analytically by considering the problem in the absence of Brownian motion, which is given by    
\begin{equation}
\label{No_brown}
     \frac{d \boldsymbol{x}}{dt}=  \frac{2a_p^2 \mu_0 \chi }{3 (3+ \chi )   \eta}  (\boldsymbol{H}_e \cdot \nabla) \boldsymbol{H}_e.
\end{equation}
Moreover, if we assume that $x=0$ and $y=0$, \eqref{No_brown} reduces to a single ODE for $z(t)$, and for a travel time of $t_{tr}$ the maximum initial $z$ value i.e. $r_{max}$ that still reaches the target is given by

\begin{equation}
\label{R_max}
    r_{max}=\left ( \frac{4 m^2 \mu_0 \chi a_p^2 t_{tr}}{(3 + \chi) \pi^2 \eta } \right) ^{\frac{1}{8}}.
\end{equation}

\section{Results and Discussion}
\label{results}
We present the following numerical analysis by solving the equations of motion. We focus on capillaries, arterioles, and arteries because they transport drugs to the tumor site, whereas venules and veins transport drugs away from the tumor site, making them ineffective for drug delivery. We use experimentally found constants for the maximum blood velocity and vessel radii in humans: $R = 4 \times 10^{-6}$ m and $u_{max} = 9.3 \times 10^{-4}$ m/s in capillaries, $R = 1.5 \times 10^{-5}$ m and $u_{max} = 3.26 \times 10^{-3}$ m/s in arterioles, and $R = 2 \times 10^{-3}$ m and $u_{max} = 0.19$ m/s in arteries \cite{muller2008high, stucker1996capillary, wang2016vessel, klarhofer2001high}. 

We define a capture as a particle whose trajectory terminates within the radius of the tumor and on the bottom half of the pipe (i.e $z < 0$). We take advantage of the vessel symmetry and release 100 particles on the right half of the pipe for each cross section. The capture rate is the number of particles captured divided by the total number of particles released. We assume no-slip conditions and that particles that contact the vessel wall are immediately absorbed, since movement after the particle contacts the wall occurs on a timescale of $10^3$ s \cite{yue2012motion}. 

\subsection{Motion in the Absence of a Magnetic Field}
\label{absentMagnet}
\begin{figure*}[!htb]
    \centering
    \includegraphics[scale = 0.14]{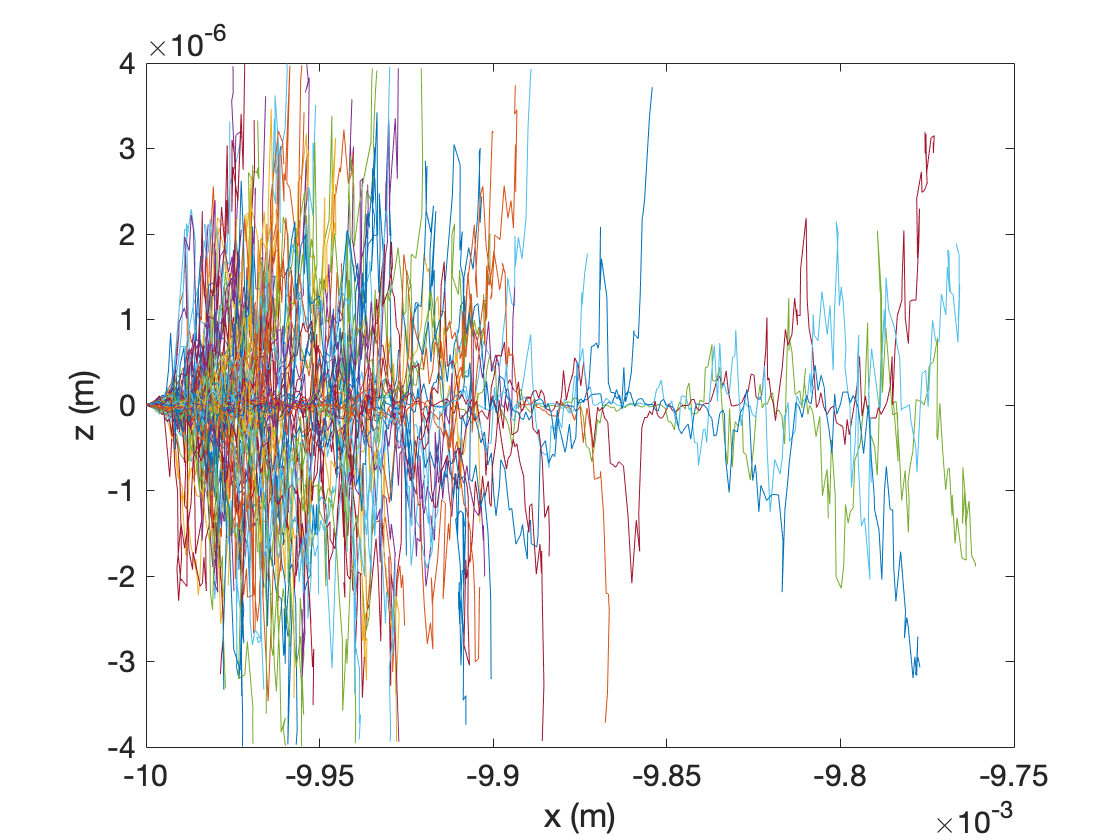}
    \includegraphics[scale = 0.14]{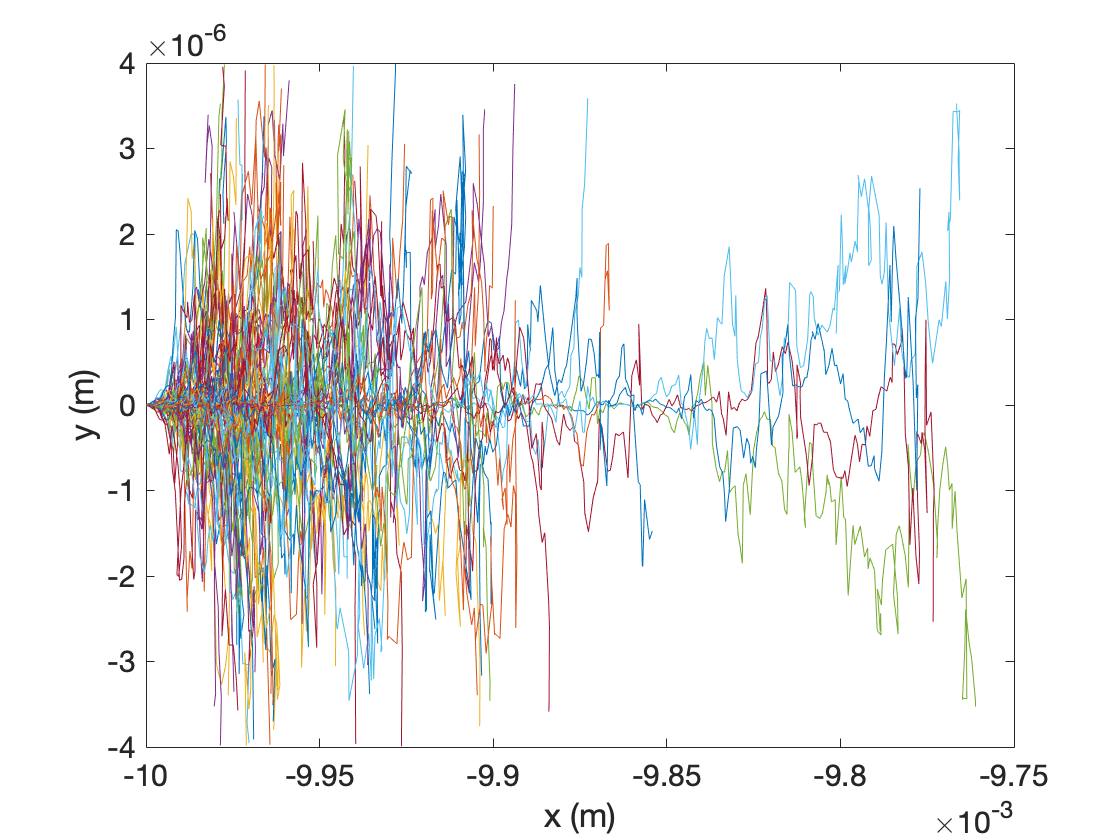}
    \includegraphics[scale = 0.14]{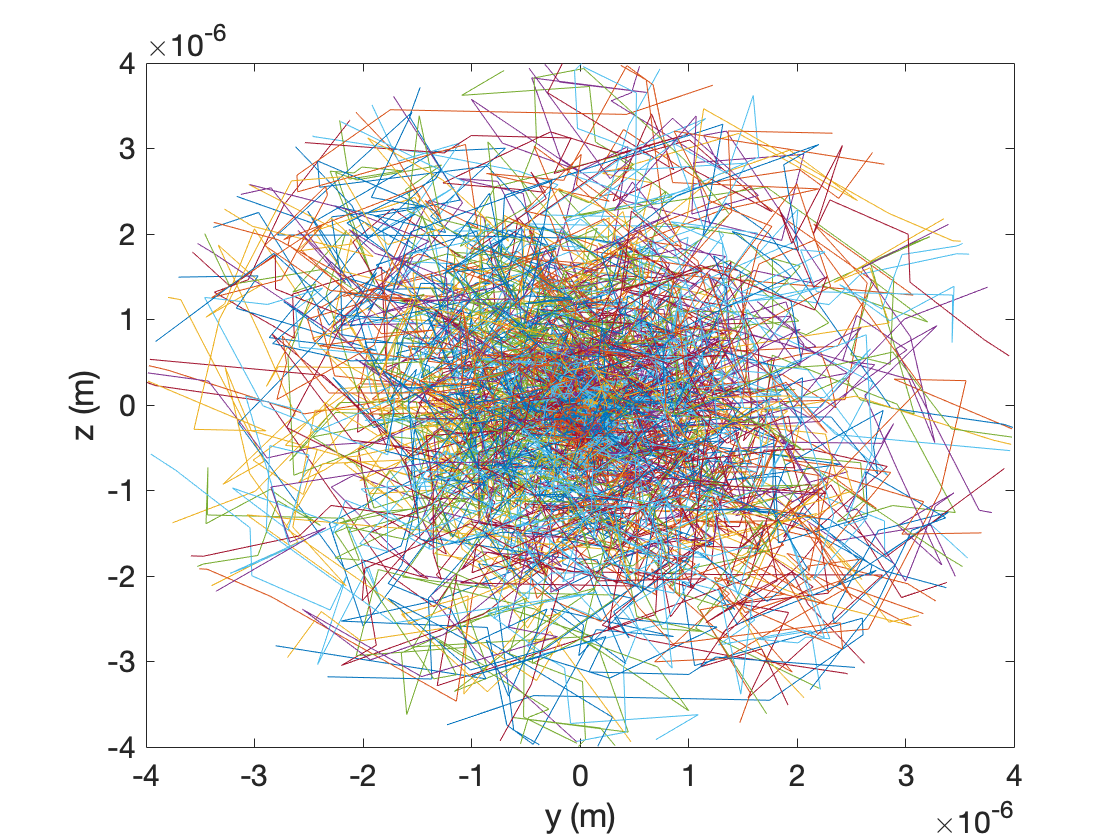}
    \caption{Particle trajectories along the $x-z$, $x-y$, and $y-z$ planes in the absence of a magnetic field in capillaries. Note the horizontal drift of particles is minimal and diffusion to the edges of the pipe is extremely rapid. Particles are released from $(-0.01, 0, 0)$; 100 trajectories are shown. $R = 4 \times 10^{-6}$ m and $u_{max} = 9.3 \times 10^{-4}$ m/s. $K_{sh} = 0.05$.}
    \label{NoMagnet}
\end{figure*}

In the absence of a magnetic field, particle diffusion due to Brownian motion and RBC collisions occurs extremely quickly, and the majority of particles released inside capillaries hit the wall within 0.01 cm as shown in Figure \ref{NoMagnet}. Similar results are observed in arterioles, but over a distance of 2 cm (not shown). This makes it exceedingly difficult to direct particles over long distances due to diffusion. However, if particles are released within the radius of the tumor, diffusion will lead to a capture rate of 0.5 even in the absence of a magnetic field. In capillaries and arterioles, the stochastic forces acting on the nanoparticles are enough to cause significant diffusion such that half the particles hit the bottom half of the vessel and the tumor, while the other half diffuses upwards and hit the top half of the vessel, missing the tumor. Although a capture rate of 0.5 is a direct consequence of our capture definition, Figure \ref{NoMagnet} shows that diffusion causes particles to hit the vessel wall extremely quickly and that nanoparticle endpoints are evenly distributed, a result independent of the capture definition. A previous review of nanoparticle delivery found that a median value of 0.7\% of the nanoparticle dose reached the tumor target \cite{wilhelm2016analysis}, but did not consider where particles were released; these results show that nanoparticle delivery can potentially be much more effective if particles are released within the tumor radius, even in the absence of a magnetic field.  Particle motion in arteries in the absence of a magnetic field is discussed in Sec.~(\ref{arteries}). 

\subsection{Point Dipole Model}
\label{Dipole Moment}
We first consider the magnetic field $\boldsymbol{H}_d$ generated by a point dipole. We use this model to approximate strong magnetic fields comparable to those generated by an electromagnet. Weaker magnetic fields generated by physical bar magnets are considered in Sec.~(\ref{Bar Magnet}). The dipole is located at  $(0, 0, -R-d)$, where the point dipole is placed some distance $d$ below the vessel. The tumor is centered at $(0, 0, -R)$ and has a radius $r_t$ of 0.01 m. We conduct a parameter study by varying the dipole moment $m$, the distance from the magnet to the vessel $d$, and the $x$ coordinate of the initial release position within the tumor radius from $-200R$ to $100R$. We discuss the effect of releasing particles outside the tumor radius in Sec.~(\ref{initRelease}).


\subsubsection{Nanoparticle Motion in Capillaries}
\paragraph{Motion with RBC Collisions}

\begin{figure}
    \centering
    \includegraphics[scale=0.2]{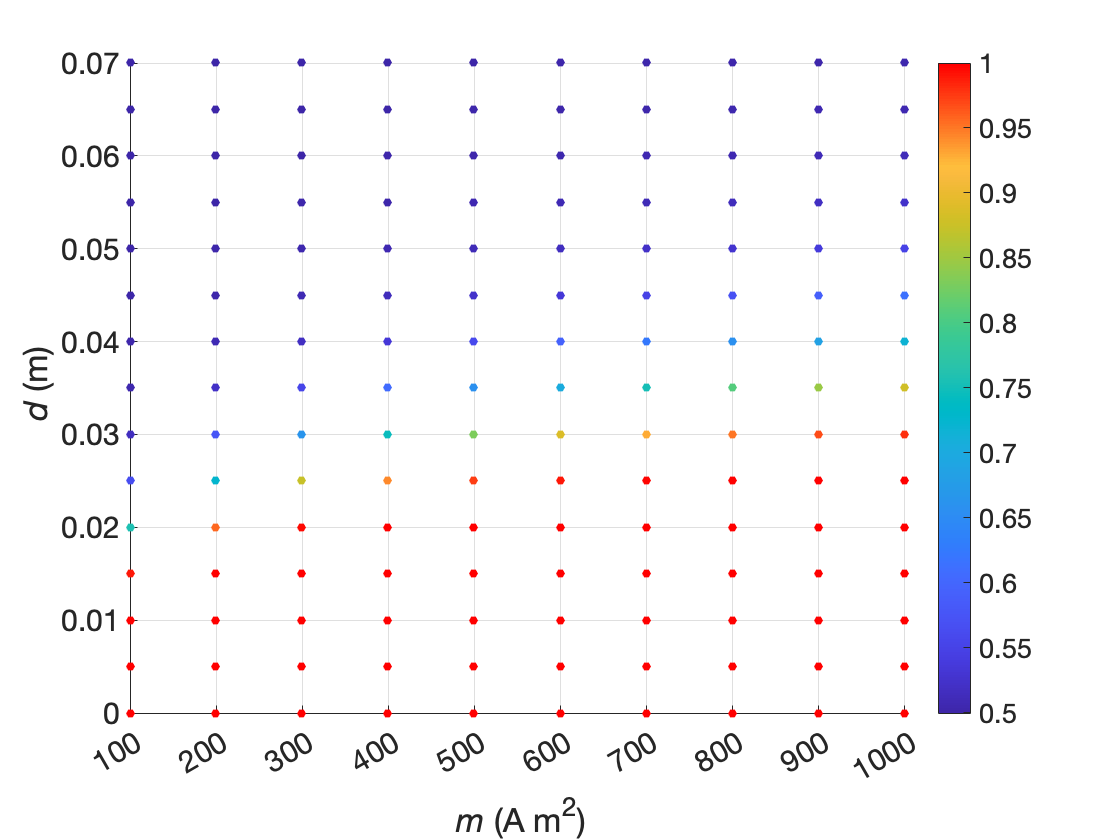}
    \caption{Capture rates in capillaries at various magnetic moments and distances from the blood vessel. Colorbar represents the capture rate, where red represents a capture rate of 1 and blue represents a capture rate of 0.5. Capture rates remain the same along multiple release points (not shown). $K_{sh} = 0.05$.}
    \label{Ksh0.05Cap}
\end{figure}

The results of changing the initial release point, the dipole moment, and the distance from the vessel on the capture rate in a capillary are shown in Figure \ref{Ksh0.05Cap}. The results show the distance from the magnet has a larger impact on capture rate than the dipole moment. From \eqref{dipoleMagneticField}, it is evident the magnetic field strength relies more heavily on $r_p$ than on $m$, and thus smaller values of $d$ correspond to lower values of $r_p$ and higher magnetic field strengths. This high magnetic field strength causes the nanoparticles to be pulled quickly downward and hit the target with minimal influence of background blood flow. At distances less than $1.5$ cm, $r_p^3$ is extremely small, so the capture rate remains at $1$ regardless of the dipole moment due to extremely high magnetic field strengths. Furthermore, this explains the wide variation in capture rate at different distances from the blood vessel and limited variation in capture rate across increasing dipole moments. 

As $d$ increases to $5.0$ cm, the capture rate across all dipole moments  below $1000$ A m\textsuperscript{2} decreases to $0.5$. 
This rapid decrease in capture rate across all dipole moments reflects the rapid decrease in the magnetic force as $d$ increases and the increased impact of diffusion. This sensitivity indicates that small changes in distance can lead to large changes in capture rate, making precision in magnet placement critical to the success of magnetic drug delivery. These results also indicate that large magnetic moments are not necessary to achieve high capture rates, but instead the effectiveness of magnetic drug delivery is determined primarily by the distance between the magnet and the blood vessel. Clinically, this means the procedure can be done with weaker electromagnets as long as the magnet can be placed within $1.5$ cm of the tumor. However, if the magnet must be placed farther away from the tumor, stronger electromagnets are capable of sustaining high capture rates up until distances of $5.0$ cm. 

Interestingly, the capture rate does not vary with the initial release position of the particles when the dipole moment and distance from magnet to vessel are kept constant. This is partly due to $r_{tumor} \gg R$ so that at distances of $-200R$ and $100R$, the particles are still released within the radius of the tumor. Due to the small radius of capillaries and weak background blood flow, particles are pulled directly downwards by the strong magnetic field and do not travel very far forwards before hitting the vessel wall. Downstream of the magnet, the magnetic field strength counteracts the background blood flow and actually leads to some retrograde particle motion. In this case, no particles are swept away by the background blood flow and miss the target; this is crucial to minimizing unintended side effects of drugs on the rest of the body.

\paragraph{Motion in the Absence of RBC Collisions}

Although red blood cell collisions are significant for larger blood vessels, their effects are greatly diminished in capillaries. Due to the extremely small radius of capillaries, red blood cells are forced to undergo slight deformation and pass through the capillary single file. Thus, particle motion in the capillary can be alternatively considered as constrained to a pipe with a length equivalent to the distance between two red blood cells and devoid of collisions. We consider this case by setting $K_{sh} = 0$ to remove red blood cell collisions, although Brownian motion is still considered.

\begin{figure}
    \centering
   \includegraphics[scale=0.2]{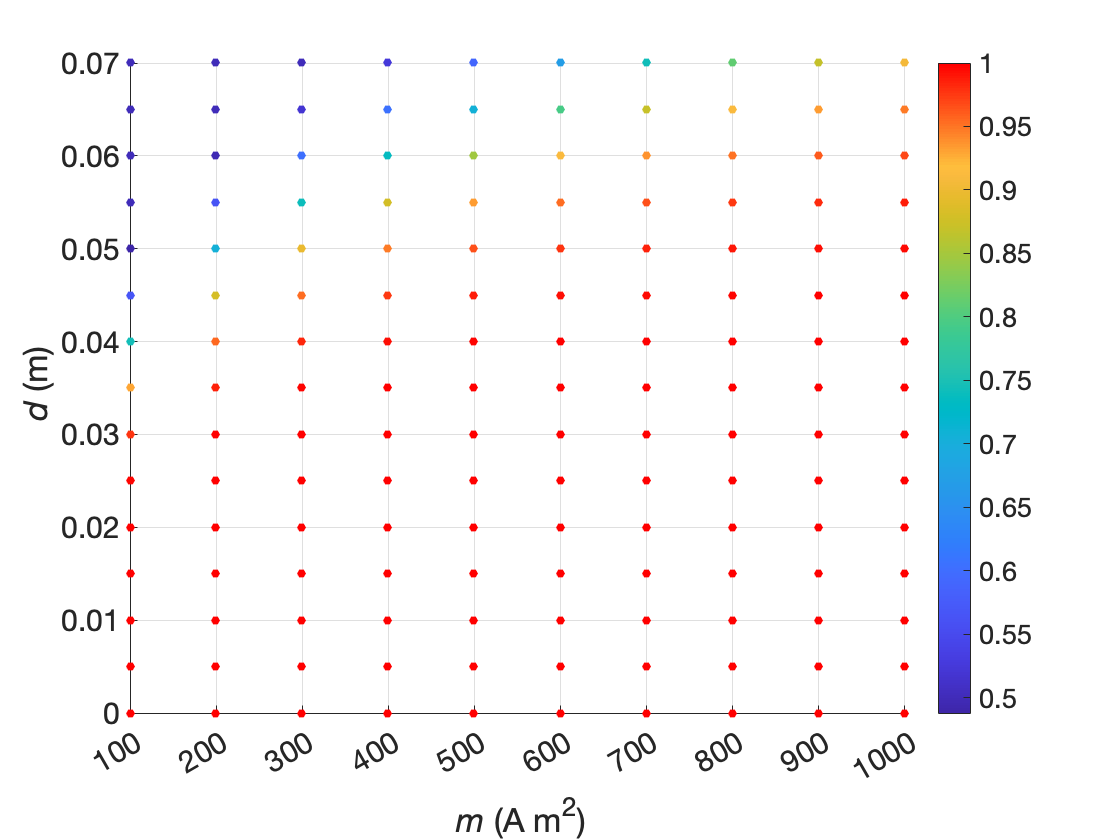}
    \caption{Capture rates in capillaries at various magnetic moments and distances from the blood vessel in the absence of RBC collisions. Colorbar represents the capture rate probability, where red represents capture rate of 1 and blue represents a capture rate of 0.5. Capture rates remain the same along multiple release points (not shown). $K_{sh} = 0$.}
    \label{NoKshCap}
    
\end{figure}

The results in Figure \ref{NoKshCap} show the effect of setting $K_{sh} = 0$ on capture rate. In this case, the capture rate remains at 1 for all magnetic moments until a distance of 3 cm. Additionally, for different magnetic moments, there is substantial variation in the maximum distance at which capture rate remains at 1. At a magnetic moment of 1000 A m\textsuperscript{2}, a capture rate of 1 can be achieved at a distance of 6.0 cm from the blood vessel, compared to a distance of 3.0 cm for a magnetic moment of 100 A m\textsuperscript{2}. This variation across magnetic moments is due to the absence of RBC collisions, leading to weaker magnetic fields to continue influencing particle motion and causing them to hit the tumor. When RBC collisions are included, the stochastic forces compete with the magnetic force and cause particles to diffuse towards the vessel wall instead of being pulled towards the tumor, causing capture rate to begin decreasing at lower distances.

Nevertheless, the distance between the magnet and the blood vessel remains the most critical factor for capture rate: similar to our previous results, there is a sensitive area where capture rate rapidly decreases from 1 to 0.5. Furthermore, the capture rates remain identical for particles released at various distances away from the tumor. This is because the magnetic field strength over the range of the tumor is high compared to the background blood flow, minimizing particle drift in the $x$ direction. 

The results shown here and in the previous section depict two extremes in which RBC collisions are either present or completely absent. Due to the complex physiology of blood flow and other factors not considered in our model such as RBC deformation in capillaries and varying RBC concentrations, the reality likely lies in between these two scenarios. We study this in the following section by analyzing changes in capture rate as a result of varying the $K_{sh}$ parameter.
\subsubsection{Nanoparticle Motion with Varying \texorpdfstring{$K_{sh}$}{Lg} }

\begin{figure}
    
    \includegraphics[scale = 0.2]{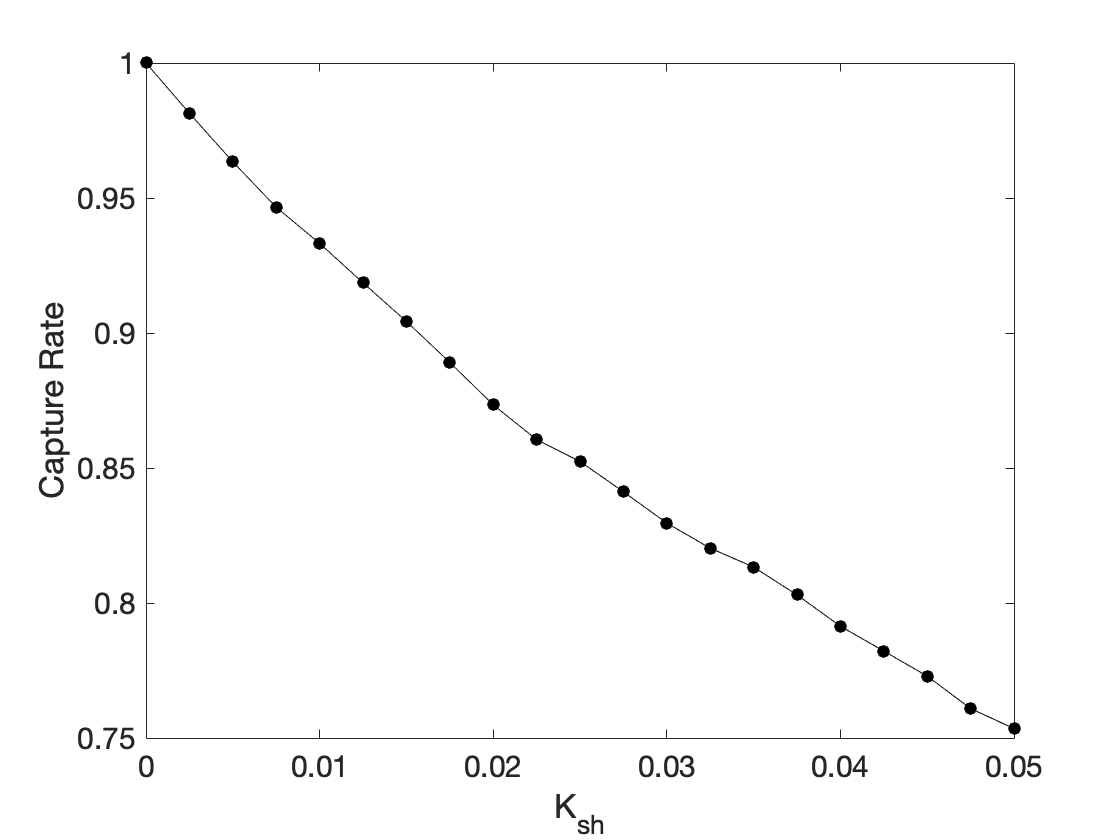}
    \caption{Capture Rate in capillaries as a function of changing $K_{sh}$. Particles are released from $-100R$ from 100 points in the cross section. $m = 700$ A m\textsuperscript{2}, $d = 3.5$ cm.}
    \label{KshChange}
\end{figure}

Figure \ref{KshChange} shows the effect of changing the $K_{sh}$ value on capture rate. The results show capture rate decreases linearly with $K_{sh}$. At high $K_{sh}$ values, greater stochasticity from RBC collisions causes many particles to diffuse towards the top of the vessel faster than the magnetic field pulls particles down towards the tumor, resulting in lower capture rates. At low $K_{sh}$ values, the magnetic field exerts a greater influence on particle trajectories due to less competing stochastic forces. Although these results are calculated for a fixed dipole moment, magnet distance, and initial release position, we expect this relationship to hold for similar values in the area where capture rate decreases quickly with distance. Points in this area are extremely sensitive to small parameter changes, making the $K_{sh}$ term very important. However, we do not expect this relationship to be present at extreme distances when the magnet is placed either very close or very far from the vessel. When the magnet is close to the vessel ($<1.5$ cm), the magnetic force is much greater than the stochastic force, causing particle trajectories to be unaffected by minor changes in $K_{sh}$. Similarly, when the magnet is far from the vessel ($>5$ cm), the magnetic force is much weaker than the stochastic force and particles already exhibit pure diffusion. While increases in $K_{sh}$ may shorten the time it takes for particles to diffuse, the capture rate will still remain at 0.5. Therefore, the value of $K_{sh}$ will have major effects on capture rate only in the sensitive area, while capture rates will remain high when the magnet distance is small.

\begin{figure}[!htb]
    \centering
   \includegraphics[scale=0.2]{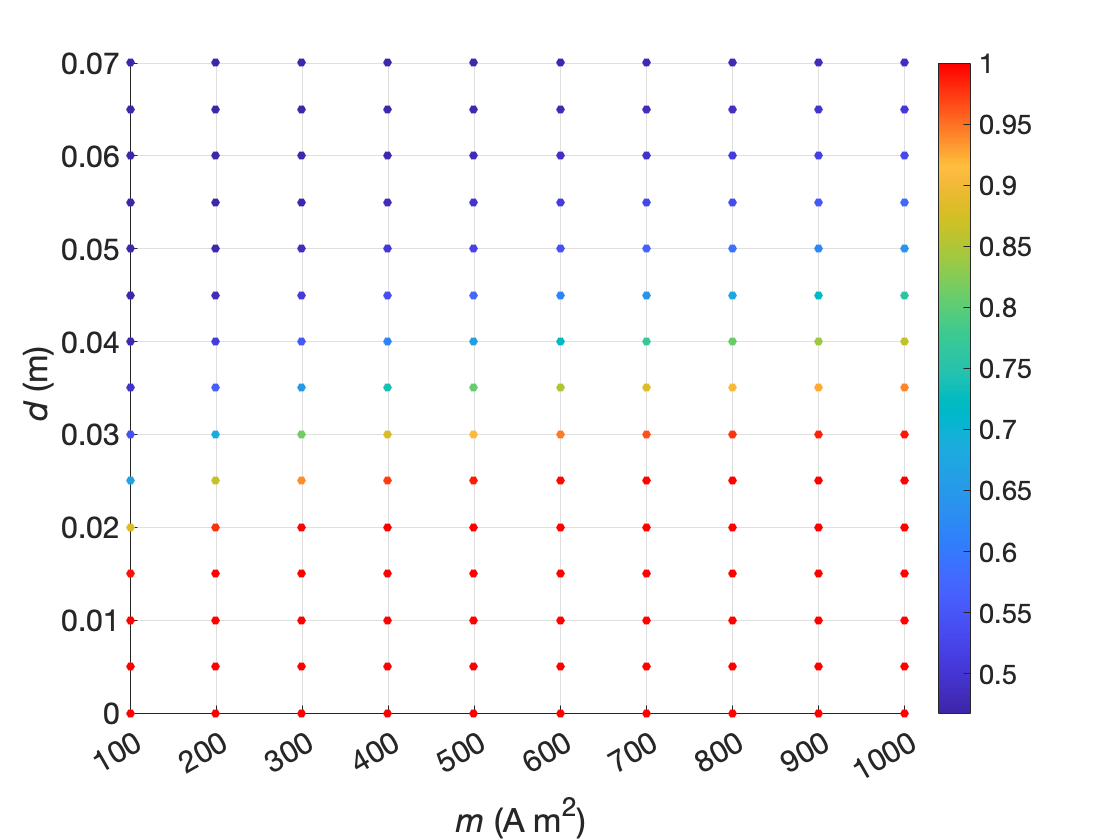}
    \caption{Capture rates in arterioles at various magnetic moments and distances from the blood vessel. Colorbar represents the capture rate, where red represents a capture rate of 1 and blue represents a capture rate of 0.5. Capture rates remain the same along multiple release points (not shown). $K_{sh} = 0.05$. }
    \label{arteriolemomentdistance}
\end{figure}
\subsubsection{Nanoparticle Motion in Arterioles with RBC Collisions}
We now consider the motion of nanoparticles in arterioles. Due to $R \gg r_{RBC}$, there is significant stochasticity resulting from red blood cell collisions and $K_{sh} = 0.05$. Results from varying the dipole moment and the magnet distance are shown in Figure \ref{arteriolemomentdistance}. Similar to motion in capillaries with $K_{sh} = 0.05$, the capture rate remains at $1$ for all dipole moments until the magnet is placed a distance of $1.5$ cm away from the vessel. The decrease in capture rate, however, occurs over a larger range than in capillaries. Due to the larger radius of arterioles, stochastic forces do not immediately cause the particles to hit the vessel walls. The greater blood velocity also competes with stochastic forces and diffusion in the $z$ direction. At larger distances, the magnetic field can thus influence particle trajectories for a slightly longer period of time and continue to capture particles before they diffuse. Interestingly, at high dipole moments the capture rates in both capillaries and arterioles are very similar. In both cases, a capture rate of 1 is not sustained at distances more than 3 cm, once again emphasizing the need for the magnet to be placed close to the tumor regardless of the dipole moment or vessel type. Overall, these results show that magnetic drug delivery can be effective in both capillaries and arterioles, and that the differences in vessel radius and blood velocity do not have a major impact on capture rate when using the point dipole model. 

\subsubsection{Nanoparticle Motion in Arteries}
\label{arteries}
\begin{figure}[!htb]
    \centering
    \includegraphics[scale=0.2]{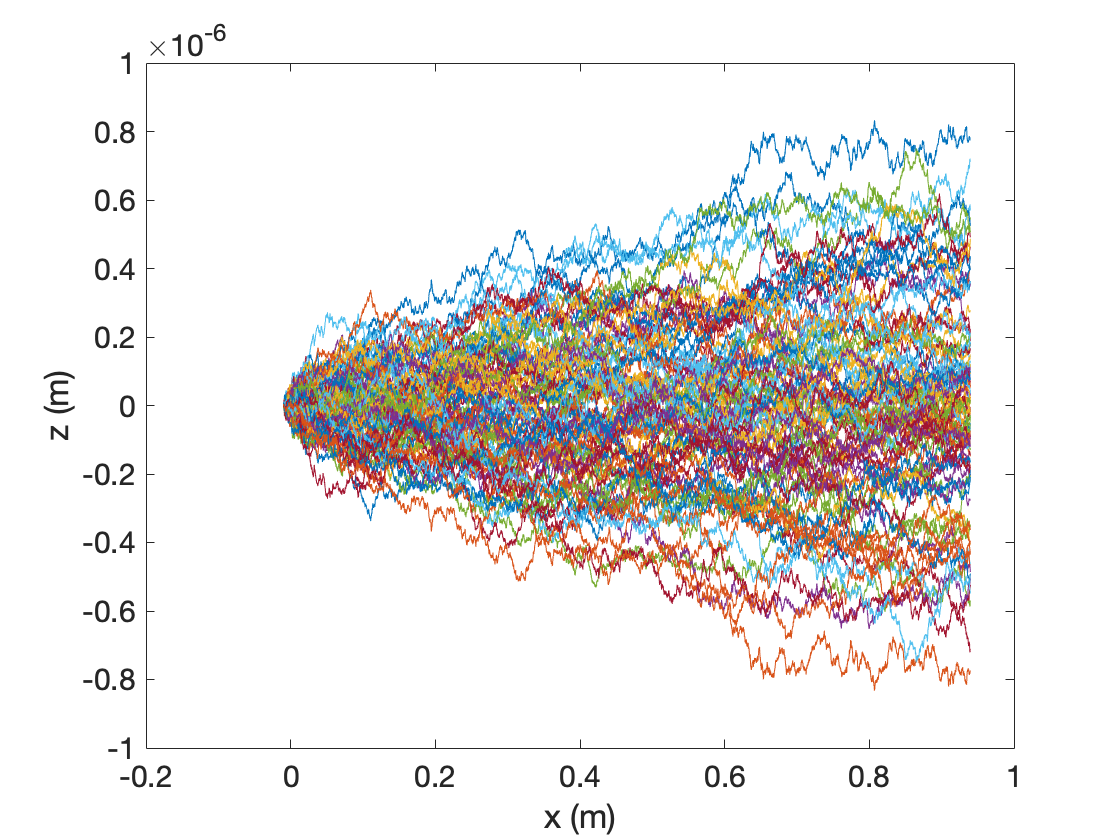}
    \includegraphics[scale=0.2]{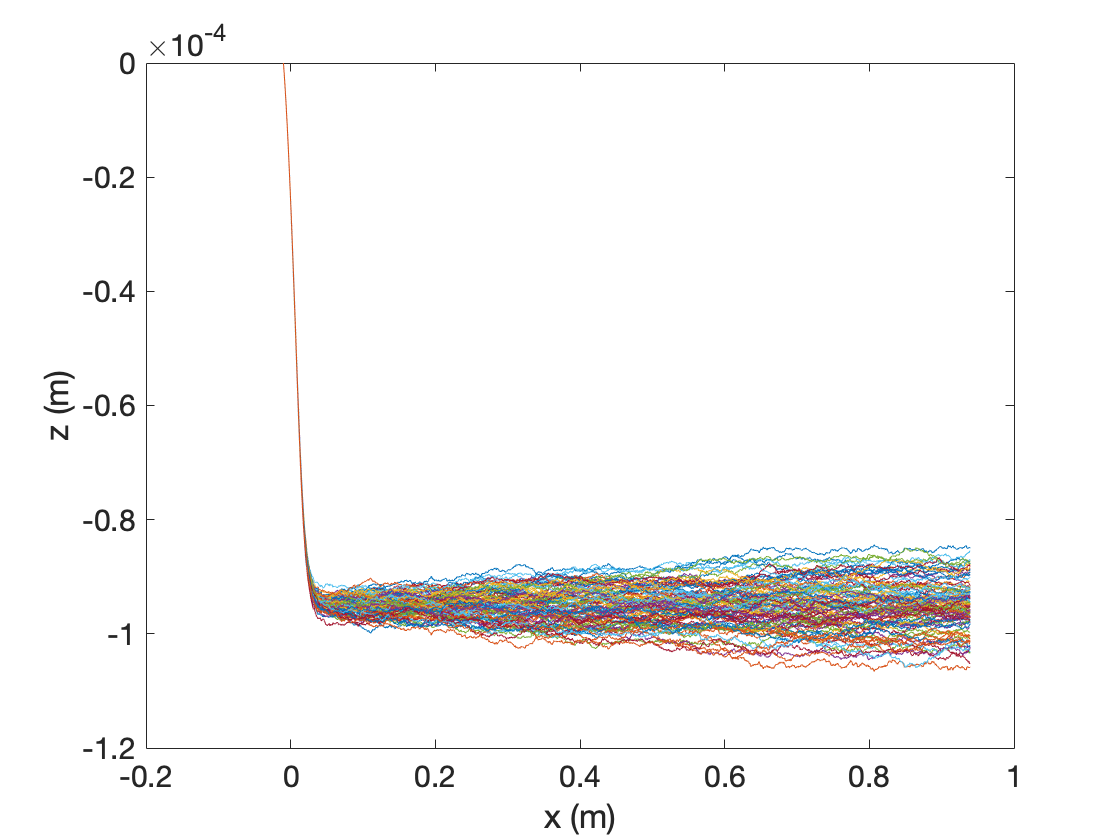}
    \includegraphics[scale=0.2]{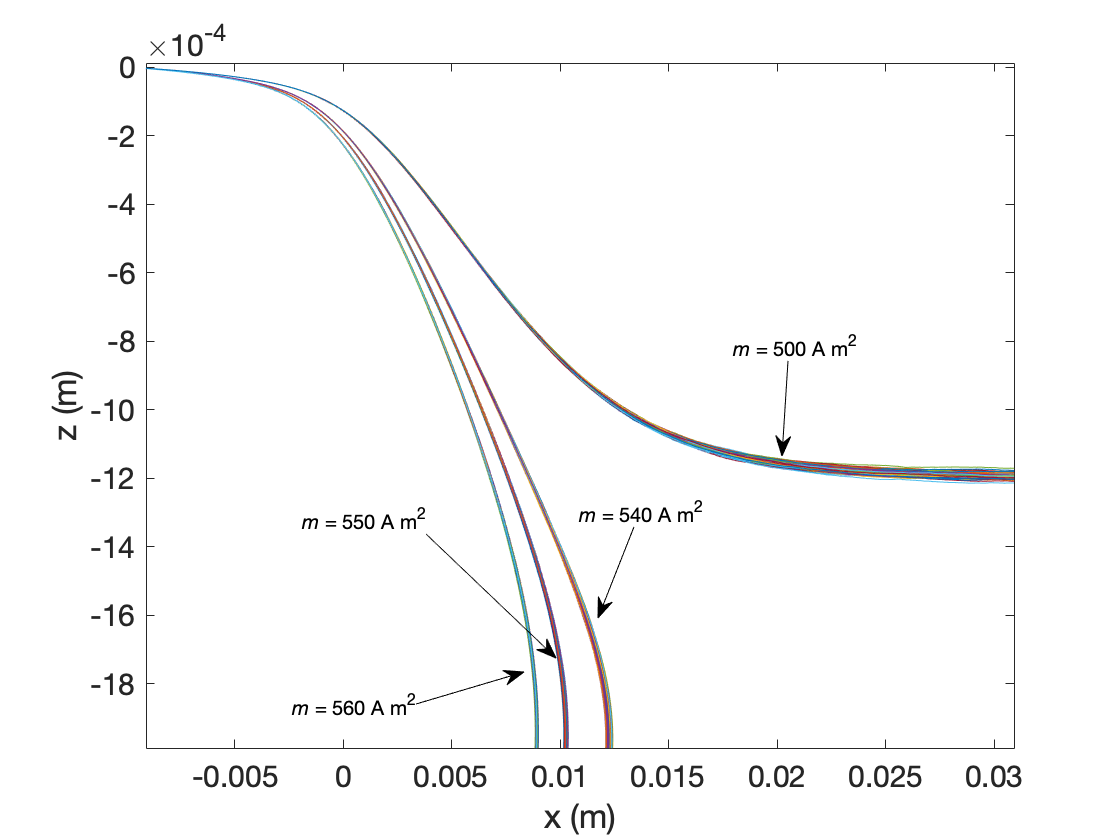}
    \caption{100 particle trajectories in arteries along the $x-z$ planes. Top panel shows trajectories in the absence of a magnetic field; middle panel shows trajectories with $d = 0.03$ m, $m = 1000$ A m\textsuperscript{2}; bottom panel shows trajectories with $d =0.02$ m, various $m$. $a_p = 500$ nm. Particles released from (-0.01, 0, 0) m.}
    \label{artery_xz}
\end{figure}

Finally, we consider nanoparticle motion in arteries, where the radius is much larger and the blood velocity is much faster than in arterioles and capillaries. Due to the strength of the background blood flow, it becomes very difficult to capture nanoparticles of size $a_p = 100$ nm. However, previous studies \cite{price2018where, ao2018polydopamine, wang2018multistage} indicate high capture rates are possible with clusters of particles. We represent this by considering particles of size $a_p = 500$ nm, making them larger and thus more responsive to the magnetic field. The results are shown in Figure \ref{artery_xz}.

In the absence of a magnetic field, we find the larger particle radius greatly decreases the effects of Brownian motion and RBC collisions. Although particles still diffuse, the velocity of the background blood flow is strong enough to prevent particles from hitting the vessel wall. Unlike capillaries and arterioles, where particles must be released very close to or within the tumor radius to be captured, larger particles in the arteries can be released much farther upstream of the tumor site without diffusing to the vessel walls. This is further discussed in Sec.~(\ref{initRelease}).

However, much stronger magnets than those needed in capillaries and arterioles are required to overcome the background blood velocity and they must be placed close to the vessel. When $m=1000$ A m\textsuperscript{2} and $d = 3$ cm, the magnetic force is not strong enough to pull the particles all the way to the tumor. Instead, the magnetic force acts on the particle trajectories for a short period of time and pulls it downwards; once the particle passes over the magnet, it resumes its normal trajectory with relatively little diffusion. Distance from the vessel to the magnet remains the most significant determinant of capture rate, since even extremely strong dipole moments are insufficient to capture the particles. 

If the magnet is placed 2 cm away from the vessel, however, a much wider range of dipole moments are sufficient to capture the particles. A dipole moment of 560 A m\textsuperscript{2} is sufficient to capture particles. However, dipole moments of 550 and 540 A m\textsuperscript{2} are able to attract all the particles to the vessel wall, but outside the radius of the tumor. The difference in trajectory endpoints is due to the high velocity of the background blood flow, causing particles to drift substantially in the $x$ direction. This further suggests an all-or-none phenomenon in arteries where particles are either completely captured or completely miss the target. At a dipole moment of 500 A m\textsuperscript{2}, the magnetic force is too weak and particle trajectories resemble what occurs when $d = 0.03$ m.
Although the lack of particle diffusion allows for more flexibility in the particle release location, magnetic drug delivery will only be effective if the magnet can be placed $<2$ cm from the tumor. Stronger dipole moments of at least $560$ A m\textsuperscript{2} are also required to capture particles, compared to $100$ A m\textsuperscript{2} in capillaries and arterioles. These results show that the high velocity of background blood flow presents a significant barrier towards the use of magnetic drug delivery in arteries.

\subsection{Bar Magnet Model}
\label{Bar Magnet}
We study the effectiveness of magnetic drug delivery at lower magnetic field strengths by replacing the point dipole with a bar magnet. The equations for the magnetic field surrounding a bar magnet are given in Appendix \ref{sec:appendix}. The schematic for this model is illustrated in Figure \ref{Schematic}, with the bar magnet placed below the blood vessel at $(0, 0, -R-d)$, where $d$ is the distance from the surface of the magnet to the blood vessel wall. The bar magnet used has a fixed size with a width of $1.8$ cm, a height of $10.0$ cm, and a length of $8.0$ cm. It represents an N52 Grade Neodymium magnet, the strongest commercially available permanent magnet \cite{fraden2004handbook}, and has a $B_r$ (residual magnetization) value of $1.48$ T. Results with magnets of other dimensions should exhibit similar trends, with larger magnets leading to higher capture rates and smaller magnets leading to lower capture rates, assuming all else is kept constant. All of the particles are released at an $x$ position of $-0.00902$ m, which is right before at the left edge of the magnet. We study the effects of varying the distance between the magnet and the blood vessel, and different levels of red blood collisions. In addition, the particle size is varied in arterioles to simulate nanoparticle clusters.

\subsubsection{Nanoparticle Motion in Capillaries}
\begin{figure}
    \centering
    \includegraphics[scale=0.2]{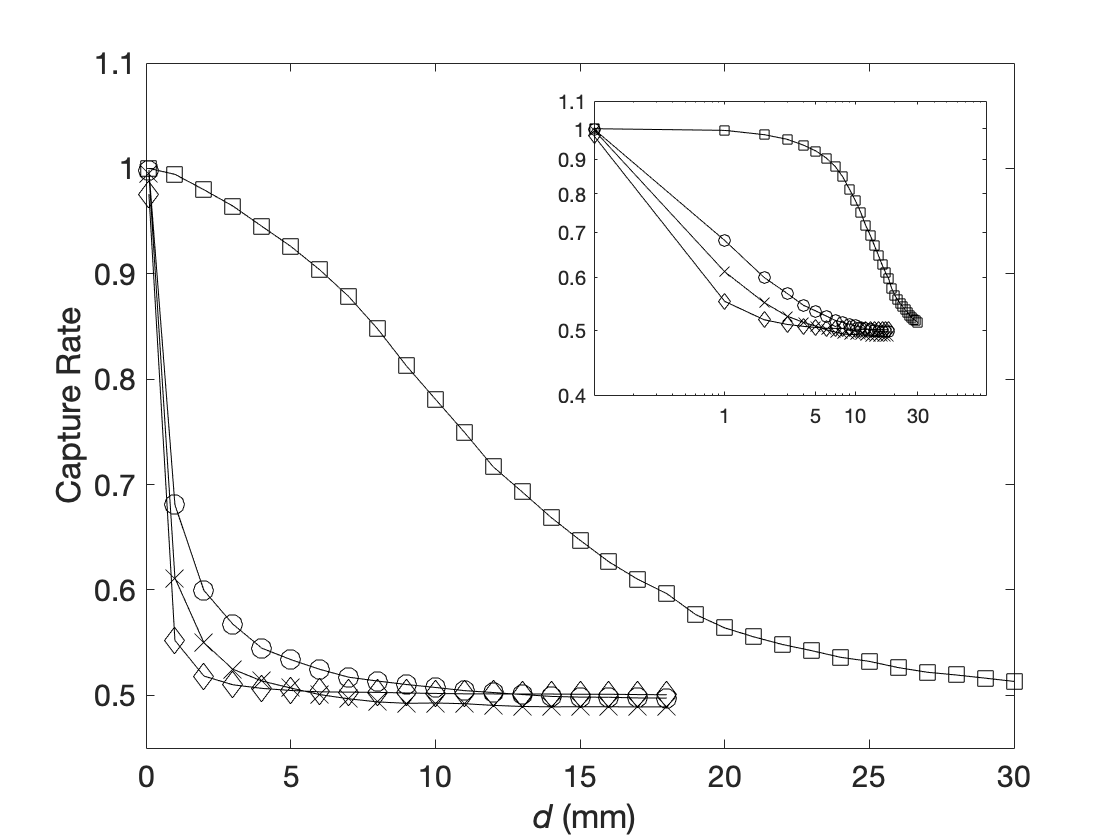}
    \caption{Capture rates at various distances from the bar magnet for different $K_{sh}$ values. (a) Squares represent capture rates in capillaries with $K_{sh} = 0$; (b) diamonds represent capture rates in capillaries with $K_{sh} = 0.05$, (c) circles represent capture rates in arterioles with $K_{sh}= 0.025$; (d) crosses represent capture rates in arterioles with $K_{sh} = 0.05$. Inset shows data graphed on a log-log scale.}
    \label{BarKshAll}
\end{figure}

\paragraph{Motion with RBC Collisions}
Figure \ref{BarKshAll} details the effect of increasing distance between the magnet and capillary wall on the capture rates of the nanoparticles for various $K_{sh}$ values. This model demonstrates that at a distance of $0$ cm, the capture rate of the nanoparticles is 1. However, as the distance is increased slightly to $0.1$ cm, the capture rate drops dramatically to $0.55$. This is caused by the rapid diffusion from Brownian motion and RBC collisions compared to the relatively weak influence of the magnetic field at a distance of $0.1$ cm.

In addition, at distances greater than $0.1$ cm, an asymptote at the capture rate of $0.5$ is present. This asymptote occurs because the magnetic field strength weakens with distance, approaching a scenario similar to what presented in Sec.~(\ref{absentMagnet}) due to Brownian motion and RBC collisions. The asymptote present in Figure \ref{BarKshAll} demonstrates that at distances of $0.1$ cm and higher, the force from the magnetic field is not sufficiently high to overcome stochastic forces,
showing the bar magnet is ineffective for magnetic drug delivery when $K_{sh} = 0.05$ (we note that $K_{sh}$ value likely lies somewhere between $0$ and $0.05$). We study the case of $K_{sh} = 0$ next.

\paragraph{Motion without RBC Collisions}
When nanoparticles are released into the capillaries and $K_{sh} = 0$, capture rates remain above 0.75 at larger distances of up to 1 cm between the magnet and capillary wall,  instead of dropping dramatically at a distance of $0.1$ cm in the $K_{sh} = 0.05$ case. In addition, Figure \ref{BarKshAll} part (a) ($K_{sh}=0$) resembles a logistic curve and continues to have an asymptote at a capture rate of $0.5$. This logistic curve provides evidence that the distance between the bar magnet and the blood vessel is the most influential factor on the capture rate. It also shows that there is a particular region of distances in which the capture rate dramatically declines. When $K_{sh} = 0.05$, this occurs between $0$ and $0.1$ cm, where the magnitude of the derivative of the graph is the greatest. When $K_{sh}=0$, the decline occurs between $0.8$ to $1.4$ cm. The decline occurs at a larger distance when $K_{sh} = 0$ because there is considerably less diffusion than when $K_{sh} = 0.05$.

These results indicate the bar magnet could be effective at distances up to $1$ cm if RBC collisions are minimal in capillaries. Future experimental results will determine whether a $K_{sh}$ of $0$ is representative of red blood cells moving single file through capillaries.



\subsubsection{Nanoparticle Motion in Arterioles with Larger Particles}
\begin{figure}
    \centering
    \includegraphics[scale=0.2]{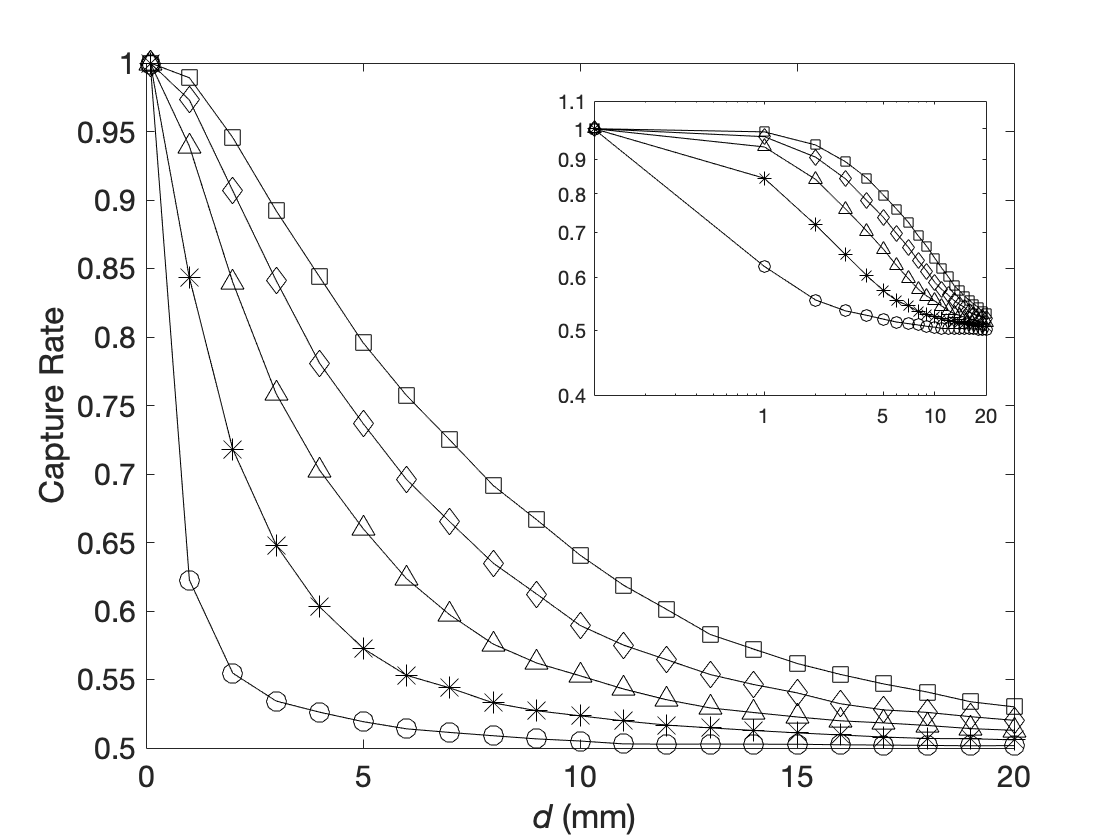}
    \caption{Capture rates in arterioles as a function of distance for different particle sizes. (a) Squares represent capture for $500$ nm particle radius. (b) Diamonds represent capture for $400$ nm particle radius. (c) Triangles represent capture for $300$ nm particle radius. (d) Asterisks represent capture for $200$ nm particle radius. (e) Circles represent capture for $100$ nm particle radius. Inset shows data on a log-log scale.}
    \label{ParticleSizeArterioles}
\end{figure}

Figure \ref{ParticleSizeArterioles} shows the effect of increasing particle size on capture rate as distance from the arteriole wall increases. 
All particle sizes had a capture rate of $1$ at $0$ cm and approached a horizontal asymptote of $0.5$ at larger distances. As distance increased, smaller particles dropped in capture rate more rapidly than particles of larger sizes did. At a distance of $0.5$ cm, the $100$ nm particle had a capture rate of about $0.52$, while the $300$ nm particle had a capture rate of $0.66$, and the $500$ nm particle had a capture rate of $0.8$, suggesting that particle size is a significant factor in capture rate.

In the previous section, the $K_{sh}$ term caused the bar magnet model to be  ineffective at distances greater than 0.1 cm when $a_p = 100$ nm. These results show that while the bar magnet model is still not as effective in arterioles as the point dipole, using a larger particle size can make the magnetic field effective at larger distances. Larger particles also decrease the influence of Brownian motion and RBC collisions. However, particle size is limited by the arteriole radius, and particles must be designed to prevent occlusion in arterioles and capillaries if any particles are not captured. Thus, using larger particles increases their responsiveness and may allow the bar magnet to be applied to arterioles very close to the skin surface.

\subsection{Optimal Particle Release Location}
\label{initRelease}
In Sec.~(\ref{Dipole Moment}), we found the capture rate does not vary within the radius of the tumor from $-200R$ to $100R$. Here we show that the optimal particle release location lies within the radius of the tumor for both the point dipole model and the bar magnet model, and that the capture rate is near 0 when particles are released outside the tumor radius. In capillaries and arterioles, it is crucial for particles to be released within the radius of the tumor to avoid early diffusion of particles to the vessel wall prior to entering the tumor radius. This limits the availability of the particle injection site and suggests the procedure will not be effective if particles are allowed to circulate in the bloodstream before reaching their target. Instead, particles must be injected directly at the tumor site. However, particle clusters may be released at farther distances in the arteries.

\subsubsection{Point Dipole Model}
 For particles released in capillaries and arterioles, we choose $m=500$ A m\textsuperscript{2} and $d=0.02$ m to be representative of particles with capture rates of 1 when released within the tumor radius, and $m=500$ A m\textsuperscript{2} and $d=0.03$ m to be representative of particles with capture rates within the sensitive area, based off of the results in Figures \ref{Ksh0.05Cap} and \ref{arteriolemomentdistance}. We release particles from the edge of the tumor at $x=0.01$ m and at $x=0.011$ m ($1$ mm before the tumor) to assess the effect of releasing particles outside the tumor.
 
\begin{figure}
    \centering
    \includegraphics[scale=0.1]{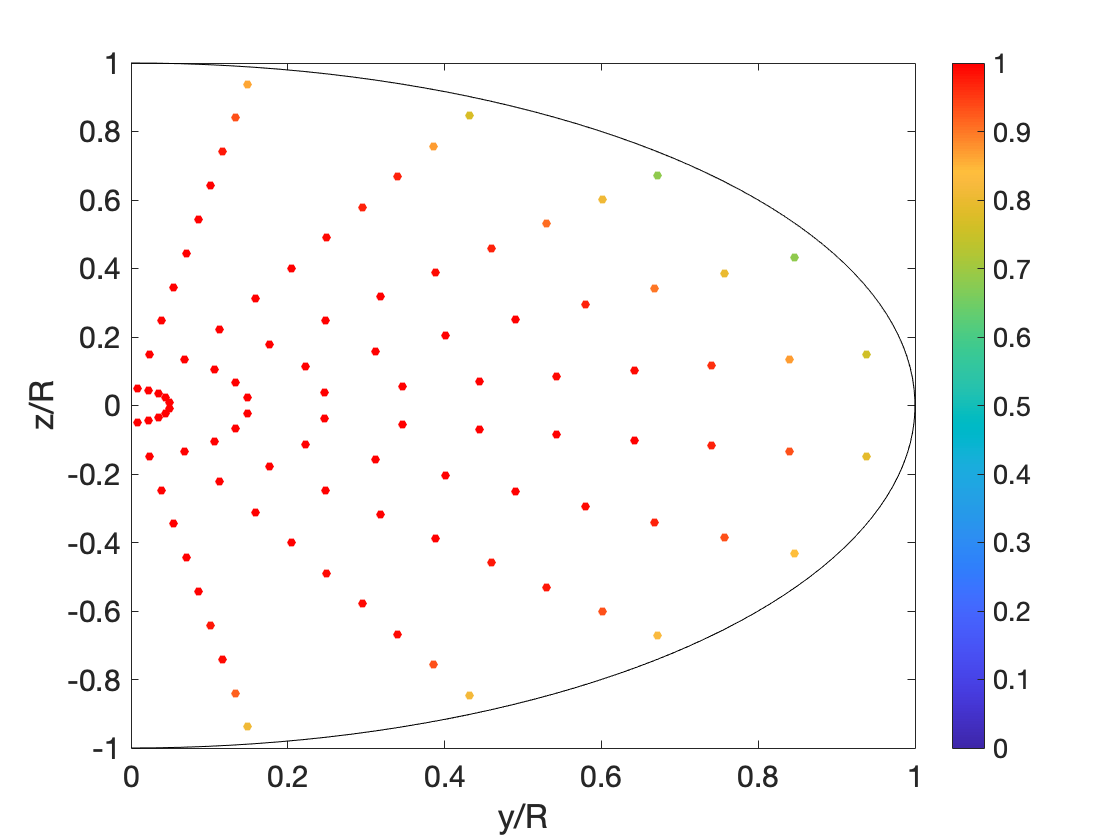}
    \includegraphics[scale=0.1]{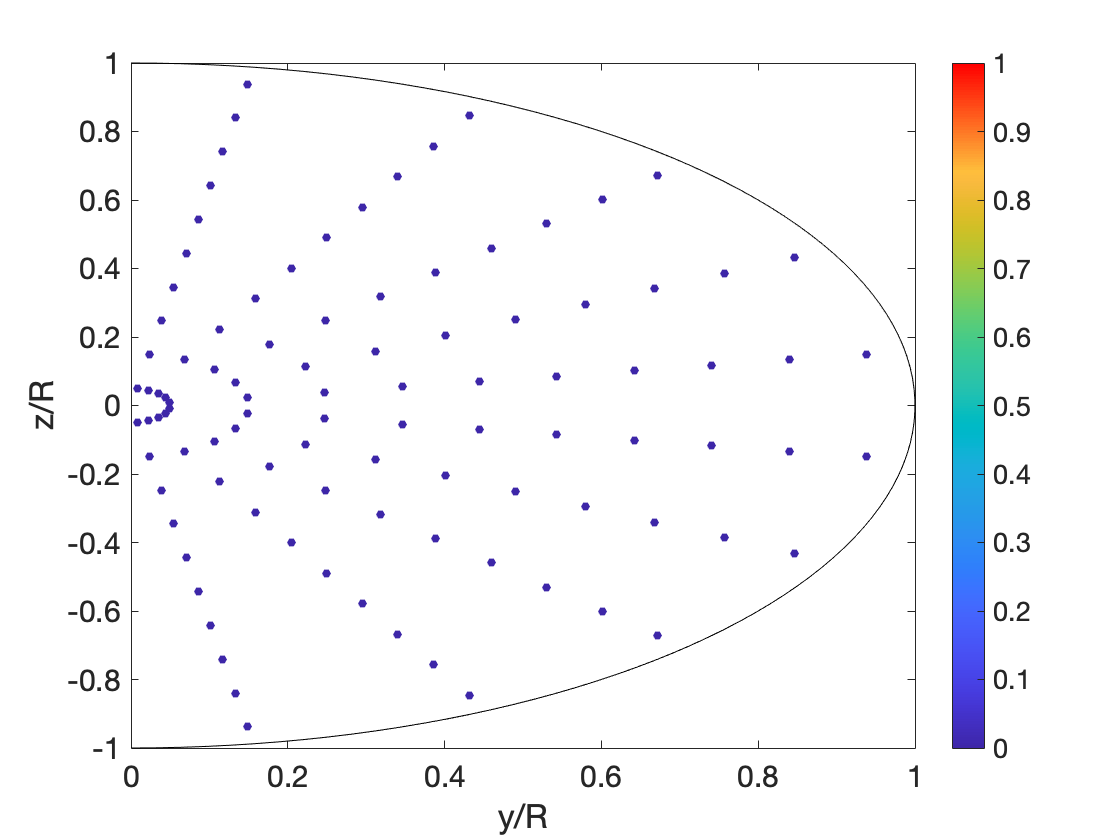}
    \includegraphics[scale=0.1]{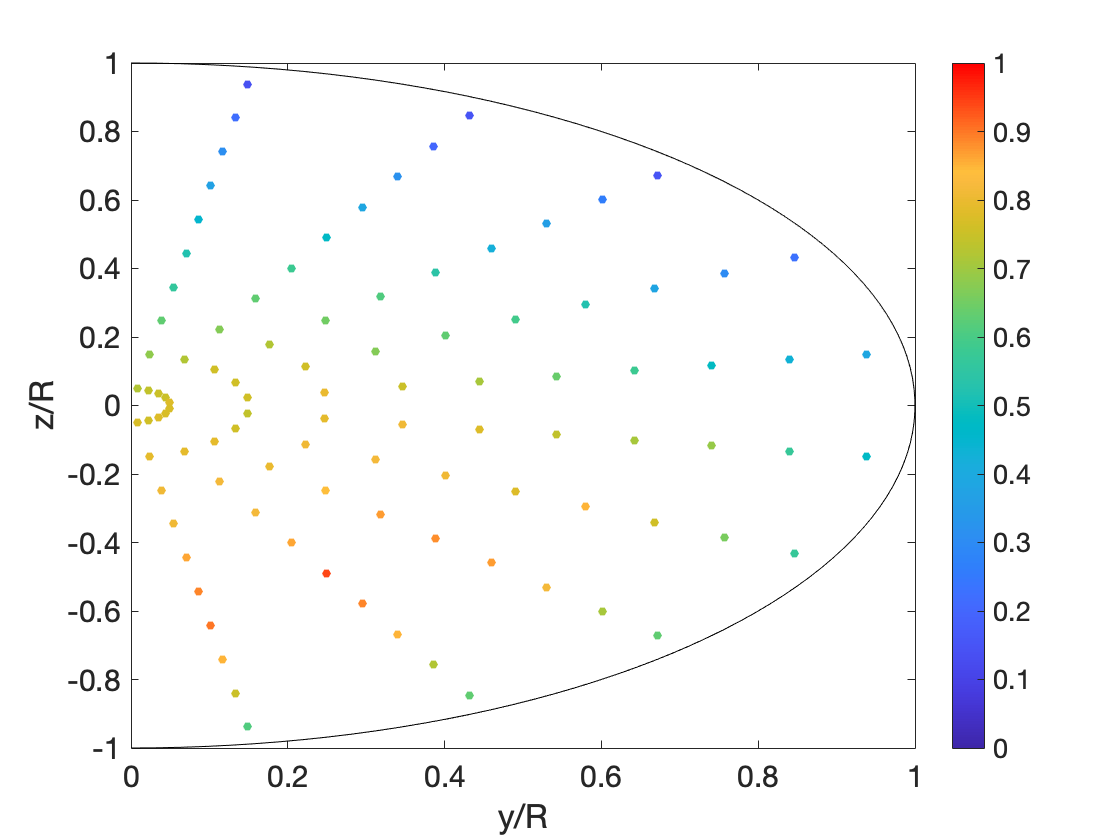}
    \includegraphics[scale=0.1]{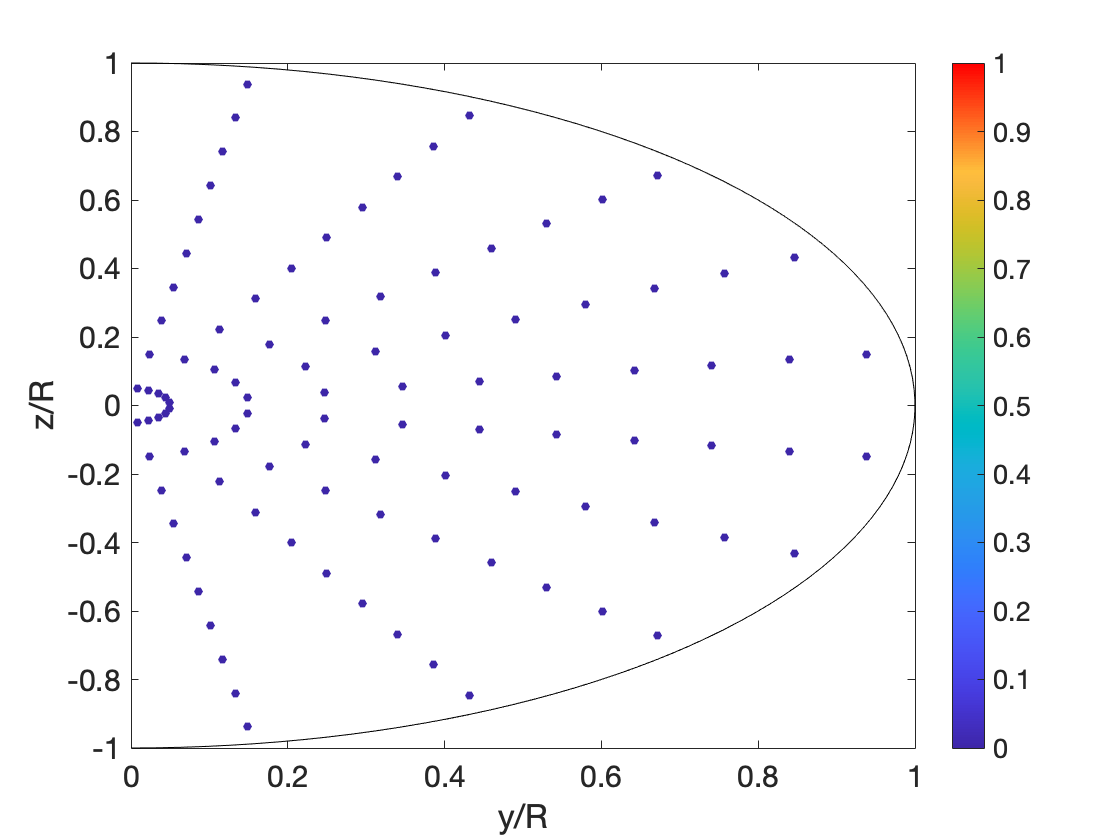}
    \caption{Top left and top right images show cross sections with $d = 0.02$ m and particles released in capillaries from $x=0.01$ m and $x=0.011$ m, respectively. Bottom left and bottom right images show cross sections with $d =0.03$ m and particles released in capillaries from $x=0.01$ m and $x=0.011$ m, respectively. Colorbar represents capture rate. $m=500$ A m\textsuperscript{2}. $K_{sh} = 0.05$.}
    \label{capillaryReleaseDipole}
\end{figure}

\begin{figure}
    \includegraphics[scale=0.2]{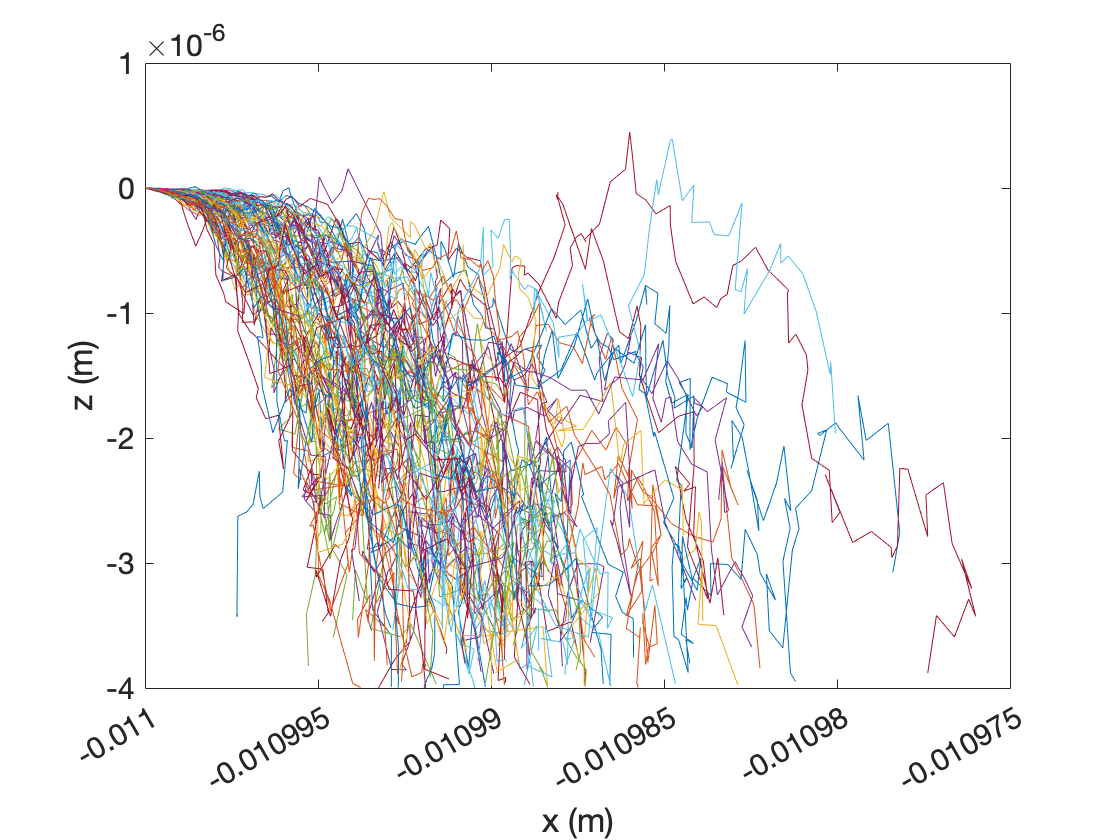}
    \includegraphics[scale=0.2]{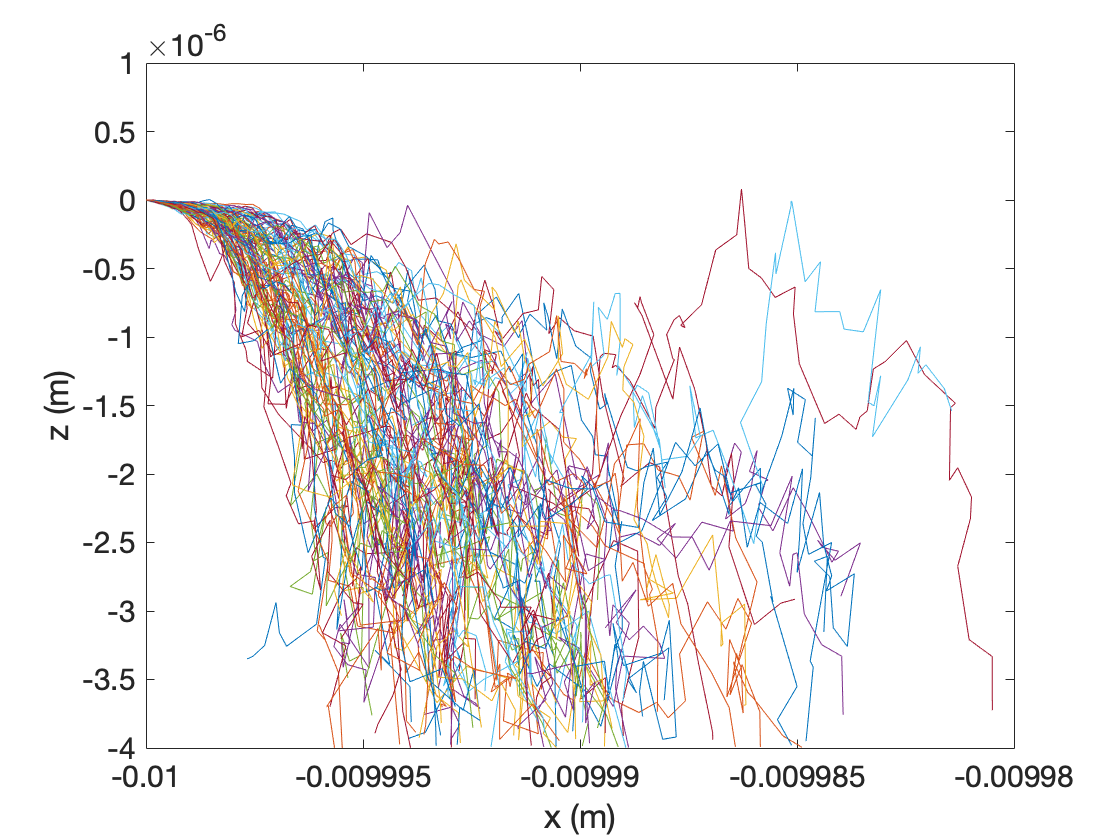}
    \caption{Two sets of particle trajectories along the $x-z$ plane for $m= 500$ m A\textsuperscript{2}, $d=0.02$ m. Top panel shows particles released from $(-0.011, 0, 0)$ with capture rate of 0; bottom panel shows particles released from $(-0.01, 0, 0)$ with capture rate of 1. The tumor is located at $x=-0.01$ m and lies to the right of the image, not shown. Colorbar represents capture rate. Particles are released in capillaries and 100 trajectories shown per release point.} 
    \label{boundary}
\end{figure}

 The effect of releasing particles outside the tumor radius in capillaries is shown in Figure \ref{capillaryReleaseDipole}. When $d=0.02$ m, the capture rate averaged across the cross section is 0.96 when particles are released at the edge of the tumor, but is 0 when particles are released 1 mm before the tumor radius. Likewise, when particles are released from $d= 0.03$ m, the average capture rates are 0.65 and 0, respectively. This difference in capture rate is due to the magnetic field strength attracting particles to the vessel wall prior to reaching the tumor radius, as shown in Figure \ref{boundary}. Particles released at both $x=0.01$ m and $x=0.011$ m experience a horizontal drift of approximately $1.5 \times 10^{-5}$ m. Particles released slightly outside the tumor radius continue to experience the full force of the magnetic field, and subsequently hit the vessel wall before reaching the tumor. 
 
\begin{figure}
    \centering
    \includegraphics[scale=0.1]{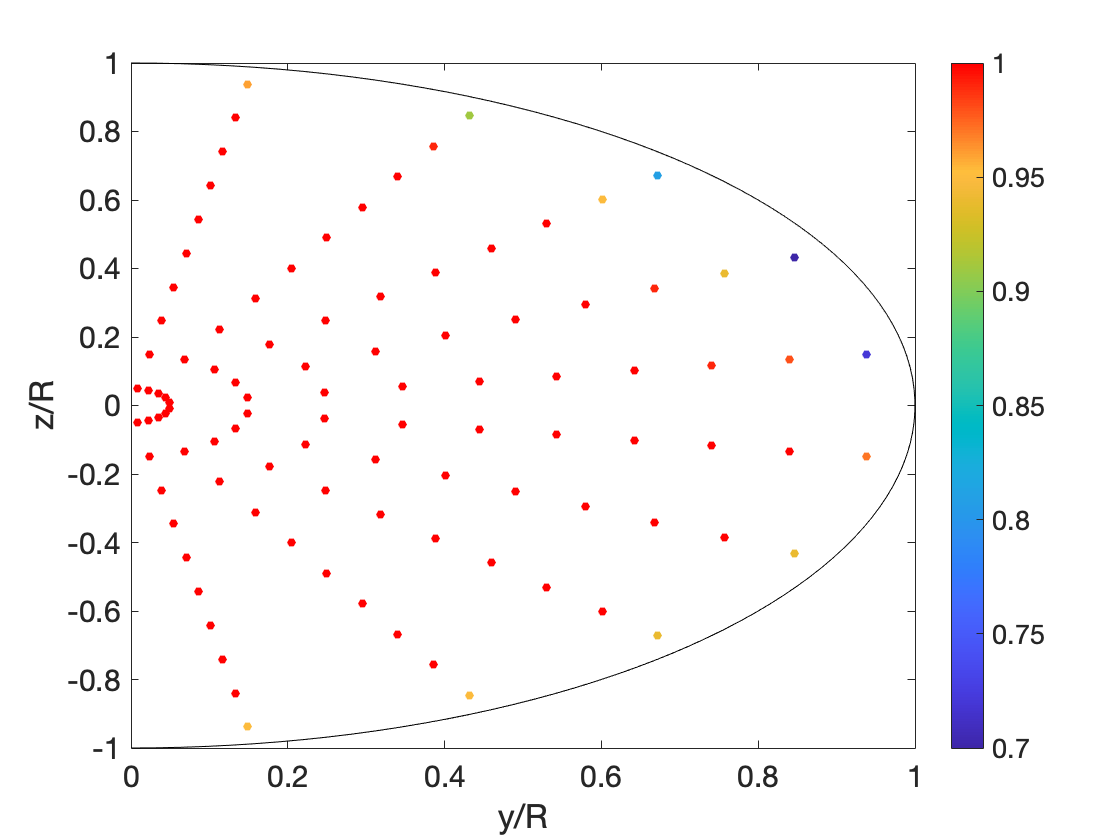}
    \includegraphics[scale=0.1]{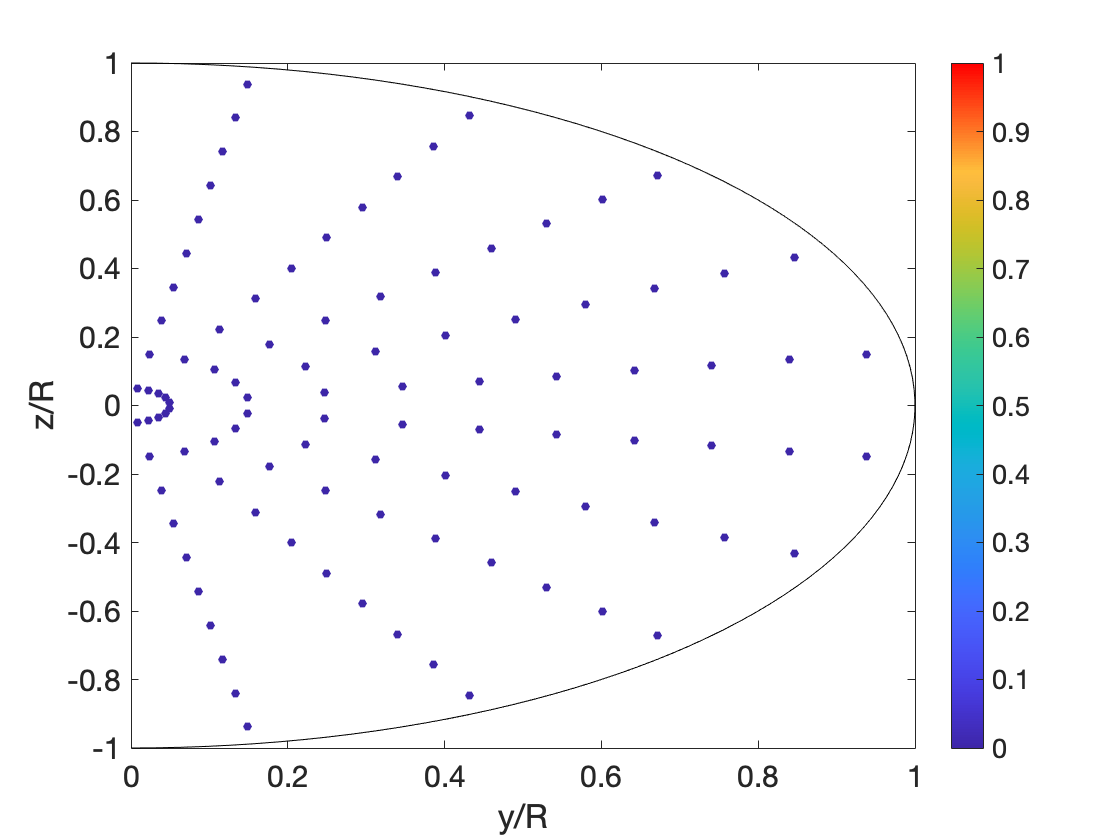}
    \includegraphics[scale=0.1]{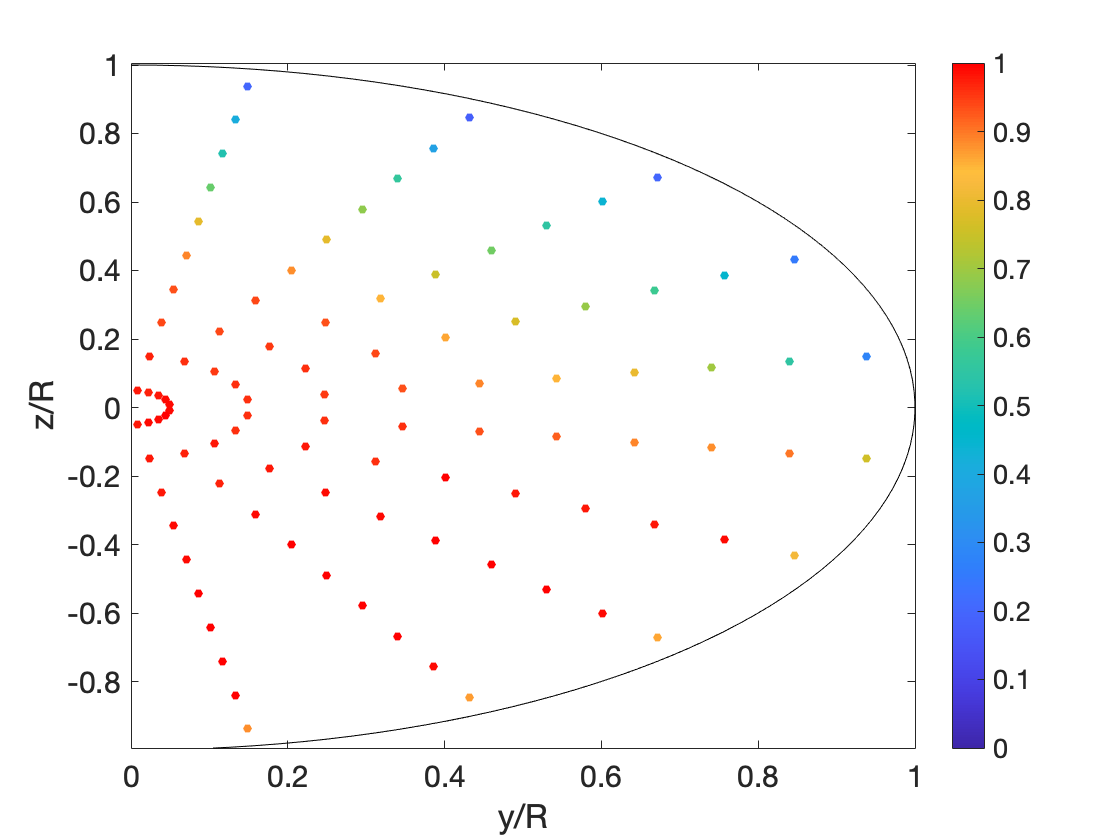}
    \includegraphics[scale=0.1]{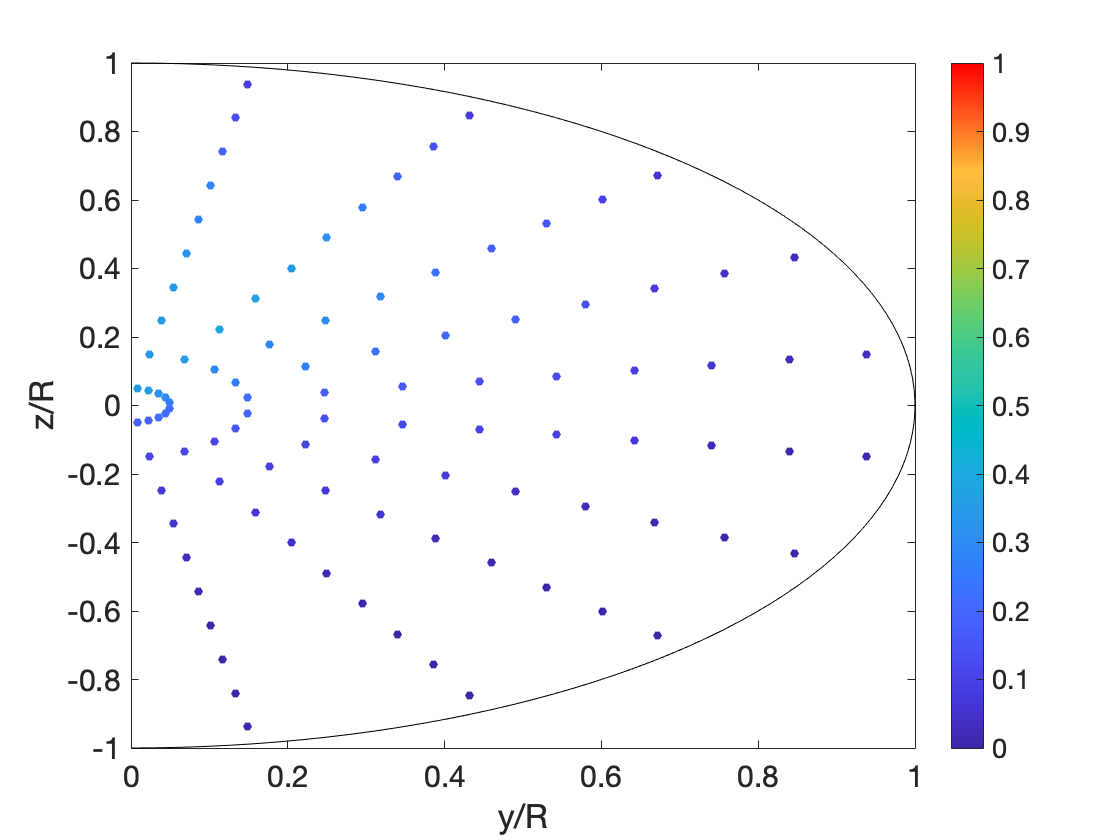}
    \caption{Top left and top right images show cross sections with $d = 0.02$ m and particles released in arterioles from $x=0.01$ m and $x=0.011$ m, respectively. Bottom left and bottom right images show pipe cross sections with $d =0.03$ m and particles released in arterioles from $x=0.01$ m and $x=0.011$ m, respectively. Colorbar represents capture rate. $m=500$ A m\textsuperscript{2}. $K_{sh} = 0.05$.}
    \label{arterioleReleaseDipole}
\end{figure}

 The same effect is present in arterioles as shown in Figure \ref{arterioleReleaseDipole}. At $d=0.02$ m, the average capture rates are 0.99 at $x=0.01$ m and 0 at $x=0.011$ m. At $d=0.03$ m, the average capture rates are 0.85 at $x=0.01$ m and 0.14 at $x=0.011$ m. We observe that particles released from $x=0.011$ m with nonzero capture rates are located along the center of the top half of the vessel; this can be attributed to both the higher background blood velocity in the center of the vessel and the larger vessel radius, allowing particles to travel farther distances and enter the tumor radius. We expect this capture rate to go to zero as particles are released farther from the tumor due to the small amount of horizontal drift. 
 
\begin{figure}
    \centering
    \includegraphics[scale=0.1]{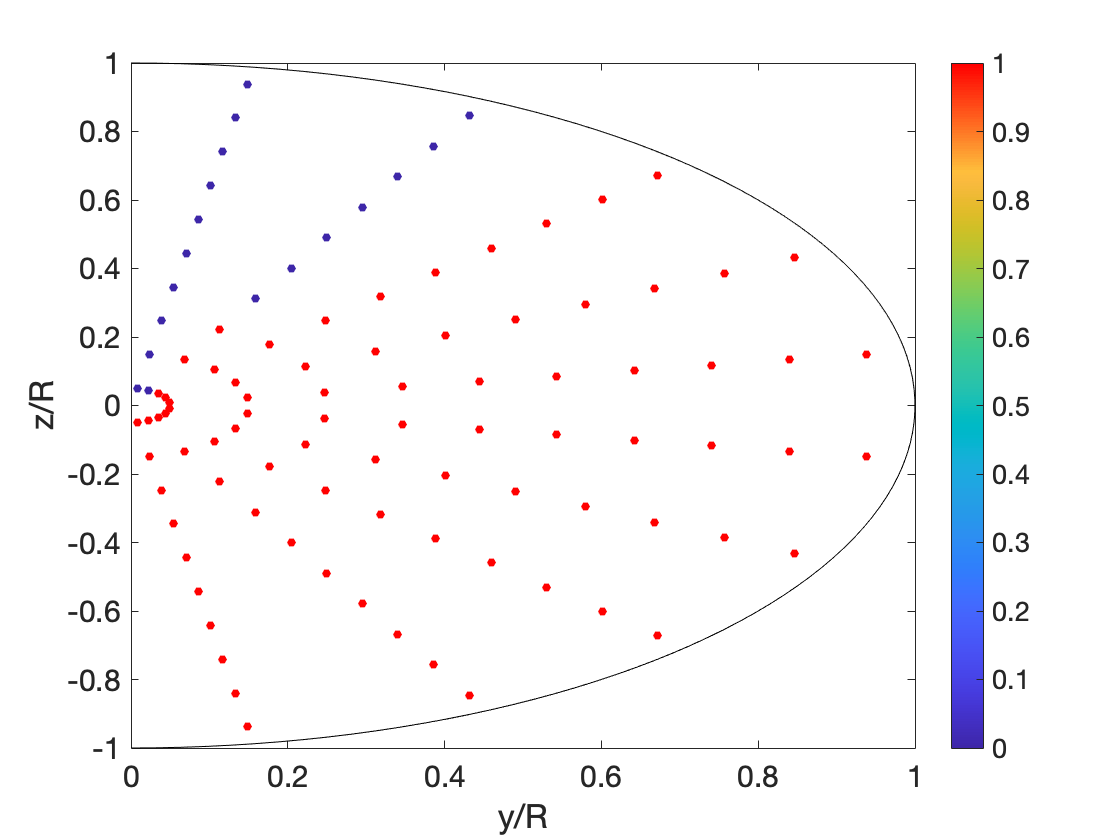}
    \includegraphics[scale=0.1]{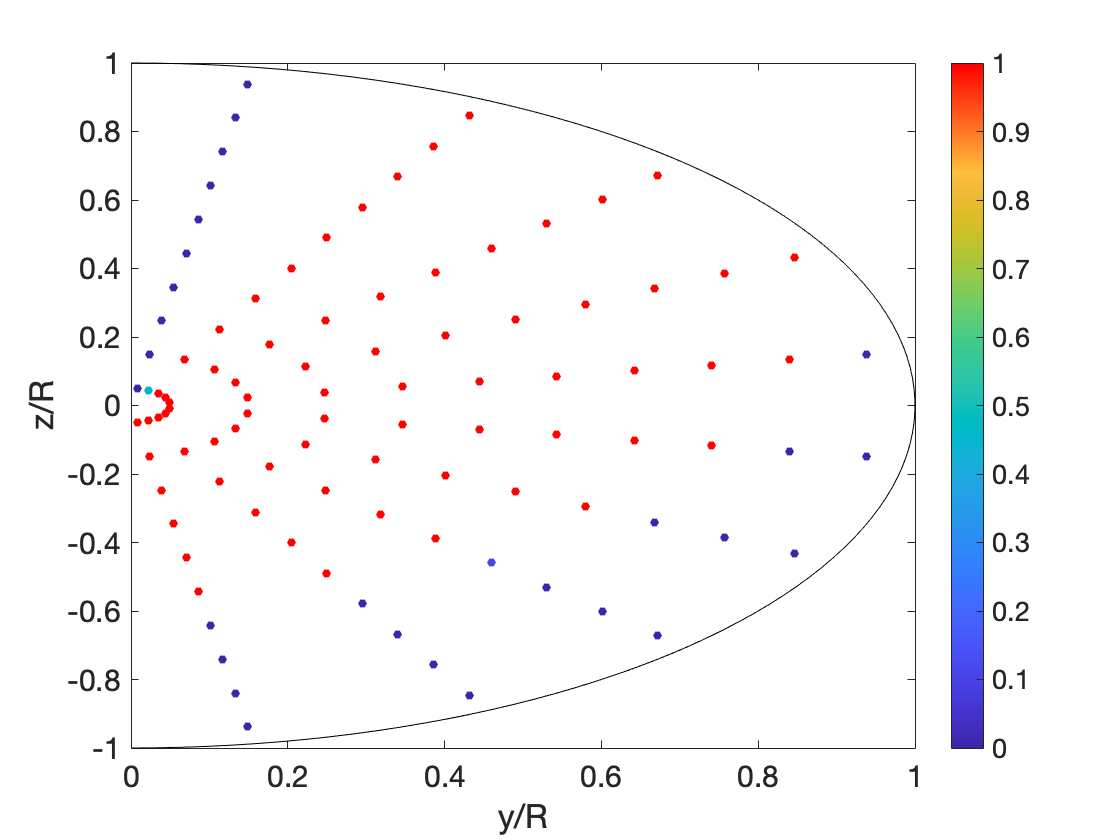}
    \caption{Particles released in arteries from $x=0.01$ m (left image) and $x=0.03$ m (right image). $m=560$ A m\textsuperscript{2}, $d=0.02$ m, $K_{sh} = 0.05$, and $a_p= 500$ nm. }
    \label{arteryReleaseDipole}
\end{figure}

Due to the high background blood velocity in arteries, we use $m = 560$ A m\textsuperscript{2} and $d=0.02$ m in accordance with the results found in Sec.~(\ref{arteries}). Figure \ref{artery_xz} shows that diffusion does not cause particles to immediately hit the vessel wall, unlike capillaries and arterioles. We extend this by showing in Figure \ref{arteryReleaseDipole} that the capture rate remains high at 0.82 and 0.72 when particle clusters of $a_p=500$ nm are released from $x=0.01$ m and $x=0.03$ m. Although the capture rate decreases, it remains relatively high and shows magnetic drug delivery can be effective in arteries when particles are released outside the tumor radius. 

\subsubsection{Bar Magnet Model}
When the bar magnet model is used, we find the optimal release location is at the edge of the magnet when RBC collisions are considered. When RBC collisions are absent, the same capture rate can be achieved anywhere within the radius of the tumor. When $d=0.001$ m, we find the magnetic field strength is too weak for the initial release location to have a significant impact on capture rate as long as particles are released within the tumor. We release particles from outside the tumor ($x=-0.011$ m), at the tumor boundary ($x=-0.01$ m), just before the edge of the bar magnet ($x=-0.00902$ m), and past the edge of the bar magnet ($x=0.008$ m) to assess multiple release locations. We simulate particle motion in both capillaries and arterioles for distances of $d=0$ m and $d = 0.001$ m, both with and without RBC collisions.
 
\begin{figure}
    \centering
    \includegraphics[scale=0.1]{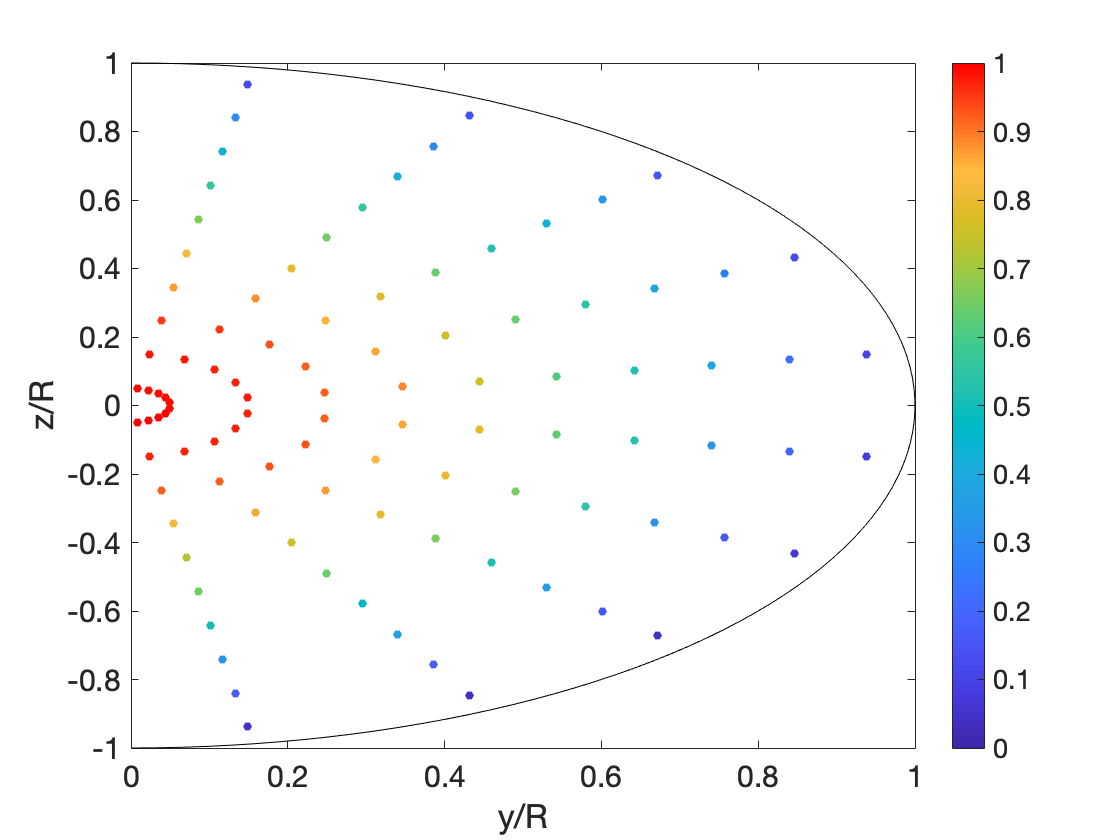}
    \includegraphics[scale=0.1]{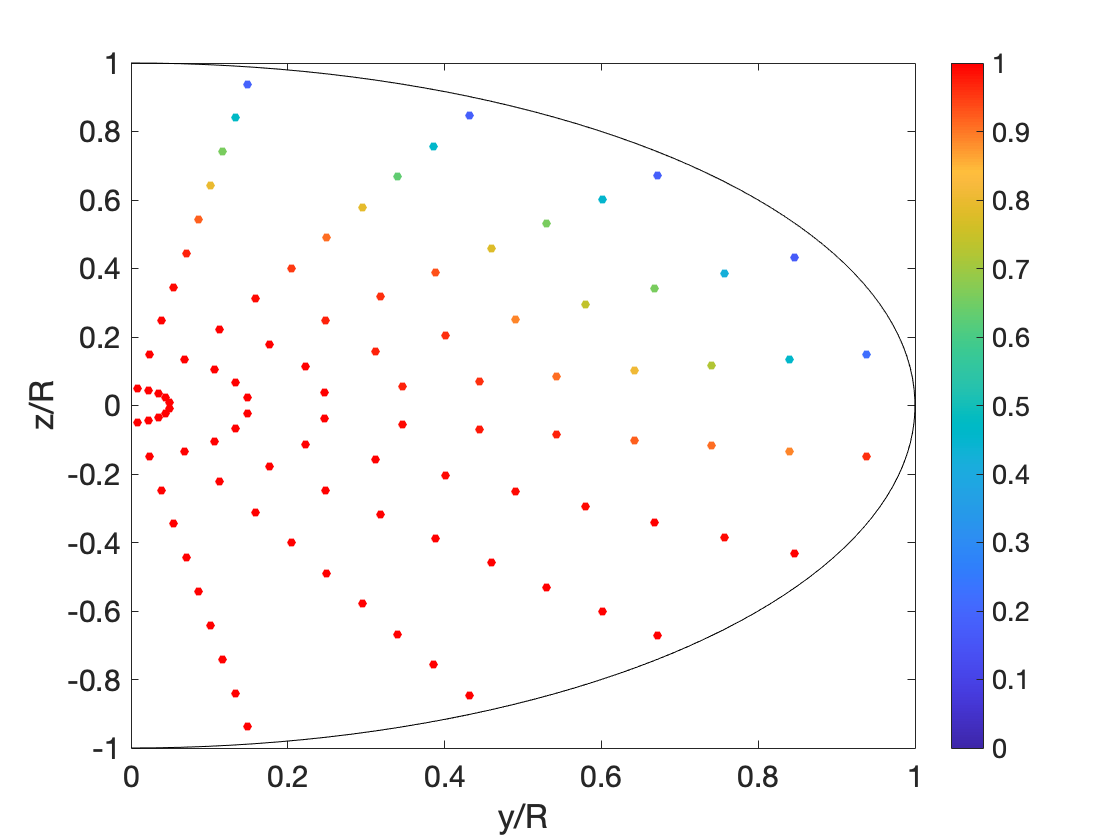}
    \includegraphics[scale=0.1]{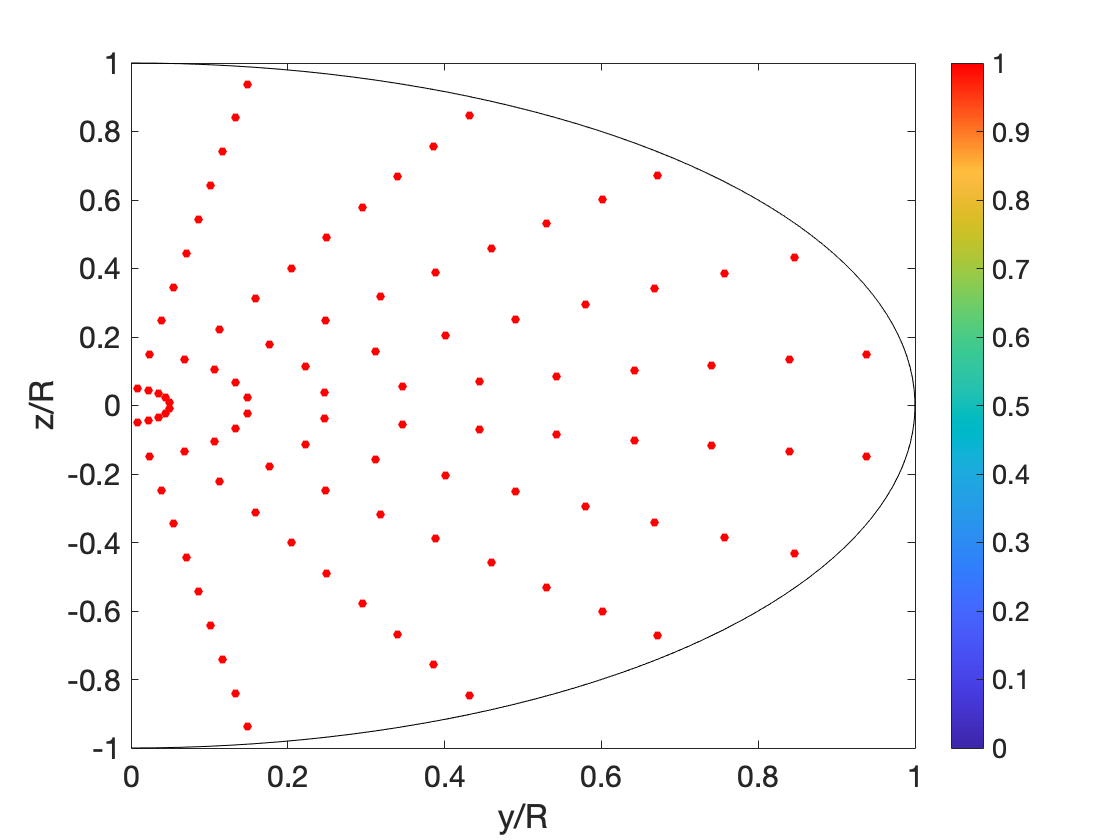}
   \includegraphics[scale=0.1]{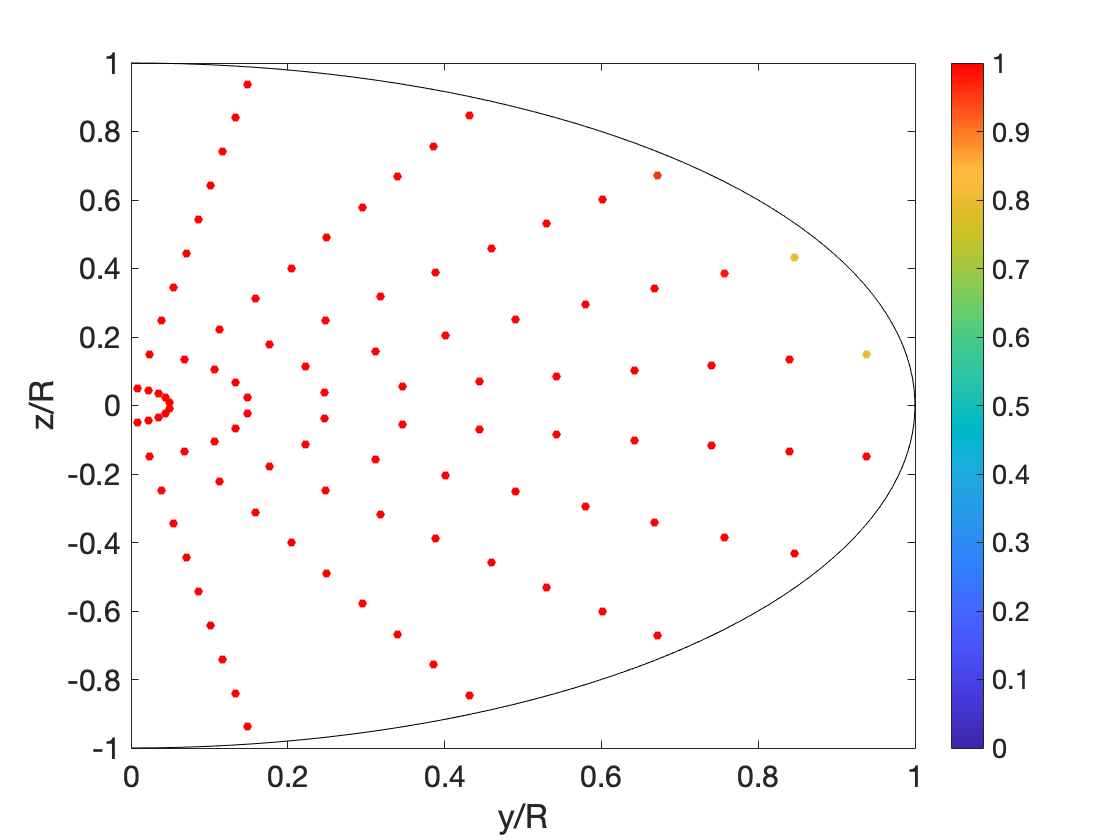}
    \caption{Capillary cross sections using the bar magnet with $K_{sh} = 0$, $d=0$ m. Top left panel shows particles released from $x=-0.011$ m, top right panel shows particles released from $x=-0.01$ m, bottom left panel shows particles released from $x=-0.00902$ m, bottom right panel shows particles released from $x=-0.008$ m. Colorbar shows particle capture rate for each point within the cross section.}
    \label{BarCapRelease0mmKsh0}
\end{figure}
\begin{figure}
    \centering
    \includegraphics[scale=0.1]{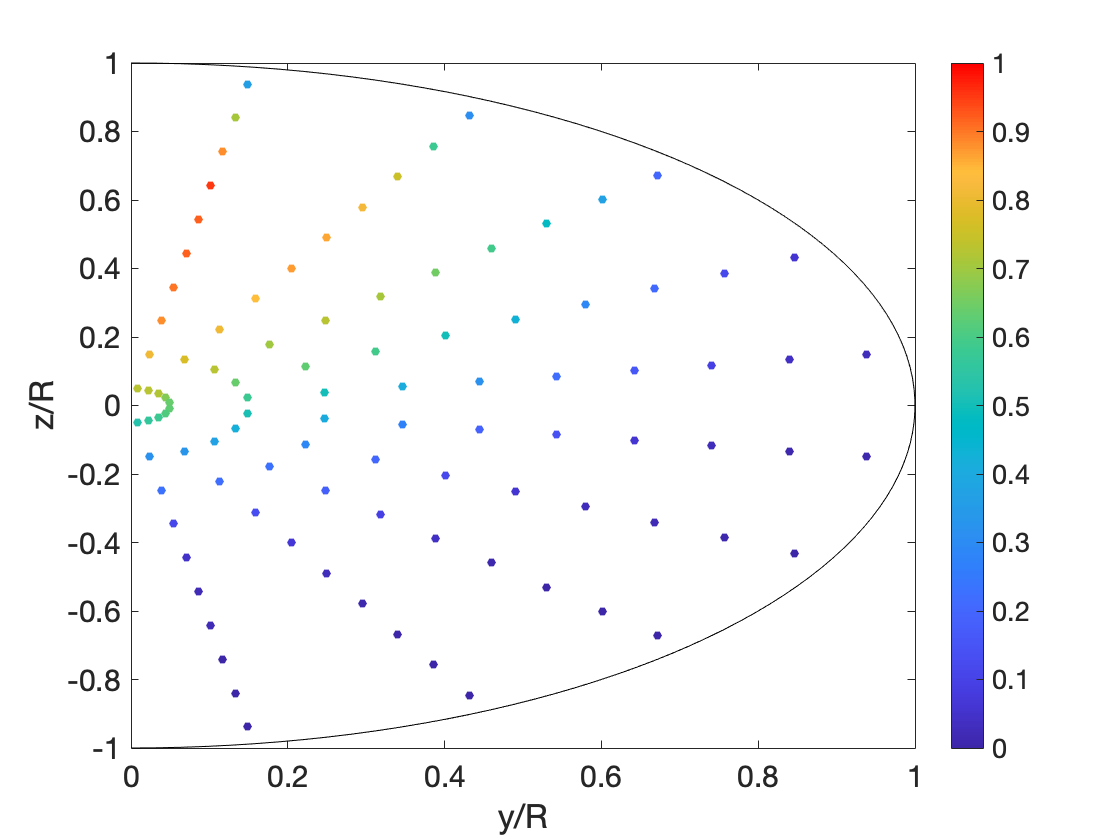}
    \includegraphics[scale=0.1]{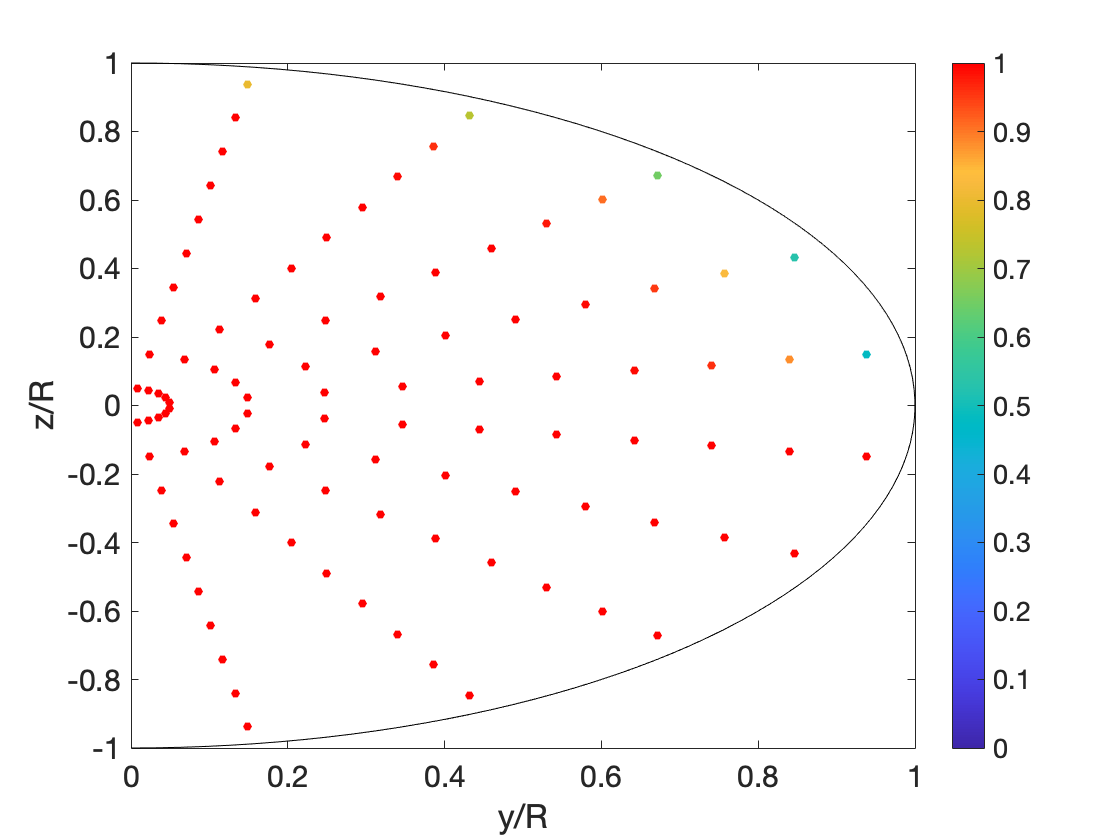}
    \includegraphics[scale=0.1]{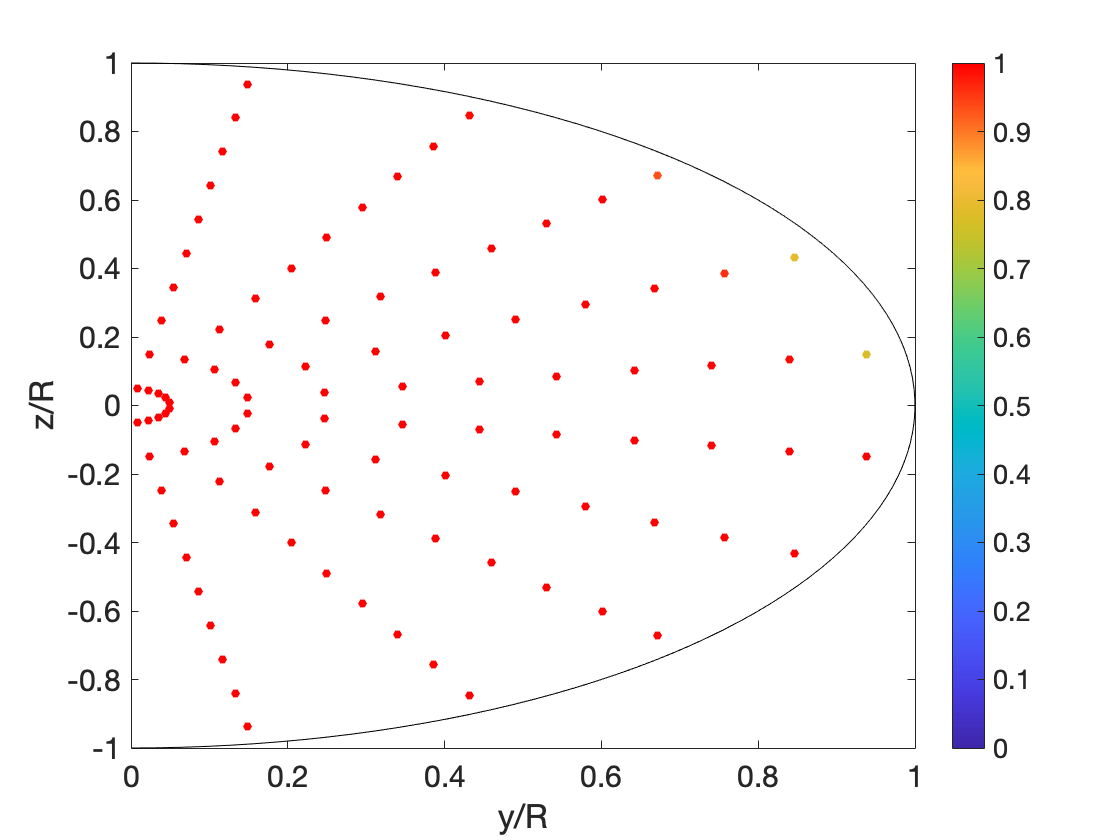}
    \includegraphics[scale=0.1]{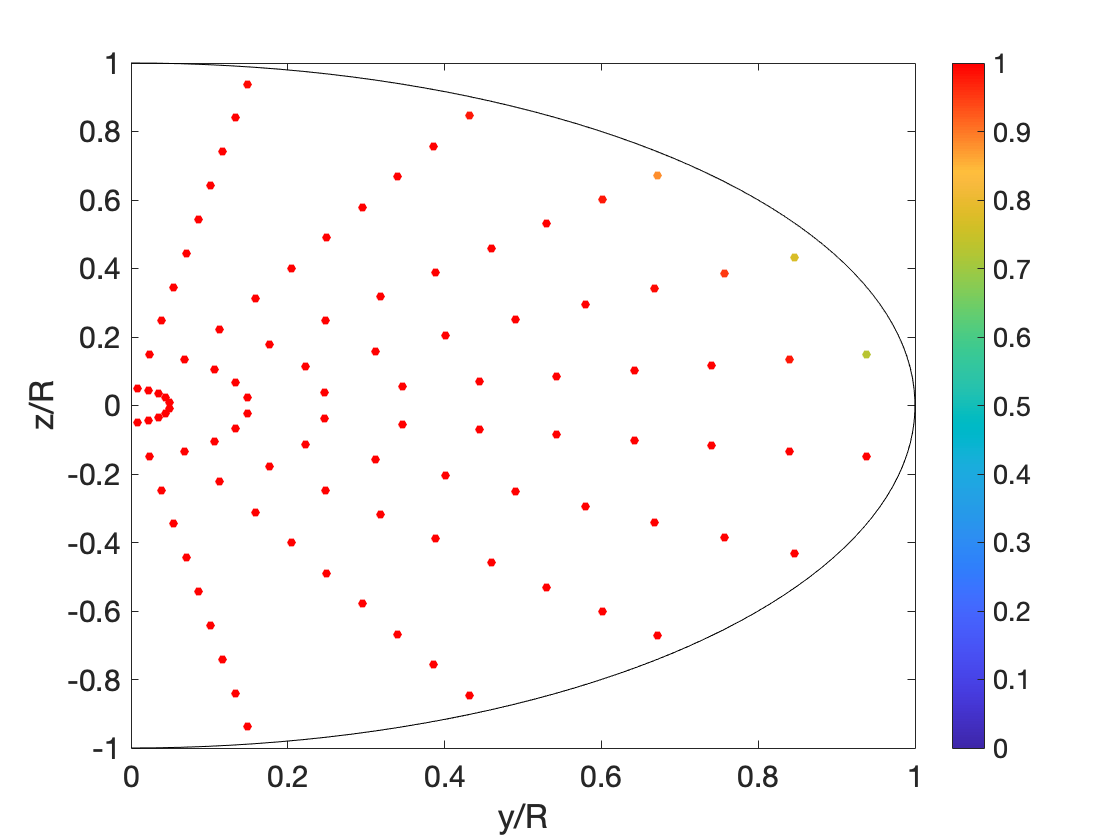}
    \caption{Capillary cross sections using the bar magnet with $K_{sh} = 0$, $d= 0.001$ m. Top left panel shows particles released from $x=-0.011$ m, top right panel shows particles released from $x=-0.01$ m, bottom left panel shows particles released from $x=-0.00902$ m, bottom right panel shows particles released from $x=-0.008$ m. Colorbar shows particle capture rate for each point within the cross section.}
    \label{BarCapRelease1mmKsh0}
\end{figure}
\begin{figure}
    \centering
    \includegraphics[scale=0.1]{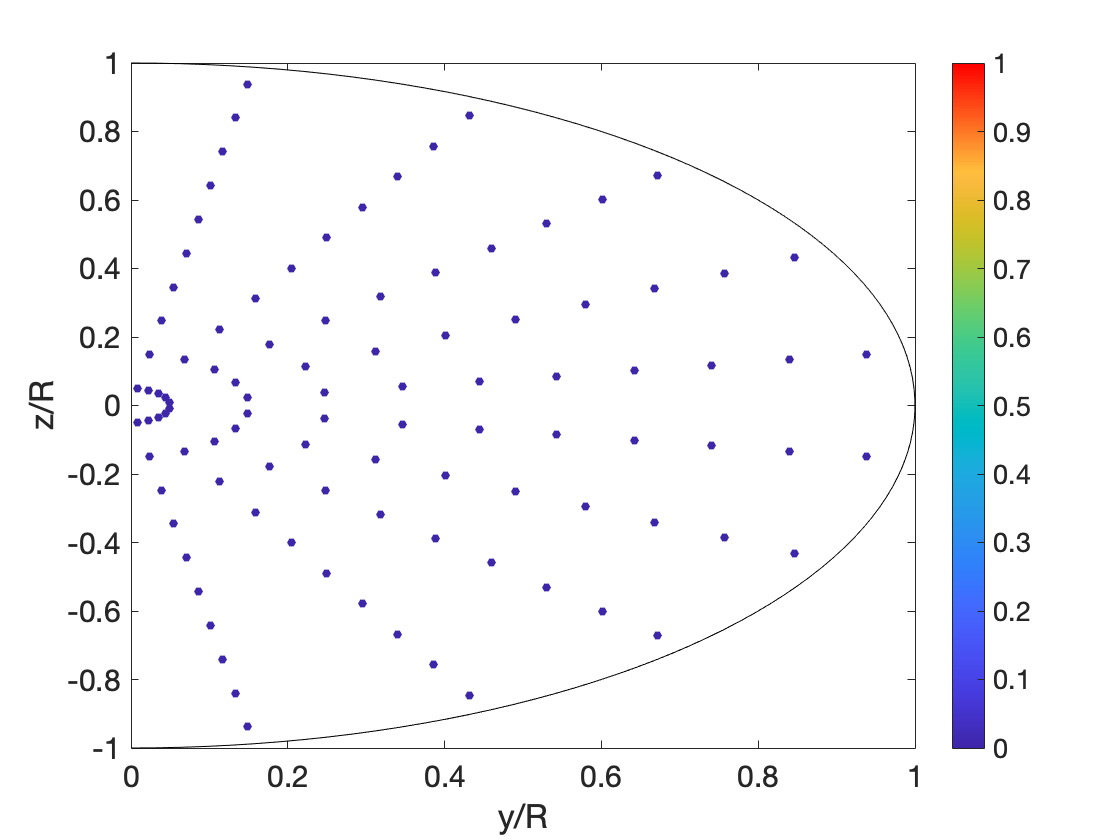}
    \includegraphics[scale=0.1]{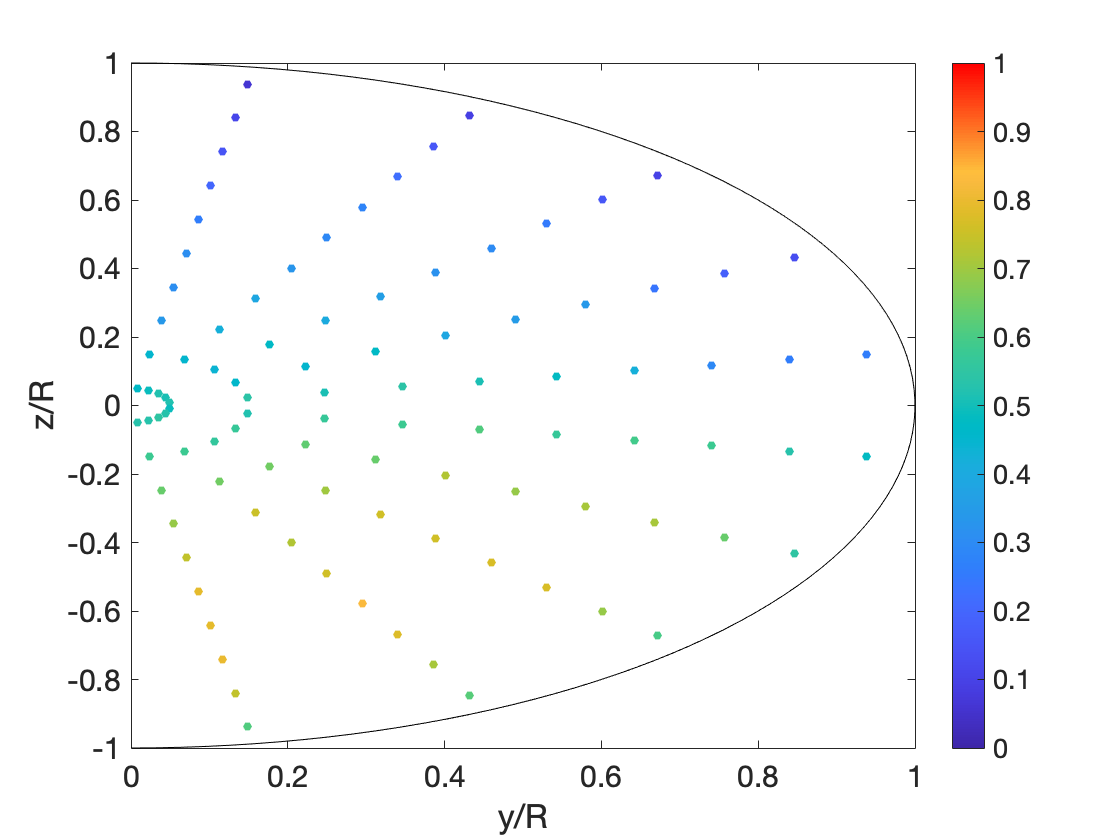}
    \includegraphics[scale=0.1]{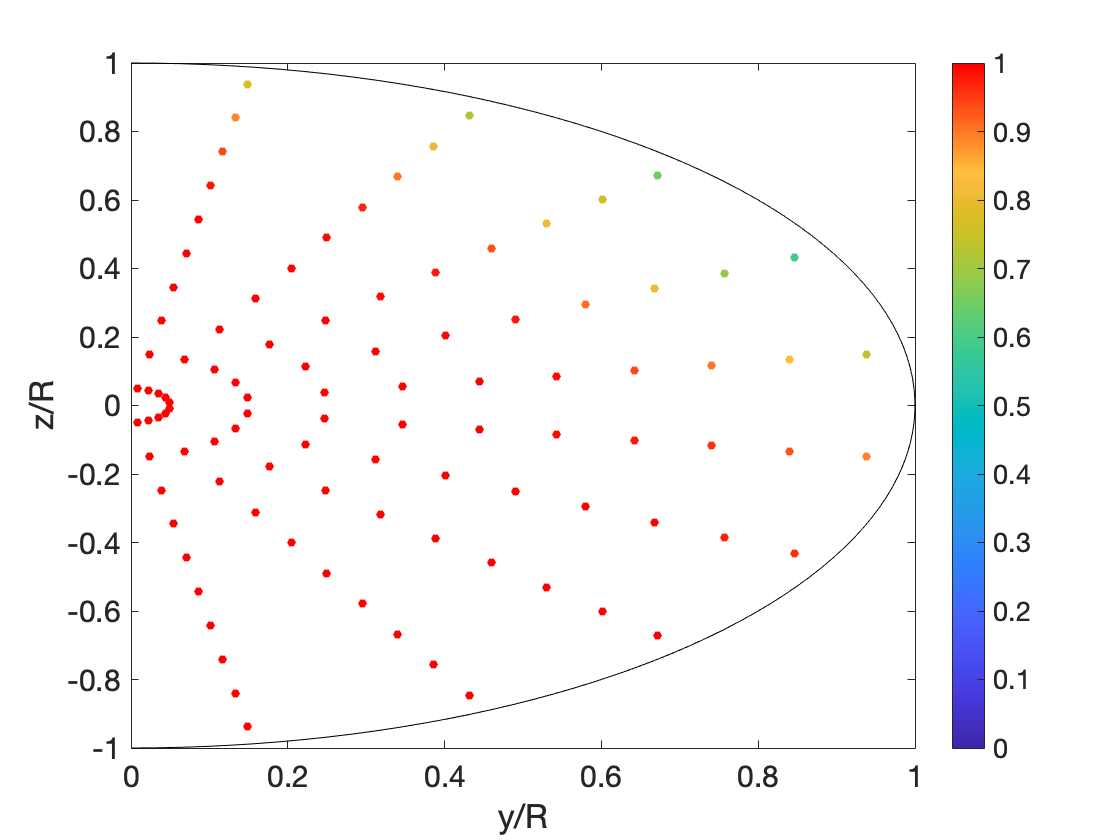}
    \includegraphics[scale=0.1]{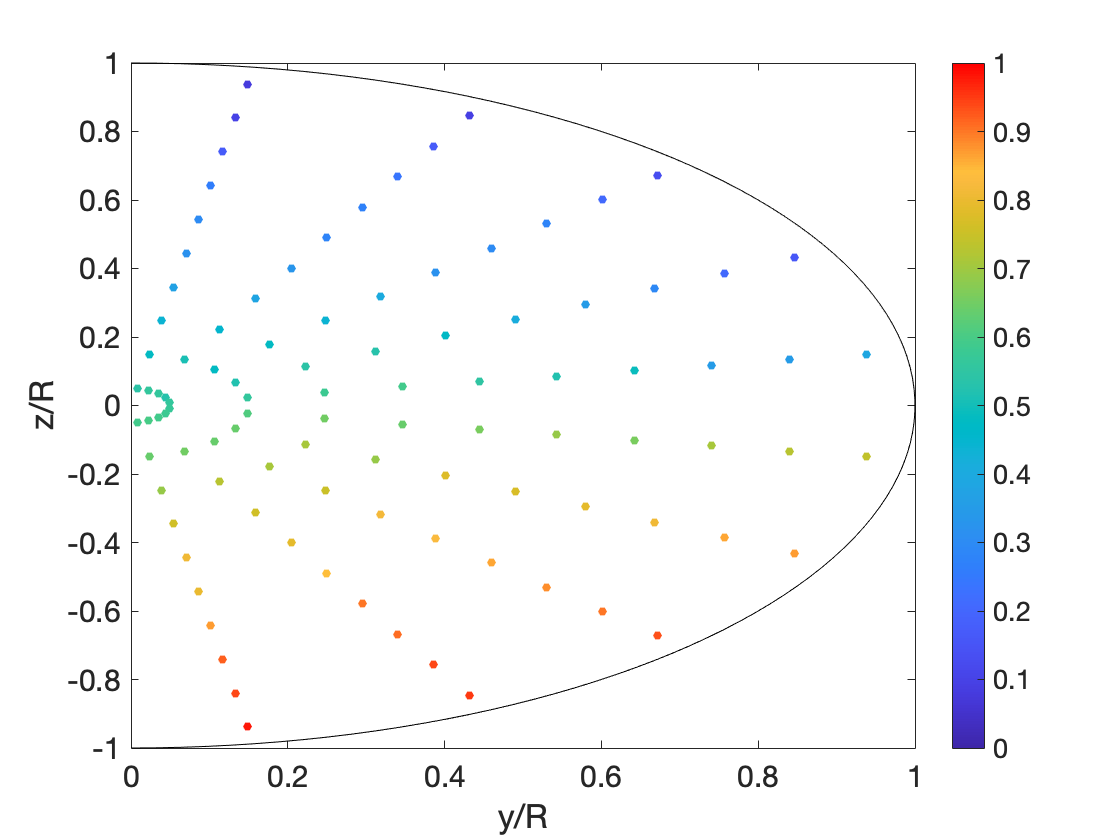}
    \caption{Capillary cross sections using the bar magnet with $K_{sh}=0.05$, $d=0$ m. Top left panel shows particles released from $x=-0.011$ m, top right panel shows particles released from $x=-0.01$ m, bottom left panel shows particles released from $x=-0.00902$ m, bottom right panel shows particles released from $x=-0.008$ m. Colorbar shows particle capture rate for each point within the cross section.}
    \label{BarCapRelease0mmKsh0.05}
\end{figure}
\begin{figure}
    \centering
    \includegraphics[scale=0.1]{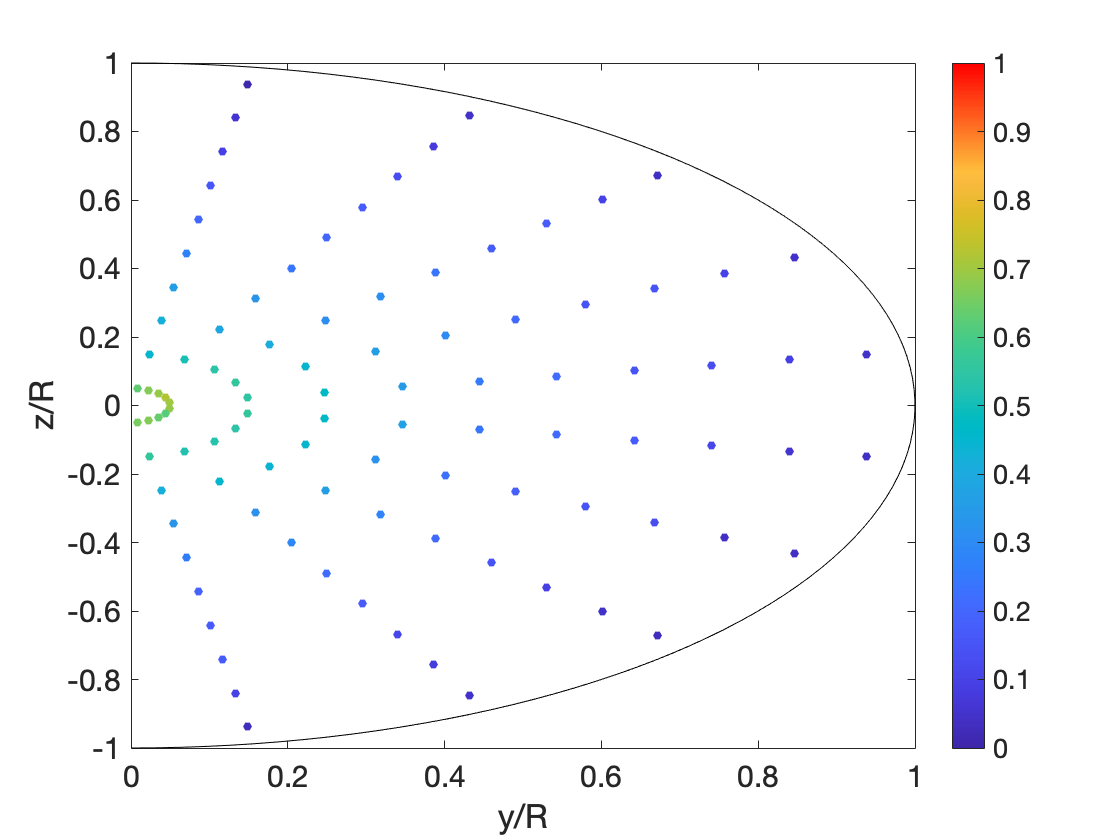}
    \includegraphics[scale=0.1]{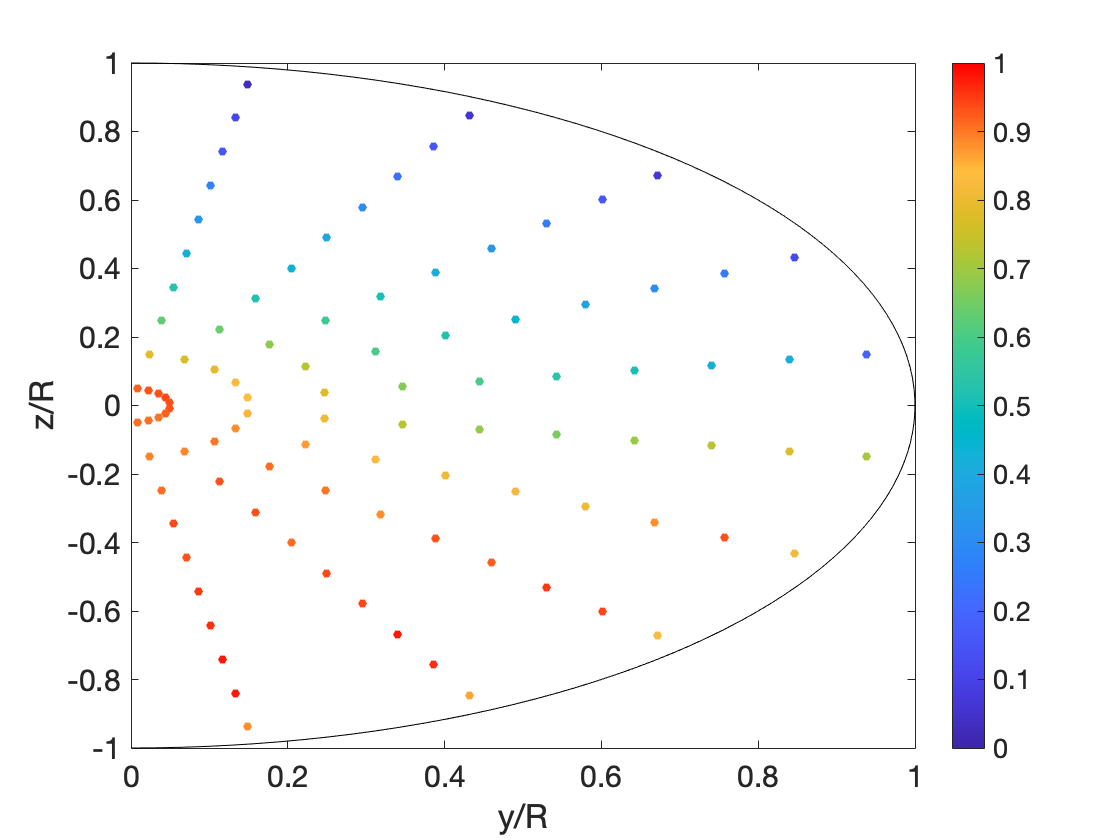}
    \includegraphics[scale=0.1]{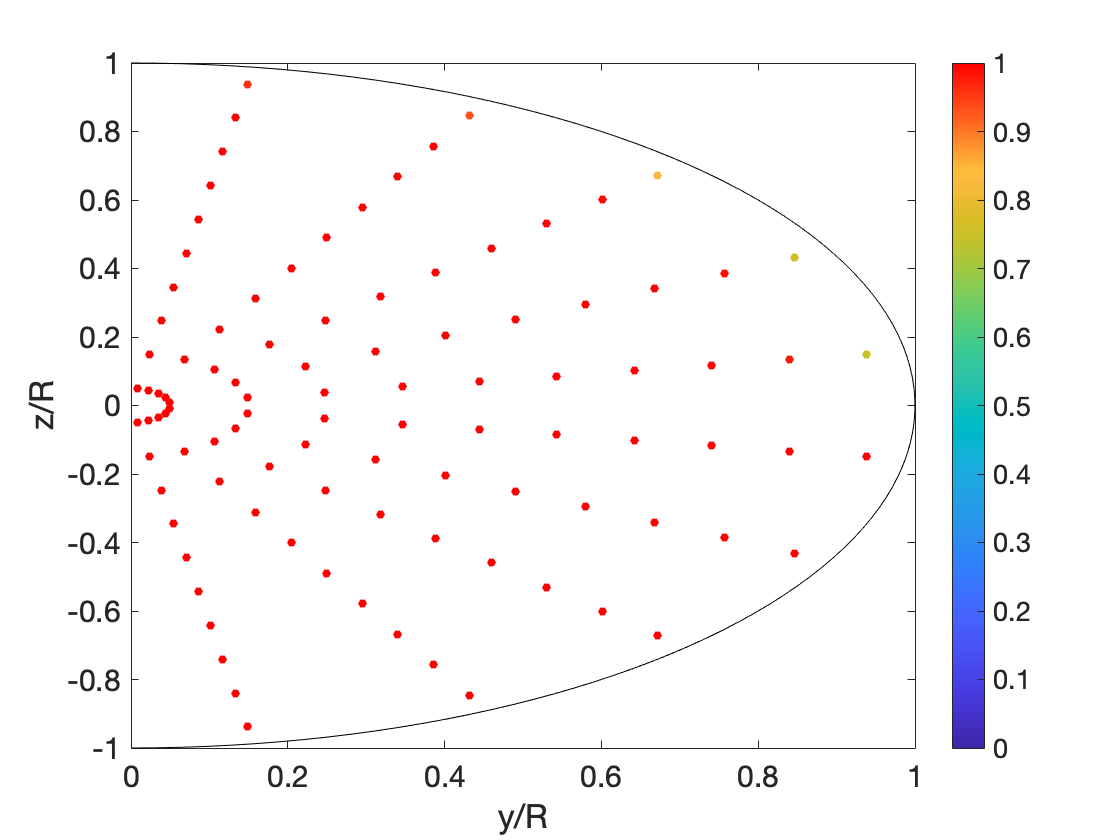}
    \includegraphics[scale=0.1]{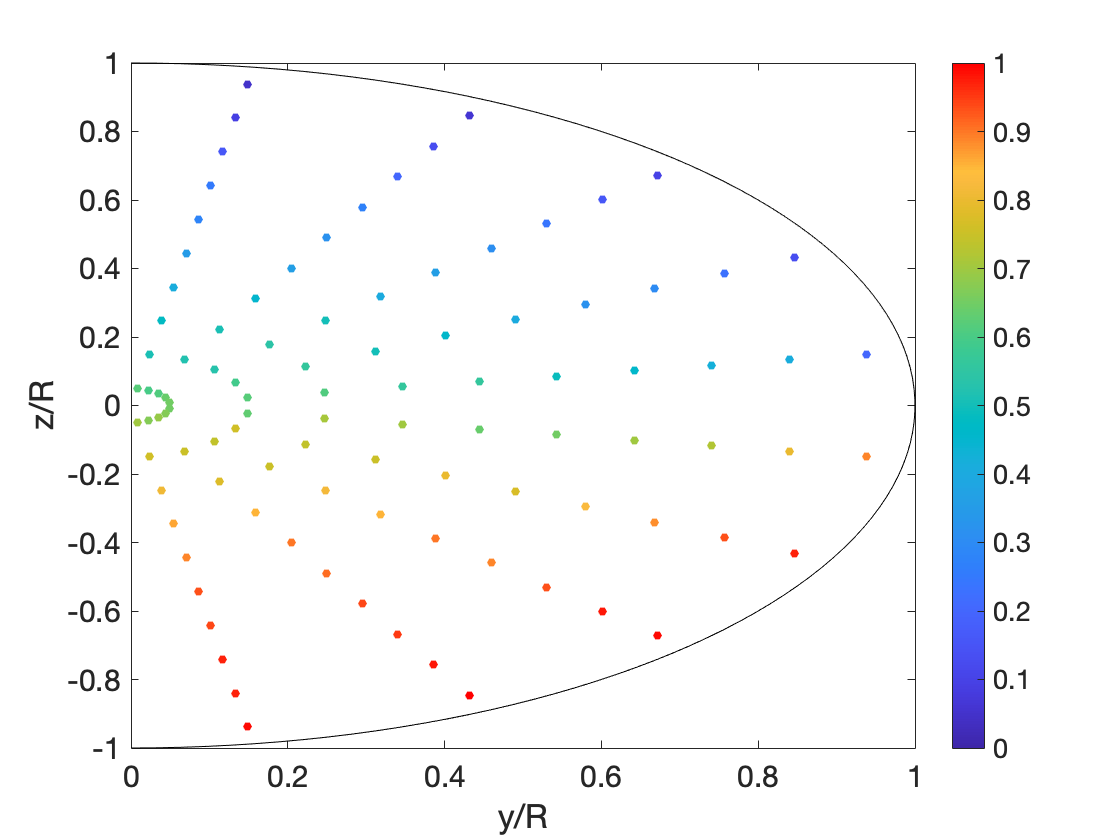}
    \caption{Arteriole cross sections using the bar magnet with $K_{sh} = 0.05$, $d=0$ m. Top left panel shows particles released from $x=-0.011$ m, top right panel shows particles released from $x=-0.01$ m, bottom left panel shows particles released from $x=-0.00902$ m, bottom right panel shows particles released from $x=-0.008$ m. Colorbar shows particle capture rate for each point within the cross section.}
    \label{BarArteriole0mm}
\end{figure}
\begin{figure}
    \centering
    \includegraphics[scale=0.1]{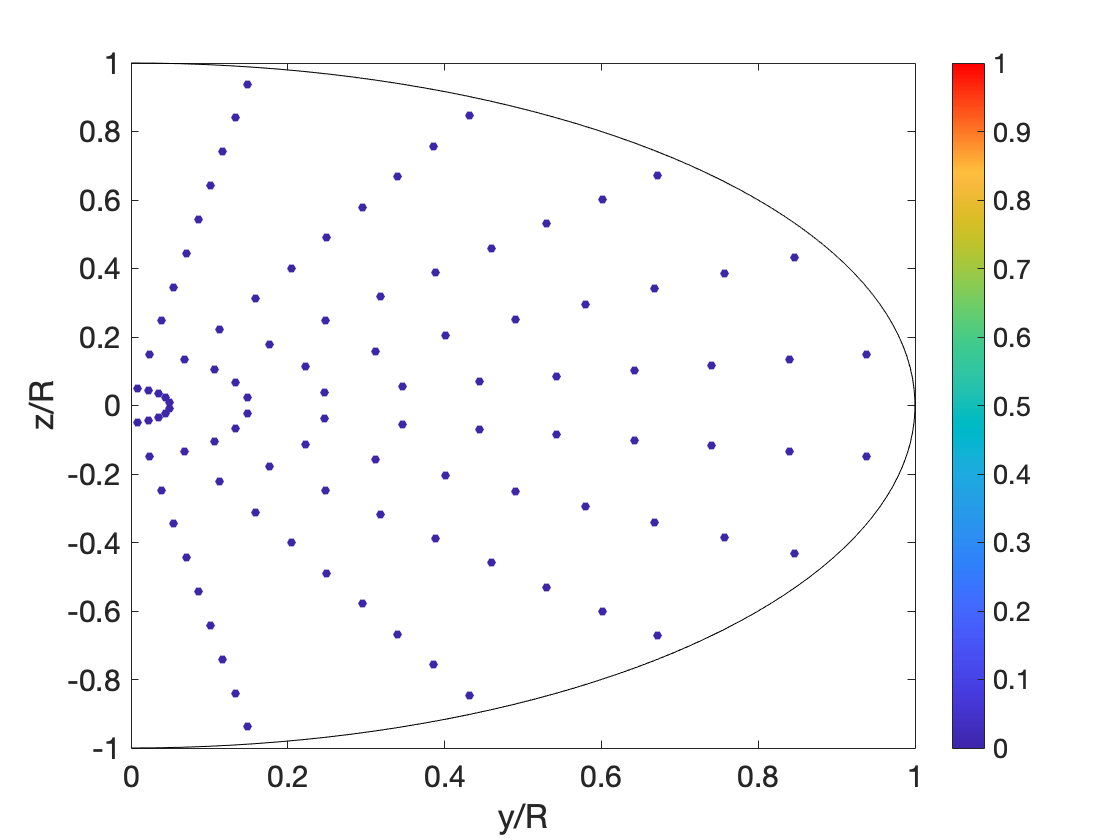}
    \includegraphics[scale=0.1]{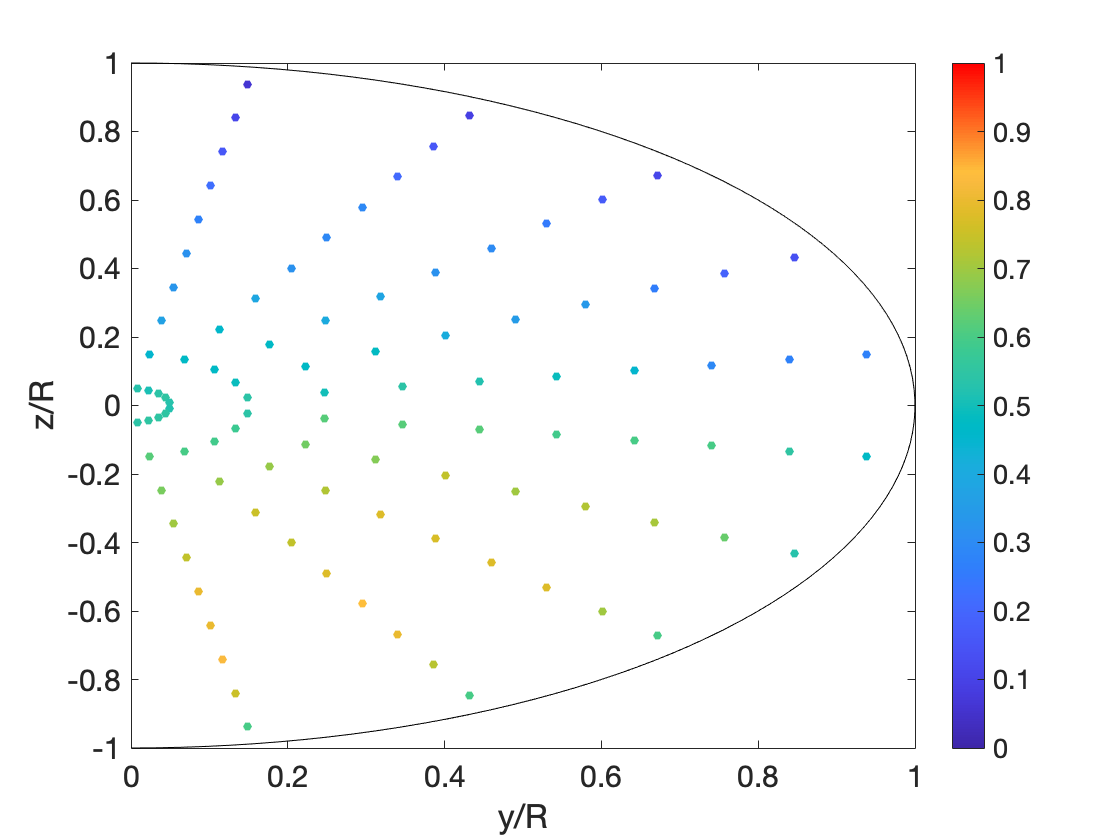}
    \includegraphics[scale=0.1]{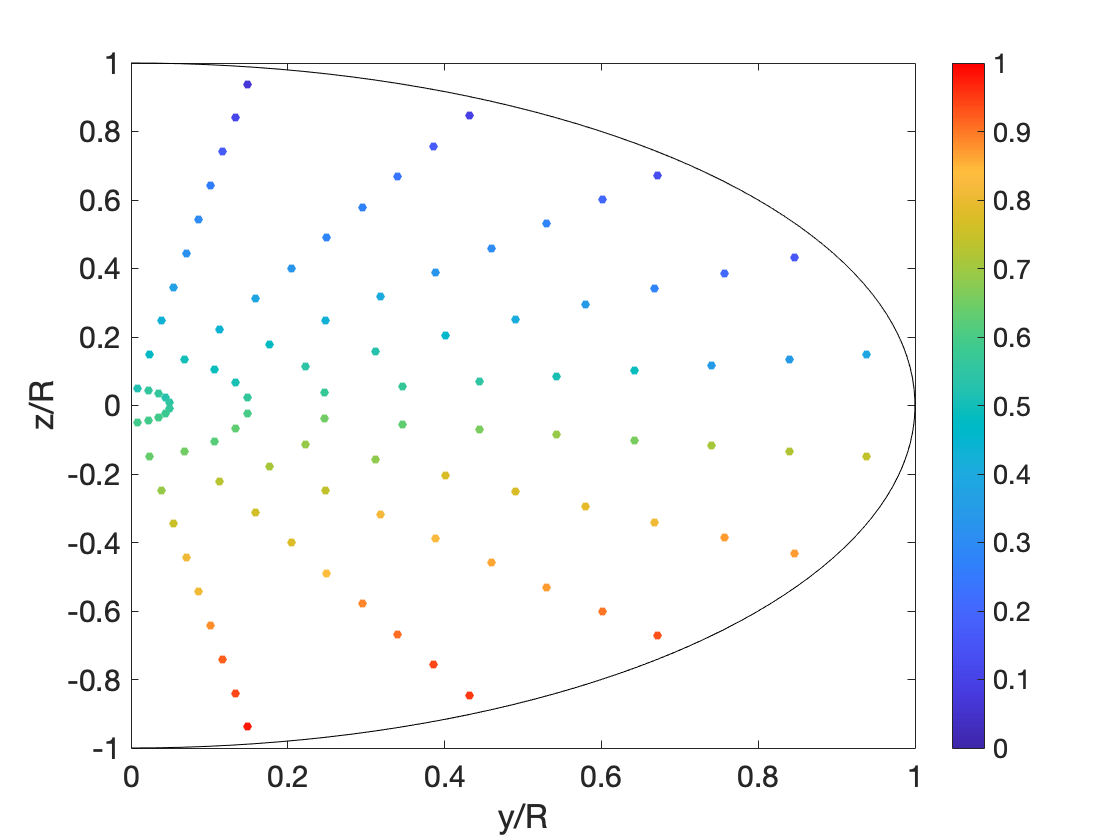}
    \includegraphics[scale=0.1]{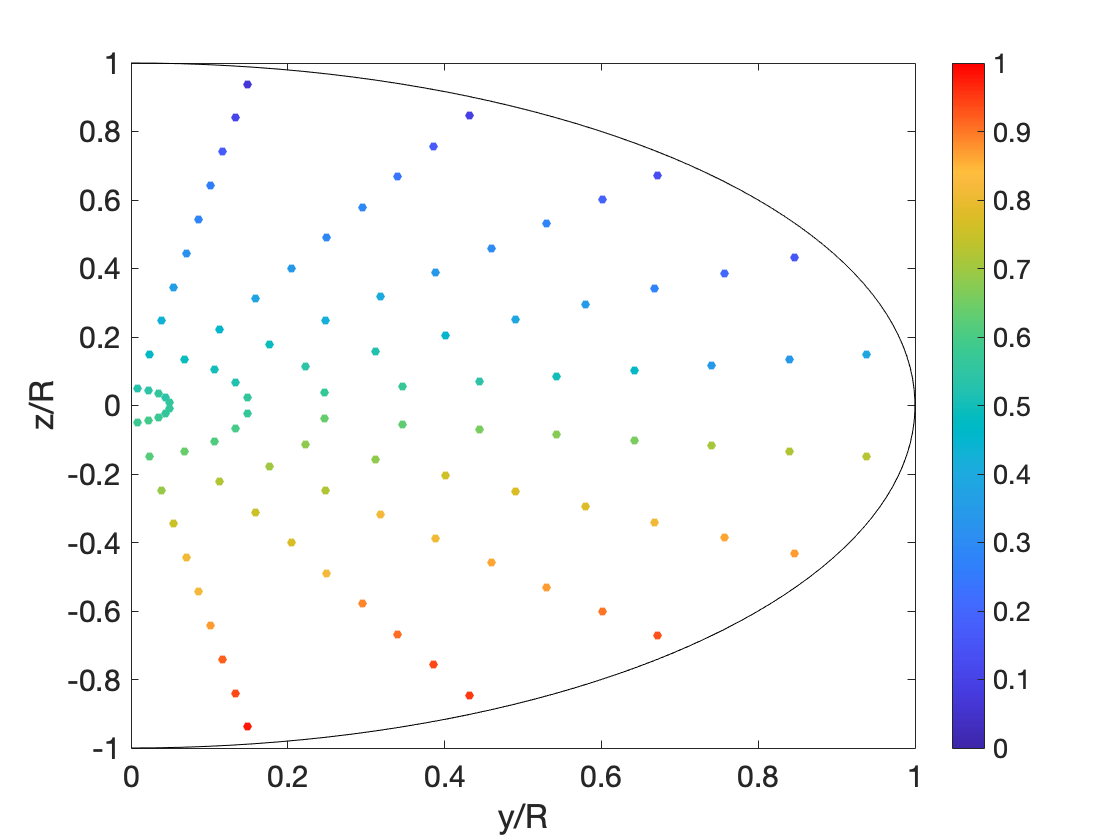}
    \caption{Capillary cross sections using the bar magnet with $K_{sh} = 0.05$, $d= 0.001$ m. Top left panel shows particles released from $x=-0.011$ m, top right panel shows particles released from $x=-0.01$ m, bottom left panel shows particles released from $x=-0.00902$ m, bottom right panel shows particles released from $x=-0.008$ m. Colorbar shows particle capture rate for each point within the cross section.}
    \label{BarCapRelease1mmKsh0.05}
\end{figure}
\begin{figure}
    \centering
    \includegraphics[scale=0.1]{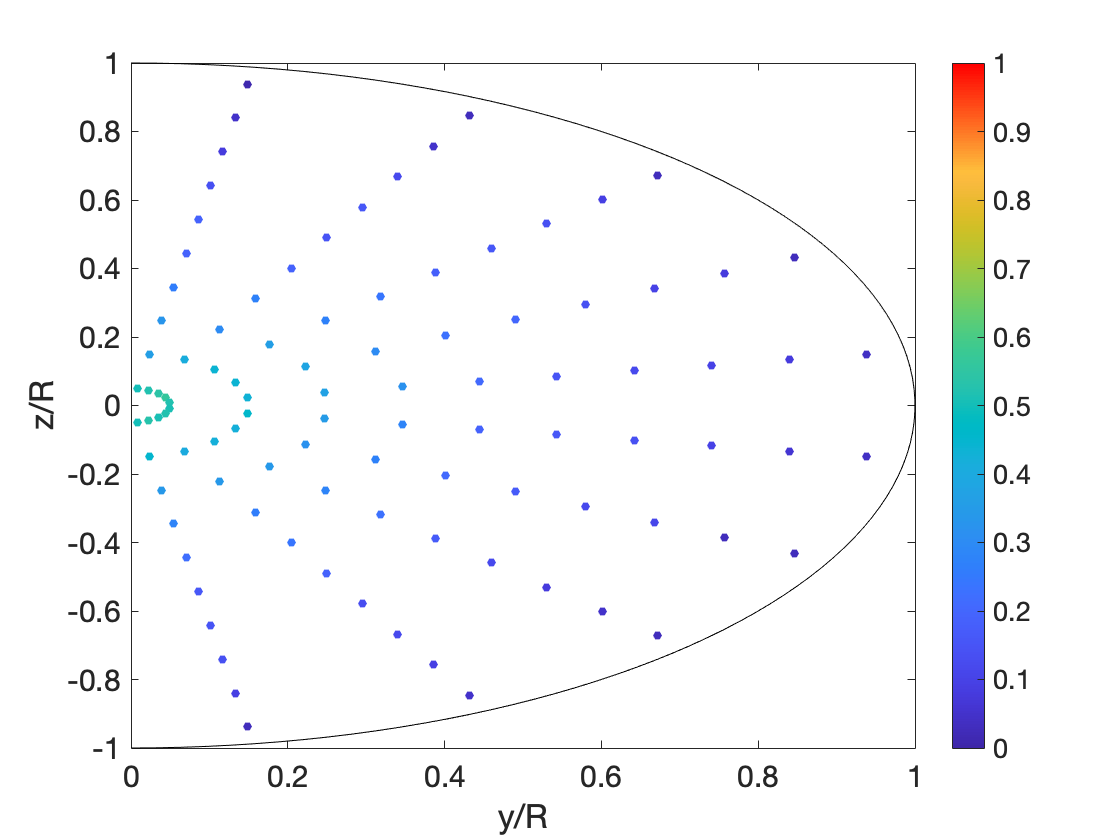}
    \includegraphics[scale=0.1]{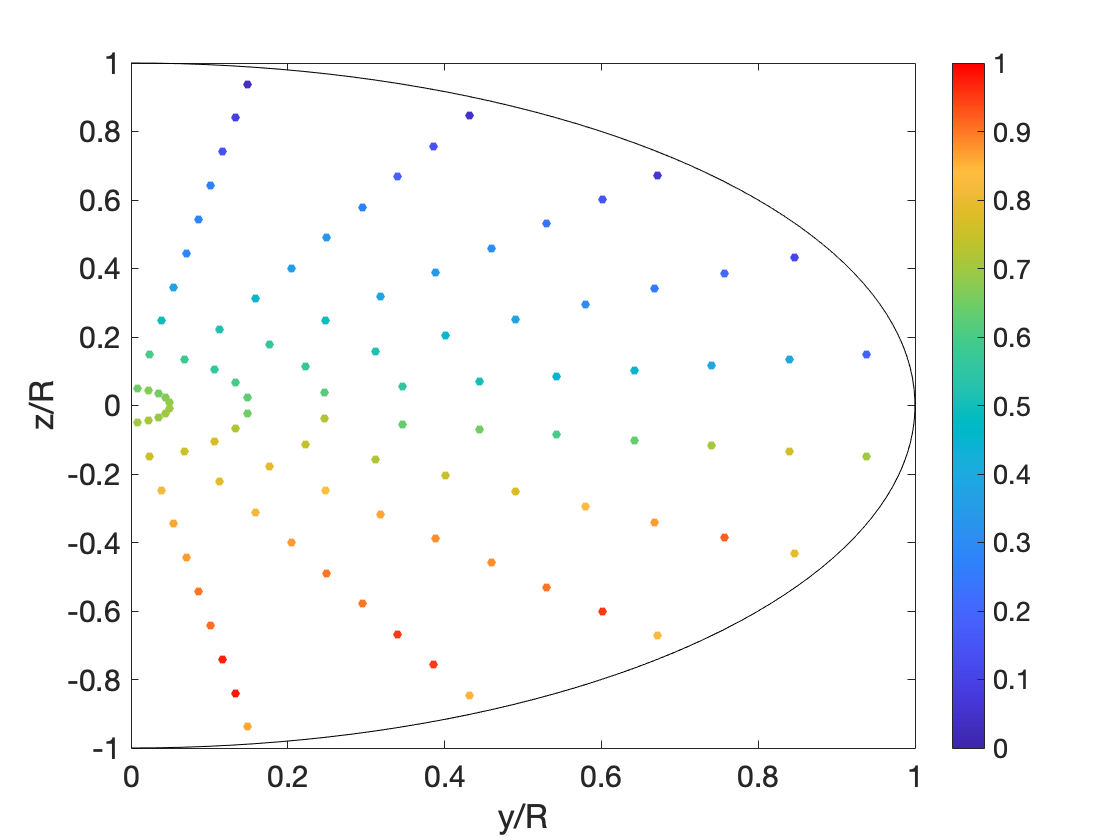}
    \includegraphics[scale=0.1]{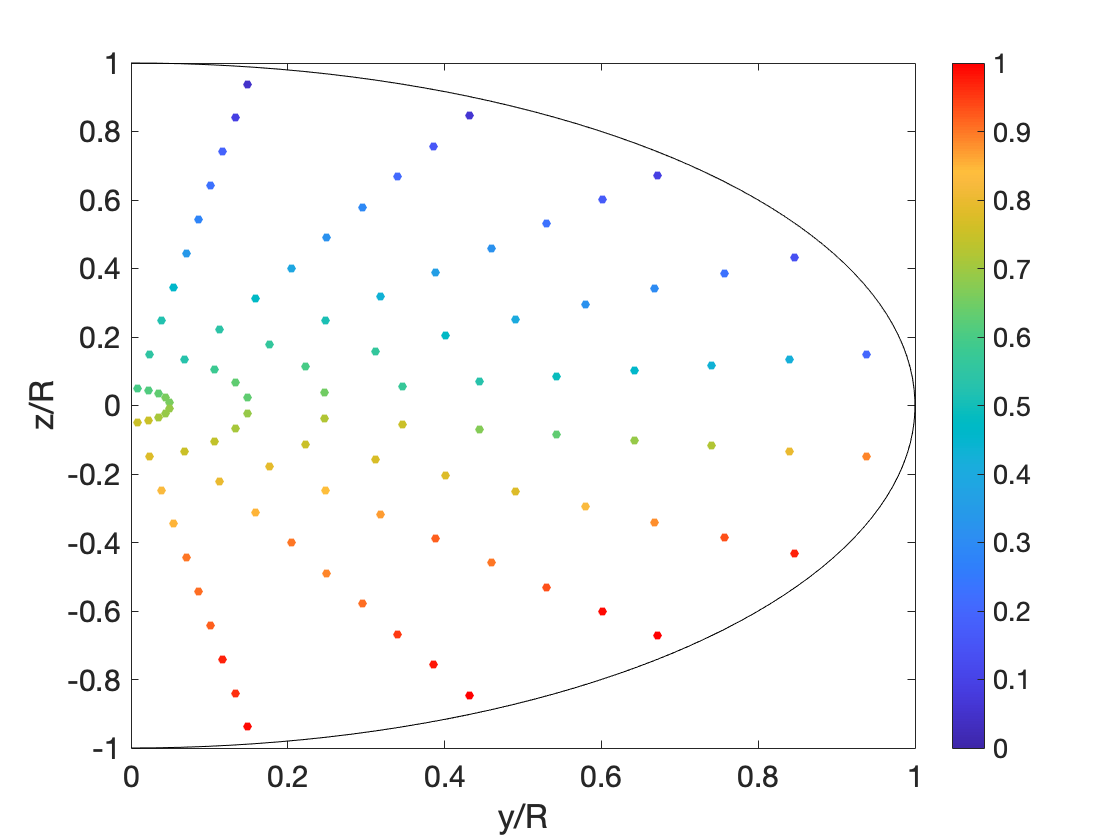}
    \includegraphics[scale=0.1]{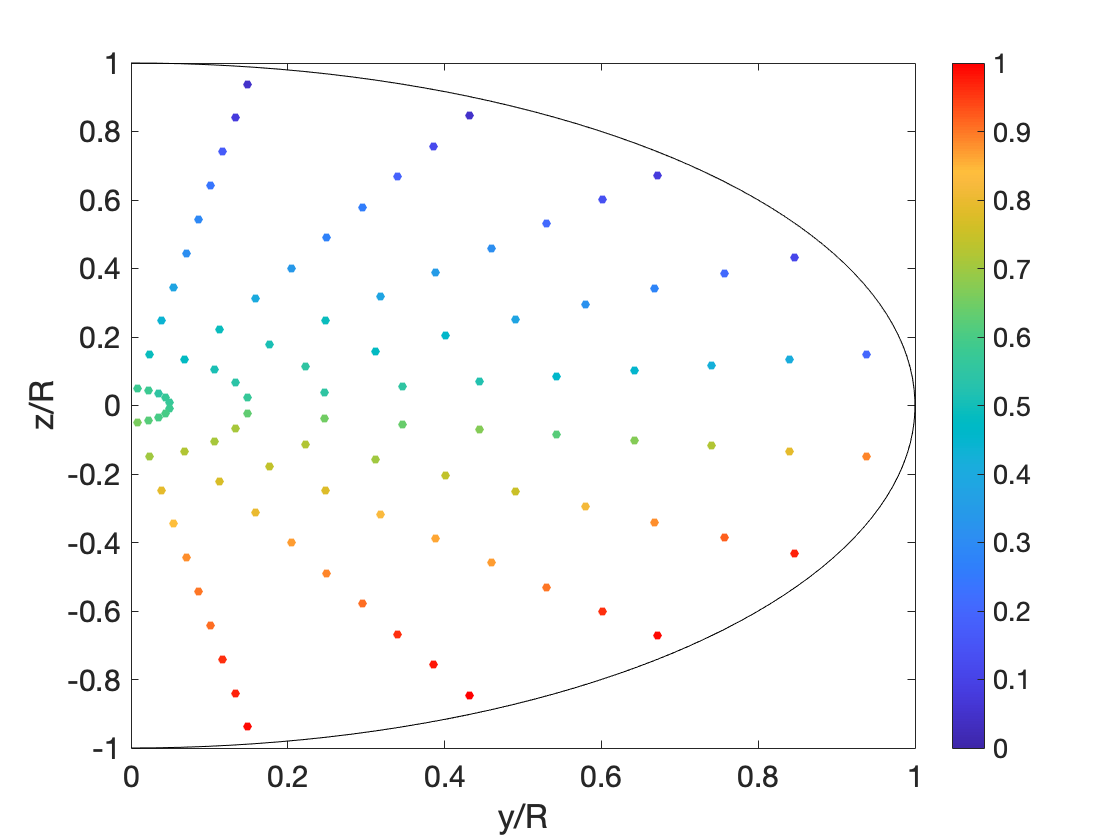}
    \caption{Arteriole cross sections using the bar magnet with $K_{sh}=0.05$, $d=0.001$ m. Top left panel shows particles released from $x=-0.011$ m, top right panel shows particles released from $x=-0.01$ m, bottom left panel shows particles released from $x=-0.00902$ m, bottom right panel shows particles released from $x=-0.008$ m. Colorbar shows particle capture rate for each point within the cross section.}
    \label{BarArteriole1mm}
\end{figure}
Figures \ref{BarCapRelease0mmKsh0} and \ref{BarCapRelease1mmKsh0} show the capture rate is close to 1 in capillaries for all particles released within the radius of the tumor when $K_{sh} = 0$. In the absence of RBC collisions, particles rapidly reach the vessel wall with minimal horizontal drift, leading to a situation similar to Figure \ref{boundary} for the dipole model. However, Figures \ref{BarCapRelease0mmKsh0.05} and \ref{BarArteriole0mm} show that when RBC collisions are present, the capture rate remains near 1 only at the edge of the bar magnet in both capillaries and arterioles. This is due to the magnetic force being the strongest at the edge of the bar magnet due to the sharp corners. At both $x=0.01$ and $x=0.008$ m, the decreased magnetic field strength is not sufficient to maintain such a high capture rate, although substantial capture still occurs. At $x= -0.011$ m, we observe a capture rate of 0 in capillaries and approximately 0.29 in arterioles. In arterioles, particles in the center of the vessel are most likely to hit the tumor because they are located farthest from the vessel wall, allowing the magnetic field to pull them into the tumor radius.

Figures \ref{BarCapRelease1mmKsh0.05} and \ref{BarArteriole1mm} show when $d$ is increased to $0.001$ m, the capture rates remain relatively constant across release points within the tumor radius. Capture rates are 0.51, 0.57, and 0.56 in capillaries and 0.6, 0.62, and 0.58 in arterioles at $x=-0.01$, $x=-0.00902$, and $x=-0.008$ m, respectively. At $d=0.001$ m, the magnetic field strength is too weak for the initial release site to have a significant impact on the capture rate. 

\section{Conclusion}
We study the motion of superparamagnetic nanoparticles in the magnetic field of both a point dipole and a physical bar magnet. We analyze particle behavior in capillaries, arterioles, and arteries, and find that while high capture rate is possible in all three vessel types, magnetic drug delivery is effective over a wide range of conditions only in capillaries and arterioles. Even in the absence of a magnetic field, we find diffusion occurs so rapidly in capillaries and arterioles that a 0.5 capture rate is still possible if nanoparticles are released at or within the tumor radius. We show that the distance from the magnet to the vessel is the most significant determinant of particle capture rate in both the point dipole and bar magnet models. Capture rates of 1 can be sustained up to a distance of 1.5 cm in both capillaries and arterioles before diminishing rapidly. The high blood velocity in arteries presents a major barrier to magnetic drug delivery, although it can be effective if larger particles of size $a_p=500$ nm are used, the magnet is placed within 2 cm of the vessel, and has a strength of at least 560 A m\textsuperscript{2}. We consider a range of $K_{sh}$ values to represent various degrees of red blood cell collisions and find that capture rate decreases linearly with $K_{sh}$ for particles located in the sensitive area; this relationship is not expected to hold at extreme distances close to or far from the magnet due to the overwhelmingly strong or weak magnetic force. When a bar magnet is used, we find the weaker magnetic field limits its effectiveness and only leads to high capture rate when placed $<1$ mm from the vessel when $K_{sh} = 0.05$. However, the bar magnet achieves high capture rate up to 1 cm when $K_{sh} = 0$, indicating a need to better understand the extent of RBC collisions in blood vessels. Finally, we find that larger particles significantly increase capture rates, making further study of particle clusters a promising direction, although potential vessel occlusion in smaller vessels must be considered and limits the maximum particle size.

Our results suggest a few key recommendations for future research and transition to clinical applications: (i) nanoparticles should be released as close to the tumor site as possible to prevent particles diffusing to the vessel wall too early; (ii) magnets should be placed as close to the tumor as possible, and that increasing the magnetic strength does not significantly extend the range of magnet placements; (iii) magnetic drug delivery should be used in capillaries and arterioles, although it is effective in limited cases in arteries only if the particle size is increased; (iv) using larger nanoparticles or clusters of nanoparticles can lead to significantly higher capture rates in arterioles. 

Finally, an underlying assumption of our model is that blood flow is analogous to pipe flow; however, this may only be an appropriate approximation for healthy individuals. For patients with atherosclerosis or other conditions that modify the vessel wall, these results may no longer be representative. Additionally, medications such as blood thinners would affect the blood velocity and viscosity profiles, and cause particle trajectories to deviate from these results. Vessel bifurcations and more complex geometries are also possible directions for future research. Nevertheless, our results are informative for future researchers because they presents clear guidelines for nanoparticle release location, magnet placement, and magnet strength. 

\section*{Acknowledgements} \label{sec:acknowledgements}
We would like to thank the Liberty Science Center Partners in Science Program, Mr. Ruben Rosario, and Dr. Kara Mann for providing high school students the opportunity to conduct research with a mentor. 
SA acknowledges the support by the Petroleum Research Fund PRF-59641-ND.

\appendix
\section{Bar Magnet Field Equations}
Here we state the equations for the magnetic field surrounding a bar magnetic with width $2a$, length $2b$, and height $L$, which were obtained from \cite{Mag_equ}. Note that here the origin is located at the top of the Bar magnet (see Figure \ref{Schematic}).
\label{sec:appendix}
\begin{multline}
    H_x = \frac{B_r}{4\pi\mu_0}\log \left[{\frac{(y+b) + [(y+b)^2 + (x-a)^2 + z^2]^\frac{1}{2}}{(y-b) + [(y-b)^2+(x-a)^2 + z^2]^\frac{1}{2}}}\right.  \\
    \times\left. \frac{(y-b) + [(y-b)^2+(x+a)^2+z^2]^\frac{1}{2}}{(y+b)+[(y+b)^2+(x+a)^2+z^2]^\frac{1}{2}}\right]  \\
    -\frac{B_r}{4\pi\mu_0}\log \left[{\frac{(y+b)+[(y+b)^2 + (x-a)^2+(z+L)^2]^\frac{1}{2}}{(y-b) +  [(y-b)^2+(x-a)^2+(z+L)^2]^\frac{1}{2}}}\right.  \\
    \times\left. \frac{(y-b)+[(y-b)^2+(x+a)^2+(z+L)^2]^\frac{1}{2}}{(y+b)+[(y+b)^2+(x+a)^2+(z+L)^2]^\frac{1}{2}}\right]
\end{multline}

\begin{multline}
    H_y = \frac{B_r}{4\pi\mu_0}\log\left[\frac{(x+a)+[(y-b)^2+(x+a)^2+z^2]^\frac{1}{2}}{(x-a)+[(y-b)^2+(x-a)^2+z^2]^\frac{1}{2}}\right. \\
    \times \left. \frac{(x-a)+[(y+b)^2+(x-a)^2+z^2]^\frac{1}{2}}{(x+a)+[(y+b)^2+(x+a)^2+z^2]^\frac{1}{2}} \right] \\
    -\frac{B_r}{4\pi\mu_0}\log\left[\frac{(x+a)+[(y-b)^2+(x+a)^2+(z+L)^2]^\frac{1}{2}}{(x-a)+[(y-b)^2+(x-a)^2+(z+L)^2]^\frac{1}{2}}\right. \\
    \times \left. \frac{(x-a)+[(y+b)^2+(x-a)^2+(z+L)^2]^\frac{1}{2}}{(x+a)+[(y+b)^2+(x+a)^2+(z+L)^2]^\frac{1}{2}}\right]
\end{multline}

\begin{multline}
    H_z=\frac{B_r}{4\pi\mu_0}\left[\tan^{-1}\left(\frac{(x+a)(y+b)}{z[(x+a)^2+(y+b)^2+z^2]^\frac{1}{2}}\right)\right. \\
    +\tan^{-1}\left(\frac{(x-a)(y-b)}{z[(x-a)^2+(y-b)^2+z^2]^\frac{1}{2}}\right) \\
    - \tan^{-1}\left(\frac{(x+a)(y-b)}{z[(x+a)^2+(y-b)^2+z^2]^\frac{1}{2}}\right)\\
    -\left.\tan^{-1}\left(\frac{(x-a)(y+b)}{z[(x-a)^2+(y+b)^2+z^2]^\frac{1}{2}}\right)\right] \\
    -\frac{B_r}{4\pi\mu_0}\left[\tan^{-1}\left(\frac{(x+a)(y+b)}{(z+L)[(x+a)^2+(y+b)^2+(z+L)^2]^\frac{1}{2}}\right)\right.  \\
    + \tan^{-1}\left(\frac{(x-a)(y-b)}{(z+L)[(x-a)^2+(y-b)^2+(z+L)^2]^\frac{1}{2}}\right)  \\
    - \tan^{-1}\left(\frac{(x+a)(y-b)}{(z+L)[(x+a)^2+(y-b)^2+(z+L)^2]^\frac{1}{2}}\right)  \\
    -\left.\tan^{-1}\left(\frac{(x-a)(y+b)}{(z+L)[(x-a)^2+(y+b)^2+(z+L)^2]^\frac{1}{2}}\right)\right]
\end{multline}

\newpage
\bibliography{Bib}

\begin{thebibliography}{10}

\bibitem{mefford2007field}
O.T. Mefford, R.C. Woodward, J.D. Goff, T.P. Vadala, T.G.S. Pierre, J.P.
  Dailey, and J.S. Riffle.
\newblock Field-induced motion of ferrofluids through immiscible viscous media:
  Testbed for restorative treatment of retinal detachment.
\newblock {\em Journal of Magnetism and Magnetic Materials}, 311(1):347--353,
  2007.

\bibitem{izci2021alternate}
M.~Izci, C.~Maksoudian, B.B. Manshian, and S.J. Soenen.
\newblock The use of alternative strategies for enhanced nanoparticle delivery
  to solid tumors.
\newblock {\em Chemical Reviews}, 121(3):1746--1803, 2021.

\bibitem{mody2014magnetic}
V.V. Mody, A.~Cox, S.~Shah, A.~Singh, W.~Bevins, and H.~Parihar.
\newblock Magnetic nanoparticle drug delivery systems for targeting tumor.
\newblock {\em Applied Nanoscience}, 4(4):385--392, 2014.

\bibitem{liu2017magnetic}
Y.~Liu, M.~Li, F.~Yang, and N.~Gu.
\newblock Magnetic drug delivery systems.
\newblock {\em Science China Materials}, 60(6):471--486, 2017.

\bibitem{yue2012motion}
P.~Yue, S.~Lee, S.~Afkhami, and Y.~Renardy.
\newblock On the motion of superparamagnetic particles in magnetic drug
  targeting.
\newblock {\em Acta Mechanica}, 223(3):505--527, 2012.

\bibitem{rukshin2017modeling}
I.~Rukshin, J.~Mohrenweiser, P.~Yue, and S.~Afkhami.
\newblock Modeling superparamagnetic particles in blood flow for applications
  in magnetic drug targeting.
\newblock {\em Fluids}, 2(2):29, 2017.

\bibitem{grief2005mathematical}
A.D. Grief and G.~Richardson.
\newblock Mathematical modelling of magnetically targeted drug delivery.
\newblock {\em Journal of Magnetism and Magnetic Materials}, 293(1):455--463,
  2005.

\bibitem{sharma2015mathematical}
S.~Sharma, V.K. Katiyar, and U.~Singh.
\newblock Mathematical modelling for trajectories of magnetic nanoparticles in
  a blood vessel under magnetic field.
\newblock {\em Journal of Magnetism and Magnetic Materials}, 379:102--107,
  2015.

\bibitem{dadfar2019iron}
S.M. Dadfar, K.~Roemhild, N.I. Drude, S.~von Stillfried, R.~Kn{\"u}chel,
  F.~Kiessling, and T.~Lammers.
\newblock Iron oxide nanoparticles: Diagnostic, therapeutic and theranostic
  applications.
\newblock {\em Advanced Drug Delivery Reviews}, 138:302--325, 2019.

\bibitem{rosensweig1985ferrohydrodynamics}
R.E. Rosensweig.
\newblock Ferrohydrodynamics cambridge univ.
\newblock {\em Press, Cambridge}, page 344, 1985.

\bibitem{dobson2006magnetic}
J.~Dobson.
\newblock Magnetic nanoparticles for drug delivery.
\newblock {\em Drug Development Research}, 67(1):55--60, 2006.

\bibitem{ulbrich2016targeted}
K.~Ulbrich, K.~Hola, V.~Subr, A.~Bakandritsos, J.~Tucek, and R.~Zboril.
\newblock Targeted drug delivery with polymers and magnetic nanoparticles:
  covalent and noncovalent approaches, release control, and clinical studies.
\newblock {\em Chemical Reviews}, 116(9):5338--5431, 2016.

\bibitem{mukherjee2020recent}
S.~Mukherjee, L.~Liang, and O.~Veiseh.
\newblock Recent advancements of magnetic nanomaterials in cancer therapy.
\newblock {\em Pharmaceutics}, 12(2):147, 2020.

\bibitem{tran2017cancer}
S.~Tran, P.~DeGiovanni, B.~Piel, and P.~Rai.
\newblock Cancer nanomedicine: a review of recent success in drug delivery.
\newblock {\em Clinical and Translational Medicine}, 6(1):1--21, 2017.

\bibitem{kheirkhah2018magnetic}
P.~Kheirkhah, S.~Denyer, A.D. Bhimani, G.D. Arnone, D.R. Esfahani, T.~Aguilar,
  J.~Zakrzewski, I.~Venugopal, N.~Habib, G.L. Gallia, et~al.
\newblock Magnetic drug targeting: a novel treatment for intramedullary spinal
  cord tumors.
\newblock {\em Scientific Reports}, 8(1):1--9, 2018.

\bibitem{ak2018vitro}
G.~Ak, H.~Yilmaz, A.~G{\"u}ne{\c{s}}, and S.~Hamarat Sanlier.
\newblock In vitro and in vivo evaluation of folate receptor-targeted a novel
  magnetic drug delivery system for ovarian cancer therapy.
\newblock {\em Artificial Cells, Nanomedicine, and Biotechnology},
  46(sup1):926--937, 2018.

\bibitem{elbialy2015doxorubicin}
N.S. Elbialy, M.M. Fathy, and W.M. Khalil.
\newblock Doxorubicin loaded magnetic gold nanoparticles for in vivo targeted
  drug delivery.
\newblock {\em International Journal of Pharmaceutics}, 490(1-2):190--199,
  2015.

\bibitem{ye2018manipulating}
H.~Ye, Z.~Shen, L.~Yu, M.~Wei, and Y.~Li.
\newblock Manipulating nanoparticle transport within blood flow through
  external forces: An exemplar of mechanics in nanomedicine.
\newblock {\em Proceedings of the Royal Society A: Mathematical, Physical and
  Engineering Sciences}, 474(2211):20170845, 2018.

\bibitem{cherry2014comprehensive}
E.M. Cherry and J.K. Eaton.
\newblock A comprehensive model of magnetic particle motion during magnetic
  drug targeting.
\newblock {\em International Journal of Multiphase Flow}, 59:173--185, 2014.

\bibitem{fanelli2021magnetic}
C.~Fanelli, K.~Kaouri, T.N. Phillips, T.G. Myers, and F.~Font.
\newblock Magnetic nanodrug delivery in non-newtonian blood flows.
\newblock {\em arXiv preprint arXiv:2102.03911}, 2021.

\bibitem{zydney1987augmented}
A.L. Zydney and C.K. Colton.
\newblock Augmented solute transport in the shear flow of a concentrated
  solution.
\newblock {\em PCH PhysicoChemical Hydrodynamics}, 10(1):77--96, 1988.

\bibitem{nacev2010magnetic}
A.~Nacev, C.~Beni, O.~Bruno, and B.~Shapiro.
\newblock Magnetic nanoparticle transport within flowing blood and into
  surrounding tissue.
\newblock {\em Nanomedicine}, 5(9):1459--1466, 2010.

\bibitem{clausiusmossotti}
C.I. Mikkelsen.
\newblock {\em Magnetic Separation and Hydrodynamic Interactions in
  Microfluidic Systems}.
\newblock PhD thesis, Technical University of Denmark, 2005.

\bibitem{picchini2007sde}
U.~Picchini.
\newblock Sde toolbox: Simulation and estimation of stochastic differential
  equations with matlab.
\newblock 2007.

\bibitem{lim2011magnetophoresis}
J.~Lim, C.~Lanni, E.R. Evarts, F.~Lanni, R.D. Tilton, and S.A. Majetich.
\newblock Magnetophoresis of nanoparticles.
\newblock {\em ACS Nano}, 5(1):217--226, 2011.

\bibitem{muller2008high}
B.~M{\"u}ller, S.~Lang, M.~Dominietto, M.~Rudin, G.~Schulz, H.~Deyhle,
  M.~Germann, F.~Pfeiffer, C.~David, and T.~Weitkamp.
\newblock High-resolution tomographic imaging of microvessels.
\newblock {\em Developments in X-Ray Tomography VI}, 7078:70780B, 2008.

\bibitem{stucker1996capillary}
M.~St{\"u}cker, V.~Baier, T.~Reuther, K.~Hoffmann, K.~Kellam, and P.~Altmeyer.
\newblock Capillary blood cell velocity in human skin capillaries located
  perpendicularly to the skin surface: measured by a new laser doppler
  anemometer.
\newblock {\em Microvascular Research}, 52(2):188--192, 1996.

\bibitem{wang2016vessel}
L.~Wang, J.~Yuan, H.~Jiang, W.~Yan, H.~R. Cintr{\'o}n-Col{\'o}n, V.~L. Perez,
  D.~C. DeBuc, W.~J. Feuer, and J.~Wang.
\newblock Vessel sampling and blood flow velocity distribution with vessel
  diameter for characterizing the human bulbar conjunctival microvasculature.
\newblock {\em Eye \& Contact Lens}, 42(2):135, 2016.

\bibitem{klarhofer2001high}
M.~Klarh{\"o}fer, B.~Csapo, C.~S. Balassy, J.~C. Szeles, and E.~Moser.
\newblock High-resolution blood flow velocity measurements in the human finger.
\newblock {\em Magnetic Resonance in Medicine: An Official Journal of the
  International Society for Magnetic Resonance in Medicine}, 45(4):716--719,
  2001.

\bibitem{wilhelm2016analysis}
S.~Wilhelm, A.~Tavares, Q.~Dai, S.~Ohta, J.~Audet, H.~Dvorak, and W.~Chan.
\newblock Analysis of nanoparticle delivery to tumours.
\newblock {\em Nature Reviews Materials}, 1:16014, 2016.

\bibitem{price2018where}
P.M. Price, W.E. Mahmoud, A.A. Al-Ghamdi, and L.M. Bronstein.
\newblock Magnetic drug delivery: Where the field is going.
\newblock {\em Frontiers in Chemistry}, 6:619, 2018.

\bibitem{ao2018polydopamine}
L.~Ao, C.~Wu, K.~Liu, W.~Wang, L.~Fang, L.~Huang, and W.~Su.
\newblock Polydopamine-derivated hierarchical nanoplatforms for efficient
  dual-modal imaging-guided combination in vivo cancer therapy.
\newblock {\em ACS Applied Materials \& Interfaces}, 10(15):12544--12552, 2018.

\bibitem{wang2018multistage}
Y.~Wang, G.~Wei, X.~Huang, J.~Zhao, X.~Guo, and S.~Zhou.
\newblock Multistage targeting strategy using magnetic composite nanoparticles
  for synergism of photothermal therapy and chemotherapy.
\newblock {\em Small}, 14(12):1702994, 2018.

\bibitem{fraden2004handbook}
J.~Fraden.
\newblock {\em Handbook of modern sensors: physics, designs, and applications}.
\newblock Springer Science \& Business Media, 2004.

\bibitem{Mag_equ}
C.M. Oldenburg, S.E. Borglin, and G.J. Moridis.
\newblock Numerical simulation of ferrofluid flow for subsurface environmental
  engineering applications.
\newblock {\em Transport in Porous Media}, 38:319–344, 1999.

\end{thebibliography}
\bibliographystyle{unsrt}

\end{document}